\documentclass[aps,prd,reprint,onecolumn,superscriptaddress,longbibliography,nofootinbib,floatfix,showpacs]{revtex4-1}
\usepackage{epsfig}
\usepackage{amsmath,amssymb,amsfonts}
\usepackage{hyperref}
\usepackage{mathrsfs}
\usepackage{bbm}
\usepackage{slashed}
\usepackage{graphicx}
\usepackage{verbatim}

\usepackage[usenames]{color}

\definecolor{darkgreen}{rgb}{0.2,0.6,0}

\newcommand{\be}{\begin{equation}}
\newcommand{\ee}{\end{equation}}
\newcommand{\bw}{\begin{widetext}}
\newcommand{\ew}{\end{widetext}}
\newcommand{\bi}{\begin{itemize}}
\newcommand{\ei}{\end{itemize}}
\newcommand{\bea}{\begin{eqnarray}}
\newcommand{\eea}{\end{eqnarray}}
\newcommand{\bra}[1]{\langle\,#1\,|}          
\newcommand{\ket}[1]{|\,#1\,\rangle}          
\newcommand{\ud}{\mathrm{d}}

\newcommand{\LCm}{{\scriptscriptstyle -}} 
\newcommand{\LCp}{{\scriptscriptstyle +}}
\newcommand{\LCpm}{{\scriptscriptstyle \pm}}
\newcommand{\LCmp}{{\scriptscriptstyle \mp}}

\newcommand{\LCperp}{{\scriptscriptstyle \perp}}

\newcommand{\LCF}{{\scriptscriptstyle\rm LCF}}

\usepackage[T1]{fontenc} \usepackage[latin1]{inputenc}

\newcommand{\fsl}[1]{\slashed{#1}}

\begin{document}
%


\title{Trident pair production in plane waves: \\ Coherence, exchange, and spacetime inhomogeneity}

\author{Victor Dinu} 
\email{dinu@barutu.fizica.unibuc.ro}
\affiliation{Department of Physics, University of Bucharest, P.O.~Box MG-11, M\u agurele 077125, Romania}

\author{Greger Torgrimsson}
\email{greger.torgrimsson@uni-jena.de}
\affiliation{Theoretisch-Physikalisches Institut, Abbe Center of Photonics,
Friedrich-Schiller-Universit\"at Jena, Max-Wien-Platz 1, D-07743 Jena, Germany}
\affiliation{Helmholtz Institute Jena, Fr\"obelstieg 3, D-07743 Jena, Germany}

\begin{abstract}
We study the trident process in inhomogeneous plane wave background fields. We obtain compact analytical expressions for all terms in the probability, including the exchange part, for an arbitrarily shaped plane wave. We evaluate the probability numerically using complex deformation of lightfront time integrals and derive various analytical approximations. Our results provide insights into the importance of the one-step and exchange parts of the probability relative to the two-step process, and into the convergence to the locally constant field approximation.  
\end{abstract}

\maketitle


\section{Introduction}

The trident process in electromagnetic background fields, $e^-\to2e^-+e^+$, is a basic process in electron-laser collisions. It was first studied in detail in the $70$'s for constant crossed fields in~\cite{Ritus:1972nf,Baier} before the advent of high-intensity lasers (see~\cite{DiPiazza:2011tq} for a review of the development of such lasers). Trident, or at least part of it, was explored in a famous experiment at SLAC~\cite{Bamber:1999zt} in the late $90$'s, with a laser of still modest intensity. At that time, complete theoretical predictions for trident were still lacking. Since then, the available laser intensities have increased by a couple of orders of magnitude, but on the theory side the situation has been improved thanks to only a few publications~\cite{Hu:2010ye,Ilderton:2010wr,King:2013osa}. The importance of trident in high-intensity laser processes therefore motivates further investigations, see also~\cite{TridentPresExhilp}.

One part of trident is a two-step process where the initial electron emits a real photon that subsequently decays into an electron-positron pair, i.e. nonlinear Compton scattering followed by nonlinear Breit-Wheeler pair production, and its contribution is given by incoherently gluing together the two corresponding rates, see~\cite{King:2013osa,Baier:1998vh}. For a recent comprehensive analysis of the two-step process see~\cite{Blackburn:2017dpn}.
One of the main questions for trident is under what conditions this two-step process is a good approximation for the total trident process. 
In the experiment at SLAC~\cite{Bamber:1999zt}, corrections to the two-step part were estimated using the Weizs\"acker-Williams approximation and in that parameter regime were found to be negligible compared to the two-step process\footnote{Note that~\cite{Bamber:1999zt} uses different notation for the these processes.}. 
Further, the two-step process is expected to dominate for sufficiently large $a_0=eE/(m\omega)$, where $E$ is the field strength and $\omega$ the frequency of the field. This is important for particle-in-cell codes, see~\cite{Gonoskov:2014mda} for a review, where higher-order processes, e.g. cascades, are described as a sequence of first-order processes.  
We study here in more detail when corrections to the two-step process can be important.

The trident amplitude has two terms due to the exchange between the identical particles in the final state, and the absolute square of the amplitude gives a cross term which is referred to as the exchange part of the probability. In~\cite{Ritus:1972nf,Baier} the direct or non-exchange part of the probability was obtained from the imaginary part of a loop diagram, i.e. using the optical theorem. The direct part is expected to dominate, e.g. for $\chi=a_0b_0\gg1$, where $b_0=kp/m^2$ is the product of a characteristic wave vector of the laser, $k_\mu$, and the initial momentum of the incoming electron, $p_\mu$. Much less is known about the exchange part. Here we will calculate both the direct and the exchange parts in order to investigate in more detail when the latter can be important. 

As in~\cite{King:2013osa,Ilderton:2010wr}, we obtain the probability by calculating the amplitude using Volkov solutions for plane-wave backgrounds. In contrast to previous analytical studies~\cite{Ritus:1972nf,Baier,King:2013osa}, we derive compact expressions for nonconstant plane waves. While we recover the results in~\cite{Ritus:1972nf,Baier,King:2013osa} in the locally constant field limit, i.e. for $a_0\gg1$, our results also allow us to see how this limit is approached and to study regimes with $a_0\sim1$. 
In addition, our analytical results offer a useful alternative starting point for numerical investigations, compared to ones based on e.g. Monte Carlo integration~\cite{Hu:2010ye,TridentPresExhilp}, as well as insights into the analytical structure of the trident probability for inhomogeneous fields. 

Another difference compared to previous studies is that we use the lightfront formalism~\cite{Brodsky:1997de,Heinzl:2000ht}, which is particularly convenient when dealing with plane-wave background fields~\cite{Neville:1971uc,Ilderton:2013dba,Dinu:2013gaa}.  
The lightfront Hamiltonian has both a ``conventional'' term, given by $jA$, and an ``instantaneous'' term. These two terms suggest another split of the amplitude and the probability based on the lightfront Hamiltonian. As we will show, this lightfront separation is not the same as the standard one-step/two-step separation as in~\cite{Ritus:1972nf,Baier,King:2013osa} (though the total probability and rate are obviously the same for both separations). While we do not propose this lightfront split as something to replace the standard separation, we will show that it is convenient from an analytical and computational point of view.

The rest of this paper is organized as follows. In Sec.~\ref{Description of formalism} we describe the basic ingredients needed to calculate the trident probability with the lightfront formalism, and compare this approach with the standard one. In Sec.~\ref{Exact analytical results} we present compact analytical expressions for the trident probability in a general inhomogeneous plane wave. In Sec.~\ref{Two-step and one-step} we compare these expressions with the probabilities for nonlinear Compton scattering and Breit-Wheeler pair production, and in Sec.~\eqref{LCFsection} we consider the locally constant field (LCF) approximation and show how the standard one- and two-step terms can be obtained from our lightfront expressions. In particular, we expand in $1/a_0\ll1$ and recover literature results for the direct part. We also calculate the exchange terms in this limit, which are new results. In Sec.~\ref{Pulsed fields with large a0 and small chi} we go on to consider $\chi\ll1$ and $a_0\gg1$ but for non-constant background fields. Then in Sec.~\ref{Sauter pulse section} and Sec.~\ref{Monochromatic field section} we consider $\chi\ll1$ but $a_0\sim1$ and obtain simple analytical approximations for a pulse and a monochromatic field. In Sec.~\eqref{Numerical method} we explain how to numerically integrate our results from Sec.~\ref{Exact analytical results}. In Sec.~\ref{LCF general chi} we consider $a_0\gg1$ for up to moderately large $\chi$ and study the importance of exchange terms. In Sec.~\eqref{Convergence to the LCF limit} we consider a pulsed field and study the convergence to the LCF approximation. We conclude in Sec.~\eqref{Conclusions section}. In Appendix~\ref{Perturbative limit} we recover known results for the perturbative limit, and in Appendix~\ref{Comparing numerical and analytical} we show the agreement between our analytical approximations and numerical results.

We use units with $c=1$ and $\hbar=1$, and measure energies in terms of the electron mass such that $m=1$.

\section{Formalism}\label{Description of formalism}

We use lightfront coordinates defined for an arbitrary vector $v_\mu$ by $v^\LCpm=2v_\LCmp=v^0\pm v^3$ and $v^\LCperp=\{v^1,v^2\}$,
for coordinates $\bar{x}=\{x^\LCm,x^\LCperp\}$ and for momenta $\bar{p}=\{p_\LCm,p_\LCperp\}$. 
We consider an arbitrary pulsed plane-wave background field given by $f_{\mu\nu}=k_\mu a'_\nu-k_\nu a'_\mu$, where $k_\mu=k_\LCp\delta_\mu^\LCp$ is a null wave vector and $a_\LCperp(kx)$ is an arbitrary function. We absorb for convenience a factor of the electron charge into the definition of the background field, i.e. $ea_\mu\to a_\mu$.
Several of our results are conveniently expressed in terms of the Lorentz momentum of an electron in a plane-wave background, which is given by
\be\label{LorentzMomentum}
\pi_\mu=p_\mu-a_\mu+\frac{2ap-a^2}{2kp}k_\mu \;.
\ee
The initial state contains an electron with momentum $p_\mu$ and spin $\sigma$, and the final state contains two electrons with $p_{1,2}^\mu$ and $\sigma_{1,2}$ and a positron with $p_3^\mu$ and $\sigma_3$. 
The trident amplitude, $M$, is obtained from
\be
\langle0|b(p_1,\sigma_1)b(p_2,\sigma_2)d(p_3,\sigma_3)Ub^\dagger(p,\sigma)|0\rangle
=:\frac{1}{k_\LCp}\bar{\delta}(p_1+p_2+p_3-p)M \;,
\ee
where $U$ is the evolution operator and $\bar{\delta}(\dots)=(2\pi)^3\delta_{\LCm,\LCperp}(\dots)$. The mode operators are normalized according to
$\{b(q,r),\bar{b}(q',r')\}=\{d(q,r),\bar{d}(q',r')\}=2p_\LCm\bar{\delta}(q-q')\delta_{rr'}$,
and we use
$\ud\tilde{p}=\theta(p_\LCm)\ud p_\LCm\ud^2p_\LCperp/(2p_\LCm(2\pi)^3)$
to denote the Lorentz-invariant momentum measure.
The initial electron is described by a sharply peaked wave packet $f(p)$,
\be
|\text{in}\rangle=\int\!\ud\tilde{p}\;f(p)b^\dagger(p\sigma)|0\rangle \qquad \int\!\ud\tilde{p}\;|f|^2=1 \;,
\ee
where the last equation ensures that $\langle\text{in}|\text{in}\rangle=1$.
The total probability averaged over the initial spin is given by
\be\label{total-P}
\mathbb{P}=\frac{1}{4}\sum\limits_\text{all spin}\int\!\ud\tilde{p}_1\ud\tilde{p}_2\ud\tilde{p}_3\Big|\int\!\ud\tilde{p}\;f\frac{1}{k_\LCp}\bar{\delta}(p_1+p_2+p_3-p)M\Big|^2
=\frac{1}{4}\sum\limits_\text{all spin}\frac{1}{kp}\int\!\ud\tilde{p}_1\ud\tilde{p}_2\frac{\theta(kp_3)}{kp_3}|M|^2 \;,
\ee
where one factor of $1/2$ comes from averaging over the spin of the initial electron and another $1/2$ is due to having identical particles in the final state, and $\bar{p}_3=\bar{p}-\bar{p}_1-\bar{p}_2$.

The total amplitude can be written $M=M^{12}-M^{21}$, where $M^{21}$ is obtained from $M^{12}$ by swapping the identical particles, i.e. by replacing $p_1\leftrightarrow p_2$ and $\sigma_1\leftrightarrow \sigma_2$. This leads to a separation of $\mathbb{P}$ into a direct and an exchange part,
$|M|^2=|M^{12}|^2+|M^{21}|^2-2\text{Re }\bar{M}^{21}M^{12}$,
where the first two terms give the direct contribution and the third term gives the exchange contribution, i.e. 
\be\label{PdirExchDef}
\mathbb{P}_{\rm dir}=\frac{1}{4}\sum\limits_\text{all spin}\frac{1}{kp}\int\!\ud\tilde{p}_1\ud\tilde{p}_2\frac{\theta(kp_3)}{kp_3}|M^{12}|^2+(1\leftrightarrow2) 
\qquad
\mathbb{P}_{\rm ex}=-\frac{1}{2}\sum\limits_\text{all spin}\frac{1}{kp}\int\!\ud\tilde{p}_1\ud\tilde{p}_2\frac{\theta(kp_3)}{kp_3}\text{Re }\bar{M}^{21}M^{12} \;.
\ee
where the second term in $\mathbb{P}_{\rm dir}$ is obtained from the first by replacing $p_1\leftrightarrow p_2$ and $\sigma_1\leftrightarrow \sigma_2$.

\subsection{Lightfront quantization}

We have derived our results using two different approaches. In both approaches the plane-wave background field is taken into account exactly using Volkov solutions in the Furry picture. In the first approach we use the combination of the Hamiltonian-based lightfront formalism~\cite{Brodsky:1997de,Heinzl:2000ht} and plane-wave backgrounds~\cite{Neville:1971uc,Ilderton:2013dba,Dinu:2013gaa}. The evolution in lightfront time, $x^\LCp$, is determined by the lightfront Hamiltonian, and the interaction part of this Hamiltonian has three terms,
\be\label{totV}
H_{\rm int}=\frac{1}{2}\int\!\ud\bar{x}\; ejA+\frac{e^2}{2}j_\LCm\frac{1}{(i\partial_\LCm)^2}j_\LCm+e^2\bar{\Psi}\fsl{A}\frac{\gamma^\LCp}{4i\partial_\LCm}\fsl{A}\Psi \;,
\ee
where the photon field is given by
\be
A_\mu(x)=\int\!\ud\tilde{l}\; a_\mu e^{-ilx}+a^\dagger_\mu e^{ilx}
\qquad
[a_\mu(l),a_\nu(l')]=-2l_\LCm\bar{\delta}(l-l')L_{\mu\nu}
\qquad
L_{\mu\nu}=g_{\mu\nu}-\frac{k_\mu l_\nu+l_\mu k_\nu}{kl}  \;,
\ee
the current is $j^\mu=\bar{\Psi}\gamma^\mu\Psi$ and the spinor field is expressed in terms of Volkov solutions as
\be
\Psi(x)=\int\!\ud\tilde{p}\; Kub\varphi+\bar{K}vd^\dagger\varphi_\LCm
\qquad
\varphi=\exp\left\{-i\left(px+\int^{kx}\frac{2ap-a^2}{2kp}\right)\right\} 
\qquad
K=1+\frac{\slashed{k}\slashed{a}}{2kp} \;,
\ee
where $\varphi_\LCm=\varphi(-p)$ and $\bar{K}=1-\slashed{k}\slashed{a}/(2kp)$. Note that all the momenta in these mode expansions for $A_\mu$ and $\Psi$ are on-shell. In particular, in this formalism all photons are on-shell and so one cannot split trident into two parts with one having on-shell and the other off-shell intermediate photons.  
Only the first two terms in \eqref{totV}, $H_{\rm int}^{(1)}$ and $H_{\rm int}^{(2)}$, contribute to trident to lowest order. The first term is familiar from ordinary quantization and its contribution to the amplitude is given by a double $x^\LCp$-integral with $x^\LCp$-ordering,
\be
\frac{1}{k_\LCp}\bar{\delta}(p_1+p_2+p_3-p)M_2:=
-\langle 0|b(p_1)b(p_2)d(p_3)\int\!\ud x_2^\LCp\ud x_1^\LCp\theta(x_2^\LCp-x_1^\LCp)H_{\rm int}^{(1)}(x_2^\LCp)H_{\rm int}^{(1)}(x_1^\LCp)b^\dagger(p)|0\rangle \;.
\ee
The second interaction term, sometimes referred to as instantaneous~\cite{Brodsky:1997de,Zhao:2013cma}, only involves one $x^\LCp$-integral and is given by
\be
\frac{1}{k_\LCp}\bar{\delta}(p_1+p_2+p_3-p)M_1:=
\bra{0}b(p_1)b(p_2)d(p_3)(-i)\int\!\ud x^\LCp H_{\rm int}^{(2)}(x^\LCp)b^\dagger(p)\ket{0} \;.
\ee
After some calculation we find
\be\label{M1and2-12}
M_1^{12}=\frac{ie^2}{2kl^2}\underset{p_2}{\bar{u}}\slashed{k}\underset{p_3}{v}\underset{p_1}{\bar{u}}\slashed{k}\underset{p}{u}\int\!\ud\phi\;\underset{p_1}{\bar{\varphi}}\underset{p_2}{\bar{\varphi}}\underset{p_3}{\varphi_\LCm}\underset{p}{\varphi}
\qquad
M_2^{12}=-\frac{e^2}{4kl}\int\!\ud\phi_2\;L_{\mu\nu}\underbrace{\bar{u}\bar{K}\bar{\varphi}}_{p_2}\gamma^\mu\underbrace{\bar{K}v\varphi_\LCm}_{p_3}e^{-\frac{il_\LCp\phi_2}{k_\LCp}}
\int^{\phi_2}\!\ud\phi_1\;\underbrace{\bar{u}\bar{K}\bar{\varphi}}_{p_1}\gamma^\nu\underbrace{Ku\varphi}_{p} e^\frac{il_\LCp\phi_1}{k_\LCp} \;,
\ee
where $\phi=kx$, $\bar{l}=\bar{p}-\bar{p}_1$ and $l_\LCp=l_\LCperp^2/4l_\LCm$. To avoid clutter we put the arguments below the function, e.g. $\underbrace{\bar{u}\bar{K}\bar{\varphi}}_{p_1}=\bar{u}(p_1)\bar{K}(p_1)\bar{\varphi}(p_1)$.

\subsection{Relation to standard approach}

There are two main differences between the approach just described and the standard approach: The former is a Hamiltonian formalism in the ligthfront gauge. These two approaches should of course give the same results, and we will show this explicitly in this section.
In the non-Hamiltonian approach, the amplitude is given by (up to an irrelevant overall phase)
\be\label{S12}
\frac{1}{k_\LCp}\bar{\delta}(p_1+p_2+p_3-p)M^{12}
=e^2\int\ud^4x\ud^4y\; \underset{p_2}{\bar{\psi}(y)}\gamma^\nu\underset{p_3}{\psi_\LCm(y)}D_{\nu\mu}(y-x)\underset{p_1}{\bar{\psi}(x)}\gamma^\mu\underset{p}{\psi(x)} \;,
\ee
where 
$\psi=Ku\varphi$, $\psi_\LCm=\bar{K}v\varphi_\LCm$,
and $D_{\nu\mu}$ is the photon propagator 
\be
D_{\nu\mu}(y-x)=i\int\frac{\ud^4l}{(2\pi)^4}D_{\nu\mu}(l)\frac{e^{-il(y-x)}}{l^2+i\epsilon} \;.
\ee
In the Feynman gauge $D_{\nu\mu}=g_{\nu\mu}$ and in lightfront gauge $D_{\nu\mu}=L_{\nu\mu}$ (see~\cite{Mantovani:2016uxq} for a recent discussion of the photon propagator in the lightfront gauge). 
Performing the trivial integrals in~\eqref{S12} gives delta functions implying
$\bar{l}=\bar{p}-\bar{p}_1$. 

To see that the two gauges give the same result, consider the contributions from the $l_\mu$-terms in $L_{\nu\mu}$, which involve $\bar{\psi}(p_1)\slashed{l}\psi(p)$ and $\bar{\psi}(p_2)\slashed{l}\psi_\LCm(p_3)$. Consider first the part with $p$ and $p_1$. By writing $l_\mu=:\pi_\mu(p,\phi)-\pi_\mu(p_1,\phi)+c(\phi)k_\mu$ and using $K\slashed{p}=\slashed{\pi}K$, we find
$\bar{u}\bar{K}(p_1)\slashed{l}Ku(p)=\bar{u}(p_1)\slashed{k}u(p)\; c(\phi)$. Since the $\phi$-dependence of the exponential is given by the integral of $c(\phi)$, we hence find a total derivative that vanishes upon integrating over $\phi$. The same holds for the part with $p_2$ and $p_3$. 
Thus, $D_{\nu\mu}=g_{\nu\mu}$ and $D_{\nu\mu}=L_{\nu\mu}$ give the same result.

\begin{figure}
\centering
\includegraphics[scale=1,trim={{0.1\textwidth} {0.02\textwidth} {0.1\textwidth} 0},clip]{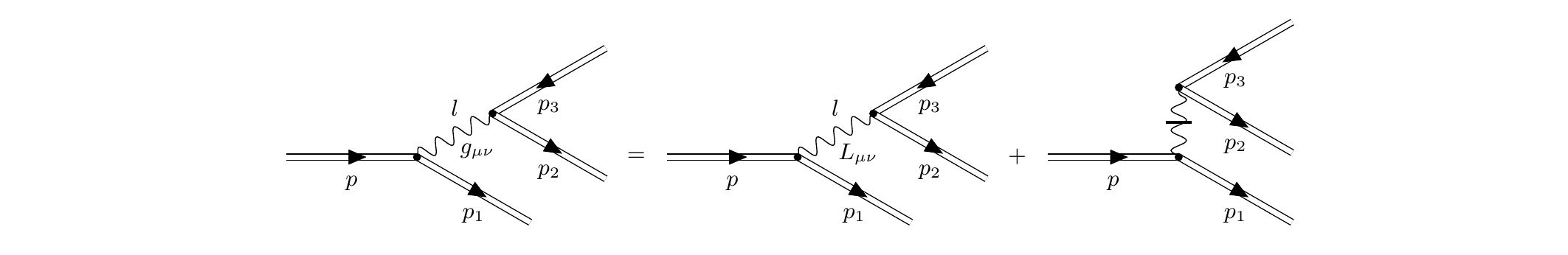}
\caption{This figure is an illustration of the lightfront separation of the trident amplitude. The diagram on the left represents the Feynman gauge (as in~\eqref{FeynmanM12}) and the two diagrams on the right represent the two lightfront terms in~\eqref{LFM12}. Note that these diagrams represent terms in the amplitude after performing the photon momentum integrals in the photon propagator, so the photon-momentum, $l_\mu$, in these diagrams is on-shell. This is why the third diagram appears. The line through the photon line in the third diagram stands for an ``instantaneous photon''~\cite{Brodsky:1997de}.}
\label{FeynmanDiagramgvsL}
\end{figure}

The next step is to reproduce the amplitude obtained in the previous section. By separating $L_{\mu\nu}$ into an on-shell and off-shell part, one finds (c.f.~\cite{Mantovani:2016uxq})
\be\label{Lsplit}
L_{\mu\nu}=L_{\mu\nu}\left(l_\LCp=\frac{l_\LCperp^2}{4l_\LCm}\right)-\frac{l^2}{kl^2}k_\mu k_\nu \;.
\ee
When performing the $l_\LCp$-integral in the propagator, the first term in~\eqref{Lsplit} gives a lightfront time-ordering step function $\theta(\phi_y-\phi_x)$, which one also finds in the Feynman gauge, see Appendix~\ref{Feynman gauge section}, while the second term in~\eqref{Lsplit} gives an ``instantaneous'' $\delta(\phi_y-\phi_x)$, which does not appear in the Feynman gauge. (See~\cite{Seipt:2012tn} for a similar separation of the fermion propagator.) We hence find 
\be\label{LFM12}
M^{12}=-i\pi\alpha\int\!\ud\phi_x\ud\phi_y\left[\frac{1}{kl}L_{\nu\mu}\theta(\phi_y-\phi_x)-\frac{2i}{kl^2}k_\nu k_\mu\delta(\phi_y-\phi_x)\right]e^{-\frac{il_\LCp}{k_\LCp}(\phi_y-\phi_x)}\overbrace{\underset{p_2}{\bar{\psi}}\gamma^\nu\underset{p_3}{\psi_\LCm}}^{\phi_y}\overbrace{\underset{p_1}{\bar{\psi}}\gamma^\mu\underset{p}{\psi}}^{\phi_x}=:M_2^{12}+M_1^{12} \;,
\ee
where the photon momentum is now on-shell, i.e. $l_\LCp=l_\LCperp^2/(4l_\LCm)$. Because of~\eqref{Lsplit}, \eqref{LFM12} contains two terms that are equivalent to one term in the Feynman gauge (compare~\eqref{LFM12} with~\eqref{FeynmanM12}), as illustrated in Fig.~\ref{FeynmanDiagramgvsL}.
In~\eqref{LFM12} we have the same $M_1$ and $M_2$ as in~\eqref{M1and2-12} (up to an irrelevant overall phase), which we obtained using the lightfront Hamiltonian formalism. In particular, the step-function term in~\eqref{LFM12} agrees with the term that comes from $H_{\rm int}^{(1)}H_{\rm int}^{(1)}$, and the delta function term~\eqref{LFM12} corresponds to the instantaneous part of the lightfront Hamiltonian, $H_{\rm int}^{(2)}$. 
While~\eqref{FeynmanM12} and~\eqref{LFM12} give the same amplitude, the lightfront separation of $M^{12}$ as in~\eqref{LFM12} can be convenient because of the vector structure of the two individual components, $M_1^{12}$ and $M_2^{12}$; we have for example $L_{\mu\nu}k^\nu=L_{\mu\nu}l^\nu=0$ and $\slashed{k}K=\slashed{k}$. 

Thus, in addition to the direct/exchange separation of the amplitude, $M=M^{12}-M^{21}$, we now separate each of these into a ``three-point vertex'' and a ``lightfront instantaneous'' part, $M^{12}=M_1^{12}+M_2^{12}$. This leads to six different contributions to the probability, 
\be\label{PdirectDef}
\{\mathbb{P}_{\rm dir}^{11},\mathbb{P}_{\rm dir}^{12},\mathbb{P}_{\rm dir}^{22}\}:=\frac{1}{4}\sum\limits_\text{all spin}\frac{1}{kp}\int\!\ud\tilde{p}_1\ud\tilde{p}_2\frac{\theta(kp_3)}{kp_3}\{|M_1^{12}|^2,2\text{Re }\bar{M}_1^{12}M_2^{12},|M_2^{12}|^2\}+(1\leftrightarrow2) \;,
\ee
\be\label{PexchangeDef}
\{\mathbb{P}_{\rm ex}^{11},\mathbb{P}_{\rm ex}^{12},\mathbb{P}_{\rm ex}^{22}\}:=-\frac{1}{2}\sum\limits_\text{all spin}\frac{1}{kp}\int\!\ud\tilde{p}_1\ud\tilde{p}_2\frac{\theta(kp_3)}{kp_3}\text{Re}\{\bar{M}_1^{21}M_1^{12},\bar{M}_1^{21}M_2^{12}+(1\leftrightarrow2),\bar{M}_2^{21}M_2^{12}\} \;.
\ee
The sum of these six terms gives the total probability. Given that $M_1$ comes from a term in the lightfront Hamiltonian that is usually referred to as instantaneous, it might be tempting to associate the corresponding terms in the probability with what one would usually call one-step terms. However, as we will show, all the terms listed in~\eqref{PdirectDef} and~\eqref{PexchangeDef} contribute to the standard one-step term, and in fact ``most'' of the standard one-step term comes from $\mathbb{P}^{22}$ (which also contains the entire standard two-step term).

\section{Exact analytical results}\label{Exact analytical results}

Rather than to keep the dependence on all the particle parameters, we content ourselves with the dependence on the longitudinal momenta, which we denote as $s_i=kp_i/kp$. 
The integrals over the transverse momentum components of the final electrons, $p_{1,\LCperp}$ and $p_{2,\LCperp}$, are Gaussian. We perform these integrals as well as the traces, coming from the spin summations, analytically for arbitrary pulsed plane-wave backgrounds. The resulting expressions contain integrals over the longitudinal momenta plus integrals over lightfront time. We use the following notation for the probability density
\be
\mathbb{P}=\int_0^1\!\ud s_1\ud s_2\theta(s_3)\mathbb{P}(s) \;,
\ee
where $s_3=1-s_1-s_2$. To make the symmetries in the following expressions manifest, we introduce $s_0=1$ in the appropriate places.

For the terms coming from the square of the ``instantaneous'' part of the amplitude, $|M_1|^2$, we find
\be\label{P11GenFin}
\{\mathbb{P}_{\rm dir}^{11}(s),\mathbb{P}_{\rm ex}^{11}(s)\}=\frac{\alpha^2}{\pi^2}\int\!\ud\phi_{12}\left\{\frac{1}{q_1^4}+\frac{1}{q_2^4},-\frac{1}{q_1^2q_2^2}\right\}\frac{-s_0s_1s_2s_3}{(\theta_{21}+i\epsilon)^2}\exp\left\{\frac{i}{2b_0}(r_1+r_2)\Theta_{21}\right\} \;,
\ee
where $b_0=kp$, $r_1:=\frac{1}{s_1}-\frac{1}{s_0}$, $r_2:=\frac{1}{s_2}+\frac{1}{s_3}$, $q_i:=1-s_i$, $\ud\phi_{12}:=\ud\phi_1\ud\phi_2$, $\theta_{ij}:=\phi_i-\phi_j$, $\Theta_{ij}:=\theta_{ij}M_{ij}^2$, and $M$ is an ``effective mass'' defined via the lightfront time average of the field as~\cite{Kibble:1975vz} 
\be
M_{ij}^2:=\langle\pi\rangle_{ij}^2=1+\langle{\bf a}^2\rangle_{ij}-\langle{\bf a}\rangle_{ij}^2
\qquad
\langle F\rangle_{ij}:=\frac{1}{\theta_{ij}}\int_{\phi_j}^{\phi_i}\!\ud\phi\; F(\phi) \;,
\ee
where $\pi_\mu$ is given by~\eqref{LorentzMomentum}.
These shorthand notations are very convenient for the more complicated terms below. To make the transverse momentum integrals converge, we have introduced an infinitesimal convergence factor $\epsilon>0$ via $\phi_2\to\phi_2+i\epsilon/2$ and $\phi_1\to\phi_1-i\epsilon/2$, which after performing the momentum integrals gives an $i\epsilon$-prescription for how to integrate around the singularity at $\theta_{21}=0$.

Next we consider the cross terms between the ``lightfront instantaneous'' and the ``three-point vertex'' parts of the amplitude.
To make the momentum integrals for $\mathbb{P}^{12}$ converge, we similarly take $\phi_2\to\phi_2+i\epsilon/2$ and $\phi_{1,3}\to\phi_{1,3}-i\epsilon/2$. We find
\be\label{P12dirGenFin}
\mathbb{P}_{\rm dir}^{12}(s)=\text{Re }i\frac{\alpha^2}{4\pi^2b_0}\int\!\ud\phi_{123}\theta(\theta_{31})
\frac{(s_0+s_1)(s_2-s_3)D_{12}}{q_1^3(\theta_{21}+i\epsilon)(\theta_{23}+i\epsilon)}\exp\left\{\frac{i}{2b_0}\left[r_1\Theta_{21}+r_2\Theta_{23}\right]\right\}+(s_1\leftrightarrow s_2)
\ee
\be\label{P12ExchGenFin}
\mathbb{P}_{\rm ex}^{12}(s)=\text{Re}\frac{-i\alpha^2}{4\pi^2b_0}\int\!\ud\phi_{123}\theta(\theta_{31})
\frac{q_1^2+[s_0s_2-s_1s_3]D_{12}}{q_1q_2^2(\theta_{21}+i\epsilon)(\theta_{23}+i\epsilon)}\exp\left\{\frac{i}{2b_0}\left[r_1\Theta_{21}+r_2\Theta_{23}\right]\right\}+(s_1\leftrightarrow s_2) \;,
\ee
where $D_{12}={\bf \Delta}_{12}\!\cdot\!{\bf \Delta}_{32}$ and
\be
{\bf\Delta}_{ij}:={\bf a}(\phi_i)-\langle{\bf a}\rangle_{ij} \;.
\ee
It turns out that the background field only enters the pre-exponential factors and the exponentials via ${\bf\Delta}_{ij}$ and $M^2_{ij}$, respectively; no other field combinations are needed. 

For the direct part of the square of the ``three-point vertex'' amplitude we find
\be\label{P22dirGenFin}
\begin{split}
	\mathbb{P}_{\rm dir}^{22}(s)=-\frac{\alpha^2}{8\pi^2b_0^2}&\int\!\ud\phi_{1234}\frac{\theta(\theta_{31})\theta(\theta_{42})}{q_1^2\theta_{21}\theta_{43}}\exp\left\{\frac{i}{2b_0}\left(r_1\Theta_{21}+r_2\Theta_{43}\right)\right\}\left\{\frac{\kappa_{01}\kappa_{23}}{4}W_{1234}+W_{1324}+W_{1423}\right.
	\\
	&\left.+\left[\frac{\kappa_{01}}{2}\left(\frac{2ib_0}{r_1\theta_{21}}+1+D_1\right)-1\right]\left[\frac{\kappa_{23}}{2}\left(\frac{2ib_0}{r_2\theta_{43}}+1+D_2\right)+1\right]-D_1D_2\right\}+(s_1\leftrightarrow s_2) \;,
\end{split}
\ee
where 
$\kappa_{ij}=(s_i/s_j)+(s_j/s_i)$, $D_1={\bf\Delta}_{12}\!\cdot\!{\bf\Delta}_{21}$, $D_2={\bf\Delta}_{34}\!\cdot\!{\bf\Delta}_{43}$, and 
\be\label{Wijdef}
W_{ijkl}:=(\mathbf{w}_{i}\!\times\!\mathbf{w}_{j})\!\cdot\!(\mathbf{w}_{k}\!\times\!\mathbf{w}_{l}) =(\mathbf{w}_{i}\!\cdot\!\mathbf{w}_{k})(\mathbf{w}_{j}\!\cdot\!\mathbf{w}_{l})-(\mathbf{w}_{i}\!\cdot\!\mathbf{w}_{l})(\mathbf{w}_{j}\!\cdot\!\mathbf{w}_{k}) \;,
\ee
where ${\bf w}_1={\bf\Delta}_{12}$, ${\bf w}_2={\bf\Delta}_{21}$, ${\bf w}_3={\bf\Delta}_{34}$ and ${\bf w}_4={\bf\Delta}_{43}$. Note that $W_{ijkl}=0$ for linear polarization. 

\begin{figure}
\centering
\includegraphics[scale=1,trim={{0.1\textwidth} {0.01\textwidth} {0.1\textwidth} 0},clip]{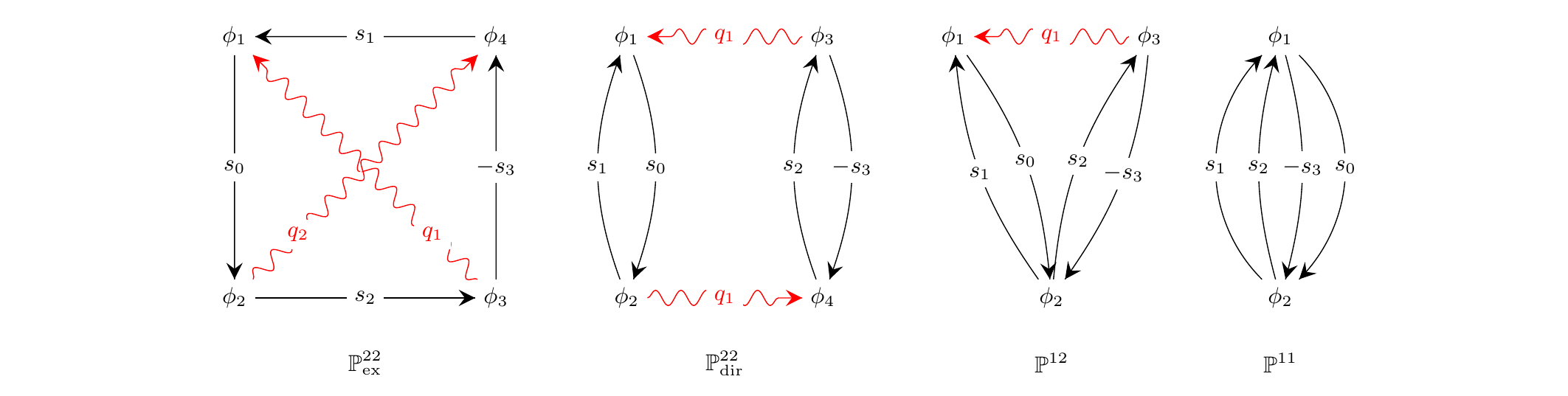}
\caption{These are diagrams on the probability level and show the momentum flow between the vertices. Black, solid lines represent fermions and red, wavy lines represent photons.}
\label{ProbabilityDiagrams}
\end{figure}

These terms have some symmetries that can be understood in terms of the probability diagrams in Fig.~\ref{ProbabilityDiagrams}. Let $\mathbb{P}_{\rm dir}^{22}(s)=\mathbb{P}_{\rm dir}^{22}(s_1;s_2)+\mathbb{P}_{\rm dir}^{22}(s_2;s_1)$, where $\mathbb{P}_{\rm dir}^{22}(s_1;s_2)$ is everything before the $(s_1\leftrightarrow s_2)$-term in~\eqref{P22dirGenFin}. $\mathbb{P}_{\rm dir}^{22}(s_1;s_2)$ is invariant under $\{s_0\leftrightarrow-s_1\}$ as well as $\{s_2\leftrightarrow s_3\}$, which correspond to reflection of one or the other of the two fermion loops in the $\mathbb{P}_{\rm dir}^{22}$ diagram in Fig.~\ref{ProbabilityDiagrams}. The integrand of $\mathbb{P}_{\rm dir}^{22}(s_1;s_2)$, apart from $\theta(\theta_{31})\theta(\theta_{42})$, is also invariant under a total reflection and a $180$-degree rotation, i.e. under $\{\phi_1\leftrightarrow\phi_3,\phi_2\leftrightarrow\phi_4,s_0\leftrightarrow-s_3,s_1\leftrightarrow s_2,q_1\to-q_1\}$ and $\{\phi_1\leftrightarrow\phi_4,\phi_2\leftrightarrow\phi_3,s_0\leftrightarrow s_2,s_1\leftrightarrow-s_3\}$. For $\mathbb{P}_{\rm dir}^{12}(s_1;s_2)$, which is everything before the $(s_1\leftrightarrow s_2)$-term in~\eqref{P12dirGenFin}, we find similar symmetries, except that $\mathbb{P}_{\rm dir}^{12}(s_1;s_2)$ changes sign under $\{s_0\leftrightarrow-s_1\}$ as well as $\{s_2\leftrightarrow s_3\}$, which implies that after integrating over the longitudinal momenta we find $\mathbb{P}_{\rm dir}^{12}=0$. Under a reflection of the $\mathbb{P}^{12}$ diagram in Fig.~\ref{ProbabilityDiagrams}, i.e. under $\{\phi_1\leftrightarrow\phi_3,s_1\leftrightarrow s_2,s_0\leftrightarrow-s_3,q_1\to-q_1,q_2\to-q_2\}$, $\mathbb{P}_{\rm dir}^{12}(s_1;s_2)$ is invariant except for $\theta(\theta_{31})\to-\theta(-\theta_{31})$. The corresponding exchange term, $\mathbb{P}_{\rm exch}^{12}(s_1;s_2)$, is invariant under the same reflection with the same change of the step function. $\mathbb{P}_{\rm exch}^{12}(s_1;s_2)$ is also invariant under $\{s_0\leftrightarrow-s_1,s_2\leftrightarrow s_3\}$, but not under $\{s_0\leftrightarrow-s_1\}$ and $\{s_2\leftrightarrow s_3\}$ separately. The direct and exchange terms of $\mathbb{P}^{11}$ have similar symmetries. Only $\mathbb{P}_{\rm dir}^{12}=0$, so for the total/integrated probability we have five nonzero terms.       

These types of symmetry considerations are particularly useful for the exchange part of the square of the ``three-point vertex'' amplitude, which is the most complicated term. After some lengthy calculation we eventually find
\be
\label{P22exchs0}
\begin{split}
\mathbb{P}_{\rm ex}^{22}(s)=\frac{\alpha^2}{16\pi^2b_0^2}&\int\!\ud\phi_{1234}\frac{\theta(\theta_{42})\theta(\theta_{31})}{s_0s_1s_2s_3d_0} 
\Bigg\{F_0+f_0-\frac{2ib_0}{d_0}(f_1+z_1)+\left[\frac{2b_0}{d_0}\right]^2z_2\Bigg\} \\
&\exp\left\{\frac{i}{2b_0}\frac{q_1q_2}{s_0s_1s_2s_3d_0}\left(\theta_{41}\theta_{23}\left[\frac{\Theta_{41}}{s_1}+\frac{\Theta_{23}}{s_2}\right]+\theta_{43}\theta_{21}\left[\frac{\Theta_{43}}{s_3}-\frac{\Theta_{21}}{s_0}\right]+\theta_{31}\theta_{42}\left[\frac{\Theta_{31}}{q_2}-\frac{\Theta_{42}}{q_1}\right]\right)\right\} \;,
\end{split}
\ee
where the various quantities are defined as follows. All the field dependent parts of the prefactor are expressed in terms of the four combinations ${\bf d}_i$, $i=1,..,4$, where
\be
{\bf d}_1=\frac{1}{d_0}\left(\frac{\theta_{23}}{s_2}\frac{\theta_{41}}{s_1}{\bf\Delta}_{14}+\frac{\theta_{21}}{s_0}\frac{\theta_{43}}{s_3}{\bf\Delta}_{12}+\frac{\theta_{42}\theta_{23}}{s_2s_3}[{\bf\Delta}_{24}-{\bf\Delta}_{23}]\right) 
\qquad
d_0=\frac{\theta_{23}}{s_2}\frac{\theta_{41}}{s_1}+\frac{\theta_{21}}{s_0}\frac{\theta_{43}}{s_3} \;.
\ee
Here we can really see the benefit of introducing $s_0$: Before setting $s_0=1$ and $s_3=1-s_1-s_2$, we can obtain the other three ${\bf d}_i$ by cyclic permutations of ${\bf d}_1$ as follows. 
We first define a permutation according to
\be\label{PermutationDefinition}
{\sf P}[\mathcal{F}]:=\mathcal{F}(\phi_1\to\phi_2\to\phi_3\to\phi_4\to\phi_1,s_1\to-s_0\to s_2\to s_3\to s_1) \;,
\ee
which corresponds to a $90$-degree counterclockwise rotation of the $\mathbb{P}_{\rm ex}^{22}$ diagram in Fig.~\ref{ProbabilityDiagrams}, followed by a change of sign of all the $s_i$.
Under this permutation we also have ${\sf P}\{q_1,q_2\}=\{q_2,-q_1\}$\footnote{This follows either directly from the $\mathbb{P}_{\rm ex}^{22}$ diagram in Fig.~\ref{ProbabilityDiagrams}, or from~\eqref{PermutationDefinition} by writing $q_1=(s_0-s_1+s_2+s_3)/2$ and $q_2=(s_0-s_2+s_1+s_3)/2$).}. 
After performing the permutation we can again set $s_0=1$ and $s_3=1-s_1-s_2$.
For example, ${\sf P}[d_0]=-d_0$. 
With this permutation we find 
\be
{\bf d}_2={\sf P}[{\bf d}_1] \qquad {\bf d}_3={\sf P}[{\bf d}_2] \qquad {\bf d}_4={\sf P}[{\bf d}_3] \qquad {\bf d}_1={\sf P}[{\bf d}_4] \;.
\ee
Now, the first term in the prefator of~\eqref{P22exchs0} is quartic in the field,
\be
F_0
=(1+{\sf P})\kappa_{03}[({\bf d}_1\!\cdot\!{\bf d}_3)({\bf d}_2\!\cdot\!{\bf d}_4)+(\mathbf{d}_{1}\!\times\! \mathbf{d}_{3})\!\cdot\!(\mathbf{d}_{2}\!\times\! \mathbf{d}_{4})] \;,
\ee
where the second term in the square brackets vanishes for linear polarization (c.f.~\eqref{Wijdef}) and can also be written as
\be
(\mathbf{d}_{1}\!\cdot\!\mathbf{d}_{2})(\mathbf{d}_{3}\!\cdot\!\mathbf{d}_{4})-(\mathbf{d}_{1}\!\cdot\! \mathbf{d}_{4})(\mathbf{d}_{3}\!\cdot\!\mathbf{d}_{2}) \;.
\ee
The next two terms in~\eqref{P22exchs0} are quadratic in the field,
\be
\begin{split}
f_0&=(1+{\sf P})\frac{1}{s_0s_1s_2s_3}(s_1q_2{\bf d}_1-s_2q_1{\bf d}_2)\!\cdot\!(s_2q_2{\bf d}_3-s_1q_1{\bf d}_4)\;, \\
f_1&=-(1+{\sf P}+{\sf P}^2+{\sf P}^3)\kappa_{03}\frac{\theta_{21}}{s_0}{\bf d}_2\!\cdot\!{\bf d}_1+(1+{\sf P})(\kappa_{03}-\kappa_{12})\frac{\theta_{42}}{q_1}{\bf d}_4\!\cdot\!{\bf d}_2 \;,
\end{split}
\ee 
and the last two terms in~\eqref{P22exchs0} do not contain the field,
\be
z_1
=(1+{\sf P}+{\sf P}^2+{\sf P}^3)\frac{-q_1^2}{s_0s_1q_2}\left(3+\frac{s_2s_3}{s_0s_1}\right)\phi_1 
\qquad
z_2
=(1+{\sf P})\kappa_{03}\left(\frac{\theta_{43}}{s_3}\frac{\theta_{21}}{s_0}+\frac{\theta_{31}}{q_2}\frac{\theta_{42}}{q_1}\right)\;.
\ee
The permutation in~\eqref{PermutationDefinition} is a symmetry of the integrand in the following sense: We have ${\sf P}\{F_0,f_0,f_1,z_1,z_2\}=\{F_0,f_0,f_1,z_1,z_2\}$ and for the exponential in~\eqref{P22exchs0} we have ${\sf P}[\exp(i...)]=\exp(-i...)$. (Note that the exponent in~\eqref{P22exchs0} reduces to that in~\eqref{P12dirGenFin} and~\eqref{P12ExchGenFin} for $\phi_4=\phi_2$.) Let $\mathcal{I}$ be the integrand in~\eqref{P22exchs0} without the step functions, then ${\sf P}[\mathcal{I}]=-\mathcal{I}^*$, but the step functions are different for each of the 4 permutations. We also have a reflection symmetry: ${\sf R}[\mathcal{F}]=\mathcal{F}(s_1\leftrightarrow s_2,\phi_1\leftrightarrow\phi_2,\phi_3\leftrightarrow\phi_4)$. We have ${\sf R}[\mathcal{I}]=\mathcal{I}^*$ and ${\sf R}$ leaves the step functions unchanged, so the integral~\eqref{P22exchs0} is invariant under ${\sf R}$.

\section{Two-step and one-step}\label{Two-step and one-step}

In this section we will show how the above results are related to the two individual processes of nonlinear Compton scattering followed by nonlinear Breit-Wheeler pair production.
We first express the corresponding probabilities in a similar way as the expressions for trident above. We assume here a linearly polarized field and an emitted photon with momentum $l_\mu$ and polarization vector $\epsilon_\LCperp=(\cos\vartheta,\sin\vartheta)$, $\epsilon_\LCp=\epsilon_\LCperp l_\LCperp/(2l_\LCm)$, $\epsilon_\LCm=0$ (so $l\epsilon=k\epsilon=0$). Averaging and summing over the spins of the incoming and outgoing electrons, respectively, gives us 
\be\label{PNLC}
\mathbb{P}_{\rm C}=\frac{i\alpha}{4\pi b_0}\int_0^1\!\ud s_1\int\!\ud\phi_1\ud\phi_2\frac{1}{\theta_{12}}\left\{\frac{\kappa_{01}}{2}\left(\frac{2i b_0}{r_1\theta_{21}}+1+D_1\right)-1+\cos(2\vartheta)D_1\right\}\exp\left\{\frac{i r_1}{2b_0}\Theta_{21}\right\} \;,
\ee
where the $i\epsilon$-prescription, $\theta_{21}+i\epsilon$, is left implicit. For pair production by the emitted photon we find after summing over the spins of the electron-positron pair
\be\label{PBW}
\mathbb{P}_{\rm BW}=\frac{i\alpha}{2\pi b_0}\int_0^{q_1}\!\ud s_2\int\!\ud\phi_3\ud\phi_4\frac{1}{q_1^2\theta_{43}}\left\{\frac{\kappa_{23}}{2}\left(\frac{2ib_0}{r_2\theta_{43}}+1+D_2\right)+1-\cos(2\vartheta)D_2\right\}\exp\left\{\frac{ir_2}{2b_0}\Theta_{43}\right\} \;,
\ee 
where, again, the singularity is avoided with $\theta_{43}+i\epsilon$. 
While~\eqref{PNLC} and~\eqref{PBW} are already suitable for comparing with our expressions above for trident, by performing partial integration for the terms proportional to $1/\theta^2$ we obtain simpler expressions,
\be\label{PNLCafterPI}
\mathbb{P}_{\rm C}=-\frac{i\alpha}{4\pi b_0}\int_0^1\!\ud s_1\int\!\ud\phi_1\ud\phi_2\frac{1}{\theta_{12}}\left\{1+\frac{\kappa_{01}}{4}[{\bf a}(\phi_2)-{\bf a}(\phi_1)]^2-\cos(2\vartheta)D_1\right\}\exp\left\{\frac{i r_1}{2b_0}\Theta_{21}\right\} \;,
\ee
and
\be\label{PBWafterPI}
\mathbb{P}_{\rm BW}=\frac{i\alpha}{2\pi b_0}\int_0^{q_1}\!\ud s_2\int\!\ud\phi_3\ud\phi_4\frac{1}{q_1^2\theta_{43}}\left\{1-\frac{\kappa_{23}}{4}[{\bf a}(\phi_4)-{\bf a}(\phi_3)]^2-\cos(2\vartheta)D_2\right\}\exp\left\{\frac{ir_2}{2b_0}\Theta_{43}\right\} \;.
\ee 
By summing~\eqref{PNLCafterPI} over two orthogonal photon polarization vectors, we recover eq.~(3) in~\cite{Dinu:2013hsd}. The reason it might not be convenient to perform similar partial integration for e.g.~\eqref{P22dirGenFin}, is that there are step functions in~\eqref{P22dirGenFin} that would then generate non-vanishing boundary terms.

For a constant field, these two expressions reduce to eq.~(22) and and~(26) in~\cite{King:2013zw} (up to some overall constants due to the difference between probabilities and rates, and ordinary and lightfront volume factors).
Without worrying too much about the formation length for the non-constant fields we consider, we follow the procedure for constant fields~\cite{King:2013osa} and glue together the probabilities for nonlinear Compton scattering and Breit-Wheeler pair production and compare the result to the trident probability. By adding the probabilities~\eqref{PNLC} and~\eqref{PBW} for parallel and perpendicular polarization, i.e. $\mathbb{P}_{\rm C}\mathbb{P}_{\rm BW}(\vartheta=0)+\mathbb{P}_{\rm C}\mathbb{P}_{\rm BW}(\vartheta=\pi/2)$, and symmetrizing with respect to $s_1\leftrightarrow s_2$ we find an expression that is identical to our $\mathbb{P}_{\rm dir}^{22}$ except for the step functions in~\eqref{P22dirGenFin}. 

In the product approach one would also include a step function associated with causality. We are therefore lead to inserting $\theta(\sigma_{43}-\sigma_{21})$, where $\sigma_{ij}=(\phi_i+\phi_j)/2$. While this step function only restricts the averages of the $\phi$-variables associated with photon emission and pair production, in the LCF regime we recover the results in~\cite{King:2013osa}, which we will show in the next section.  
So, to separate the one-step and the two-step parts of~\eqref{P22dirGenFin} we rewrite the step functions there as
\be\label{StepsForStepsSep}
\theta(\theta_{42})\theta(\theta_{31})=\theta(\sigma_{43}-\sigma_{21})\left\{1-\theta\left(\frac{|\theta_{43}-\theta_{21}|}{2}-[\sigma_{43}-\sigma_{21}]\right)\right\} \;,
\ee 
and separate~\eqref{P22dirGenFin} as $\mathbb{P}_{\rm dir}^{22}=\mathbb{P}_{\rm dir}^{22\to2}+\mathbb{P}_{\rm dir}^{22\to1}$, where $\mathbb{P}_{\rm dir}^{22\to2}$ and $\mathbb{P}_{\rm dir}^{22\to1}$ are obtained by replacing $\theta(\theta_{42})\theta(\theta_{31})$ in~\eqref{P22dirGenFin} with the first and second term in~\eqref{StepsForStepsSep}, respectively. Thus, we separate the total probability as $\mathbb{P}=\mathbb{P}_2+\mathbb{P}_1$, where
\be\label{PintermsofR}
\begin{split}
\mathbb{P}_2(s)&:=\int\!\ud\sigma_{21}\ud\sigma_{43}\theta(\sigma_{43}-\sigma_{21})R_2(\sigma_{21},\sigma_{43},s):=\mathbb{P}_{\rm dir}^{22\to2}(s)
\\
\mathbb{P}_1(s)&:=\int\!\ud\phi\; R_1(\phi,s):=(\mathbb{P}^{11}+\mathbb{P}^{12}+\mathbb{P}_{\rm ex}^{22}+\mathbb{P}_{\rm dir}^{22\to1})(s) \;.
\end{split}
\ee
We will refer to $R_{1,2}$ as rates. All the six lightfront terms contribute to $R_1(s)$, though $\mathbb{P}_{\rm dir}^{12}$ gives zero contribution after integrating over the longitudinal momenta. 
The two-step rate can be obtained from the probabilities of nonlinear Compton and Breit-Wheeler as
\be\label{R2intermsofRCandRBW}
R_2=\sum_{\vartheta=0,\frac{\pi}{2}}R_{\rm C}(\sigma_{21})R_{\rm BW}(\sigma_{43}) \qquad 
\int\!\ud\sigma_{21}R_{\rm C}(\sigma_{21}):=\eqref{PNLC} \qquad
\int\!\ud\sigma_{43}R_{\rm BW}(\sigma_{43}):=\eqref{PBW} \;.
\ee 
The argument of $R_1(\phi)$ is $\phi=(\phi_1+\phi_2+\phi_3+\phi_4)/4$ with $\phi_4=\phi_2$ for $\mathbb{P}_{12}$ and $\{\phi_4=\phi_2,\phi_3=\phi_1\}$ for $\mathbb{P}_{11}$. 
Because of the scaling of these rates, it is natural to consider instead 
\be\label{propertimeRates}
\mathcal{R}_1:=b_0R_1 \qquad \mathcal{R}_2:=b_0^2R_2 \;,
\ee
which can be seen as the rates corresponding to the propertime $\tau$, as in a plane wave it is simply related to lightfront time via $\phi=kx=kp\tau=b_0\tau$.
While we have motivated the definition of $\mathbb{P}_2$ with its relation to the product of $\mathbb{P}_{\rm C}$ and $\mathbb{P}_{\rm BW}$, the separation $\mathbb{P}=\mathbb{P}_2+\mathbb{P}_1$ might still seem ambiguous (see also the discussion in~\citep{MorozovNarozhnyi}). However, we will provide further motivation for the separation~\eqref{PintermsofR} by making an expansion in $1/a_0$. In particular, the two leading orders agree with literature results for constant fields.

\section{$a_0\gg1$ and the locally constant field approximation}
\label{LCFsection}

As mentioned, $\mathbb{P}^{22}$ should not be confused with the standard two-step obtained by gluing together the rates of nonlinear Compton and Breit-Wheeler as described e.g. in~\cite{King:2013osa}. Indeed, $\mathbb{P}^{22}$ contributes to both $\mathbb{P}_2$ and $\mathbb{P}_1$ in~\eqref{PintermsofR} and we will show in this section that it is $\mathbb{P}_2$ and $\mathbb{P}_1$ that to leading order correspond to the standard two-step and one-step parts, respectively.   
To do so we consider a linearly polarized field $a(\phi)=a_0 f(\phi)$, where the maximum of $f$ and $f'$ are around unity,  
with $a_0\gg1$ and make an expansion in $1/a_0$. As is well known, this limit takes us to the LCF approximation, for which there are analytical results in the literature to compare with. However, since we do not treat the background as constant from the start, our approach also allows us to obtain corrections to the leading order and we can avoid (very large) volume factors.
We expand the probability in $1/a_0$ as
\be
\mathbb{P}=a_0^2 P_2+a_0 P_1+P_0+\dots
\ee
where, as we will demonstrate below, $P_2$ and $P_1$ correspond to a two-step and a one-step term, respectively. Each term in this expansion is a nontrivial function of $\chi$, which is treated as independent of $a_0$ in this expansion. Since $a_0=E/\omega$ the expansion in $a_0$ is a derivative expansion. 
The higher orders are usually not considered in the literature, but can be obtained with this approach. In order to provide further motivation for the separation~\eqref{PintermsofR} it is important to note that $\mathbb{P}_2=a_0^2 P_2+\mathcal{O}(a_0^0)$, so only $\mathbb{P}_2$ contributes to $a_0^2 P_2$ and only $\mathbb{P}_1$ contributes to $a_0 P_1$. In other words, the next-to-leading-order correction to $\mathbb{P}_2$ does not mix with the leading order of $\mathbb{P}_1$.

\subsection{Two-step}

We begin with the two-step part, $\mathbb{P}_2=a_0^2 P_2+\mathcal{O}(a_0^0)$.
We change variables to $\sigma_{43}$, $\theta_{43}$, $\sigma_{21}$ and $\theta_{21}$. 
In order to expand in $1/a_0$ we rescale $\theta_{21}\to\theta_{21}/a_0$ and $\theta_{43}\to\theta_{43}/a_0$. To leading order, 
the entire $\theta$ integral can be performed and expressed in terms of Airy functions. We find (note that $a_0/\chi=1/b_0$)
\be\label{TwoStepAiry}
a_0^2 P_2(s)=-\frac{a_0^2\alpha^2}{\chi^2 q_1^2}\int\ud\sigma_{43}\ud\sigma_{21}\theta(\sigma_{43}-\sigma_{21})\left\{\left({\rm Ai}_1(\xi_1)+\kappa_{01}\frac{{\rm Ai}'(\xi_1)}{\xi_1}\right)\left({\rm Ai}_1(\xi_2)-\kappa_{23}\frac{{\rm Ai}'(\xi_2)}{\xi_2}\right)+\frac{{\rm Ai}'(\xi_1)}{\xi_1}\frac{{\rm Ai}'(\xi_2)}{\xi_2}\right\} \;,
\ee
where the arguments are given by $\xi_1=[r_1/\chi(\sigma_{21})]^\frac{2}{3}$ and $\xi_2=[r_2/\chi(\sigma_{43})]^\frac{2}{3}$,
where $\chi(\sigma)=\chi|f'(\sigma)|$ is the local value of $\chi$,
and the Airy integral is defined as in~\cite{King:2013osa},
\be
{\rm Ai}_1(\xi):=\int_\xi^\infty\!\ud v\,{\rm Ai}(v) \;.
\ee
$a_0^2 P_2$ in~\eqref{TwoStepAiry} is the natural generalization of e.g. eq.~(16) and~(17) in~\cite{King:2013osa} to non-constant fields. 
In~\cite{King:2013osa} the trident probability was calculated for a constant field, and the term quadratic in the volume $\Delta\phi$ was shown to be equal to what one obtains by gluing together the rates for nonlinear Compton and Breit-Wheeler. In~\cite{King:2013osa} it was also explained how this product approach is generalized to the LCF approximation. Our~\eqref{TwoStepAiry} is in fact equal to the LCF expression given in~\cite{King:2013osa}. To see this, we rewrite~\eqref{TwoStepAiry} as  
\be
a_0^2 P_2(s)=\int\ud\sigma_{43}\ud\sigma_{21}\theta(\sigma_{43}-\sigma_{21})\frac{1}{2}\sum_{\lambda=\pm1}\frac{-\alpha}{b_0}\left({\rm Ai}_1(\xi_1)+[\lambda+\kappa_{01}]\frac{{\rm Ai}'(\xi_1)}{\xi_1}\right)\frac{\alpha}{b_0 q_1^2}\left({\rm Ai}_1(\xi_2)+[\lambda-\kappa_{23}]\frac{{\rm Ai}'(\xi_2)}{\xi_2}\right) \;,
\ee
where the first and second factors correspond to nonlinear Compton and Breit-Wheeler, respectively, and the sum is over the photon polarization; these two factors are equal to eq.~(20) in~\cite{King:2013osa} but evaluated at two different $\phi$, as in eq.~(33) in~\cite{King:2013osa}. 
This agreement is of course not a surprise given that we have already shown that $\mathbb{P}_2$ can be expressed as~\eqref{R2intermsofRCandRBW} for non-constant fields and without expanding in $1/a_0$. 
Note that $b_0^2\eqref{TwoStepAiry}$ only depends on the field via $\chi(\sigma_{21})$ and $\chi(\sigma_{43})$, which is why it is natural in the LCF regime to consider the proper-time rates as defined in~\eqref{propertimeRates}.

Here we are mostly interested in regions of parameter space where one has important contributions from one-step terms, to which we now turn. 

\subsection{One-step}

Next we consider $\mathbb{P}_1$ in~\eqref{PintermsofR}. All the six lightfront terms in \eqref{PdirectDef} and~\eqref{PexchangeDef} contribute to $\mathbb{P}_1$ ($\mathbb{P}_{\rm dir}^{22}$ via $\mathbb{P}_{\rm dir}^{22\to1}$) and all terms in~\eqref{PintermsofR} are $\mathcal{O}(a_0)$. Together these $\mathcal{O}(a_0)$ terms give what is usually referred to as the one-step part of the probability.  

For $\mathbb{P}_{\rm dir}^{22\to1}$ we change variables according to
\be\label{phiitophivarphithetaeta}
\sigma_{21}=\phi-\frac{\varphi}{2} \qquad \sigma_{43}=\phi+\frac{\varphi}{2} \qquad
\theta_{21}=\theta-\frac{\eta}{2} \qquad \theta_{43}=\theta+\frac{\eta}{2} \;,
\ee
which in terms of the original variables are given by
\be
\phi=\frac{1}{4}(\phi_4+\phi_3+\phi_2+\phi_1) \qquad \varphi=\frac{1}{2}(\phi_4+\phi_3-[\phi_2+\phi_1]) \qquad \theta=\frac{1}{2}(\phi_4-\phi_3+\phi_2-\phi_1) \qquad \eta=\phi_4-\phi_3-[\phi_2-\phi_1] \;.
\ee
As for the two-step term we rescale $\theta\to\theta/a_0$ and $\eta\to\eta/a_0$. As the second term in~\eqref{StepsForStepsSep} shows, this also forces $\varphi$ to be small, so we also rescale $\varphi\to\varphi/a_0$. To leading order, i.e. $\mathcal{O}(a_0)$, the $\varphi$ integrand is constant and we find
\be\label{P22to1}
\begin{split}
\mathbb{P}_{\rm dir}^{22\to1}(s)=&\frac{\alpha^2a_0}{8\pi^2\chi^2}\text{Re}\int\!\ud\phi\ud\theta_{21}\ud\theta_{43}\theta(\theta_{43}-\theta_{21})\frac{\theta_{43}-\theta_{21}}{q_1^2\theta_{21}\theta_{43}}\exp\left\{\frac{i}{2\chi}\left(r_1\Theta_{21}+r_2\Theta_{43}\right)\right\}
\\
&\left\{\left[\frac{\kappa_{01}}{2}\left(\frac{2i\chi}{r_1\theta_{21}}+1+D_1\right)-1\right]\left[\frac{\kappa_{23}}{2}\left(\frac{2i\chi}{r_2\theta_{43}}+1+D_2\right)+1\right]-D_1D_2
\right\}+(s_1\leftrightarrow s_2) \;,
\end{split}
\ee
where here $D_1=-(\theta_{21}f'(\phi)/2)^2$, $D_2=-(\theta_{43}f'(\phi)/2)^2$, $M_{21}=1+\theta_{21}^2f'(\phi)^2/12$ and $M_{43}=1+\theta_{43}^2f'(\phi)^2/12$. The leading order of $\mathbb{P}^{11}$, $\mathbb{P}^{12}$ and $\mathbb{P}_{\rm ex}^{22}$ can be obtained in a similar way.

In the previous subsection we showed that $\mathbb{P}_2$ agrees to leading order with literature results for the two-step term for arbitrary values of $\chi$. However, while there are exact analytical expression also for the one-step terms, see~\cite{Baier,Ritus:1972nf,Baier:1998vh}, these are not easy to compare with, so we will instead consider $\chi\ll1$ and show that our $\mathcal{O}(a_0)$ terms agree with the results in~\cite{Baier,Ritus:1972nf,Baier:1998vh} in this regime. 

\subsection{Numerical spectrum}

\begin{figure}
\includegraphics[width=0.32\textwidth]{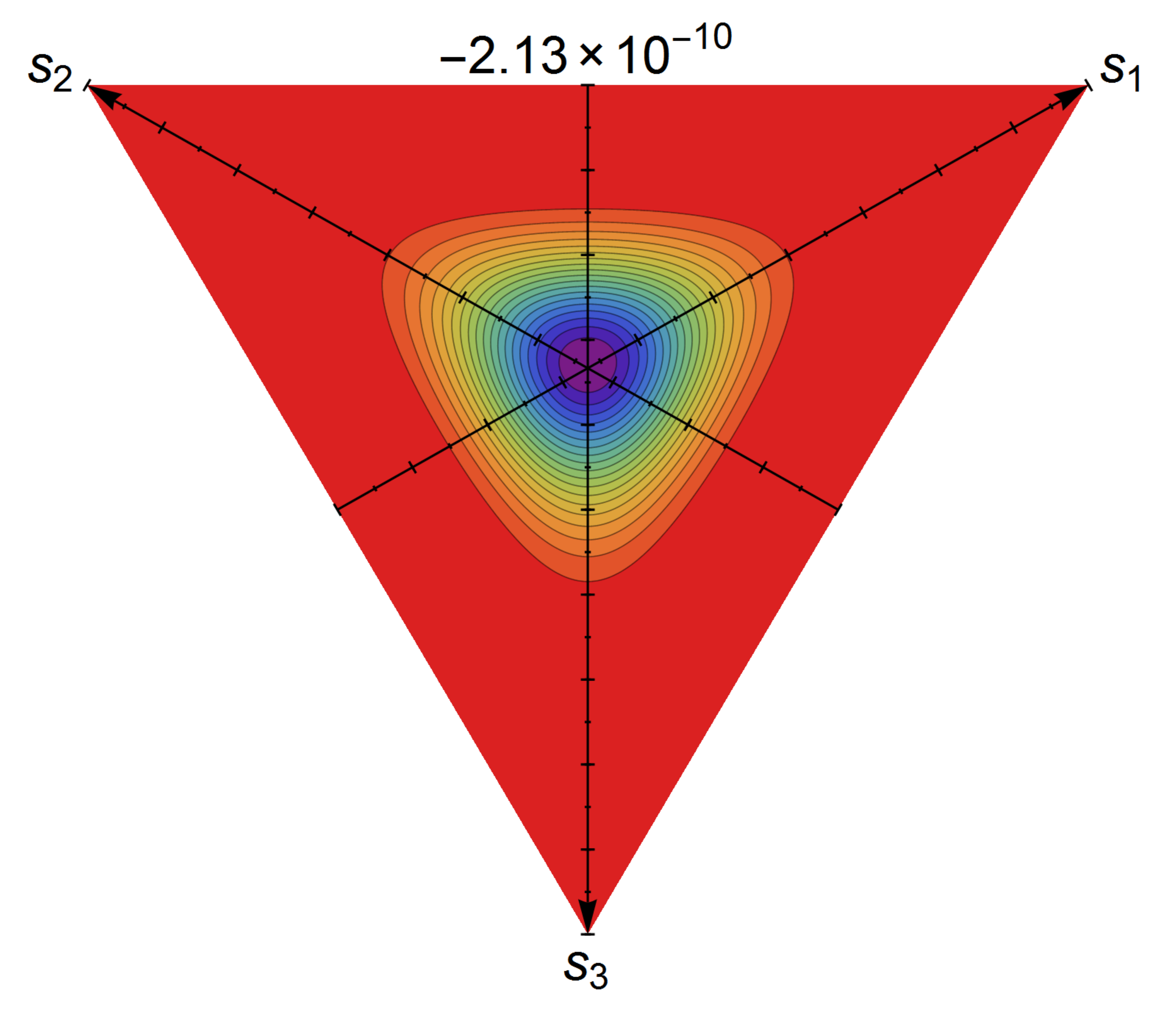}
\includegraphics[width=0.32\textwidth]{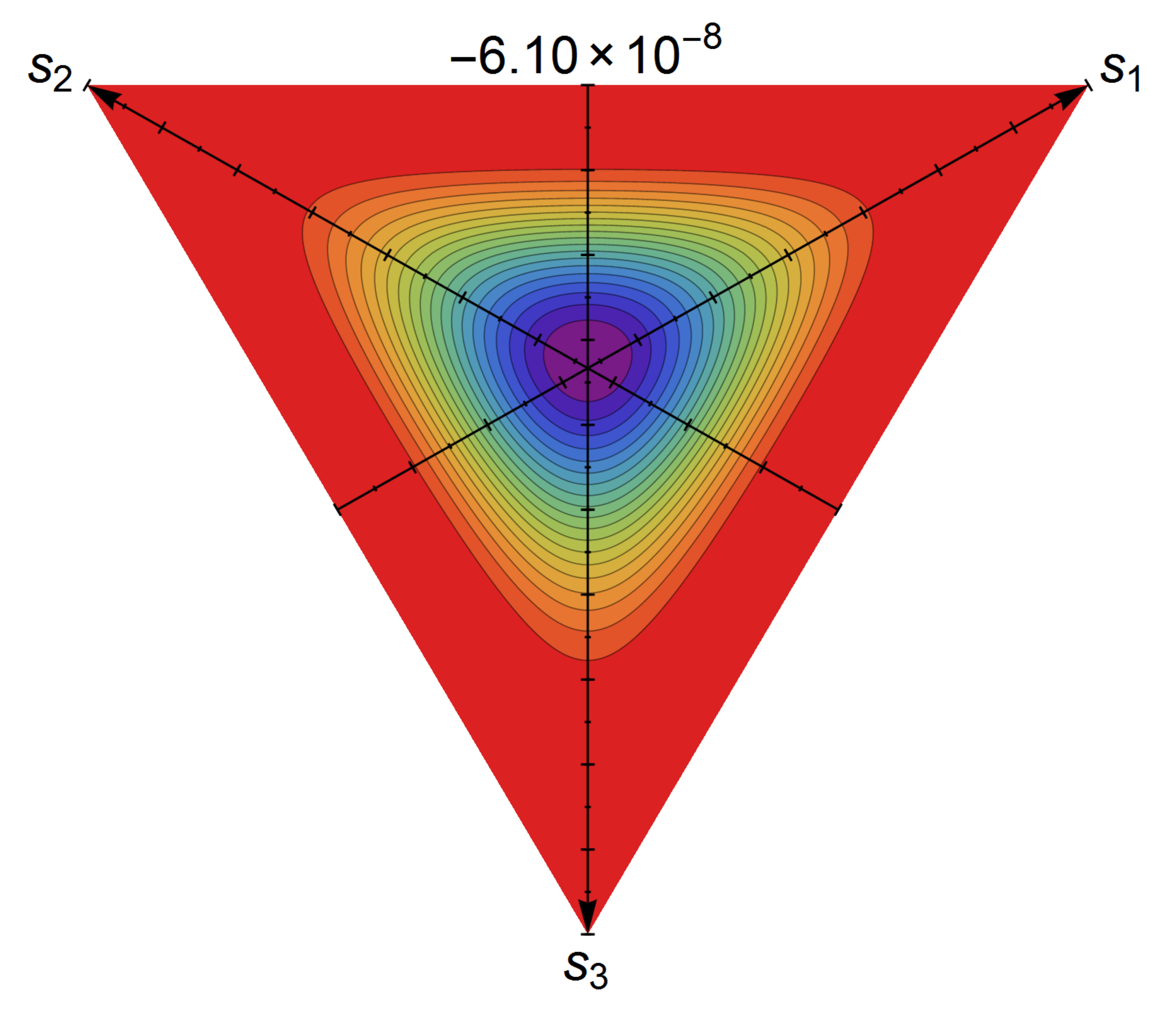}
\includegraphics[width=0.32\textwidth]{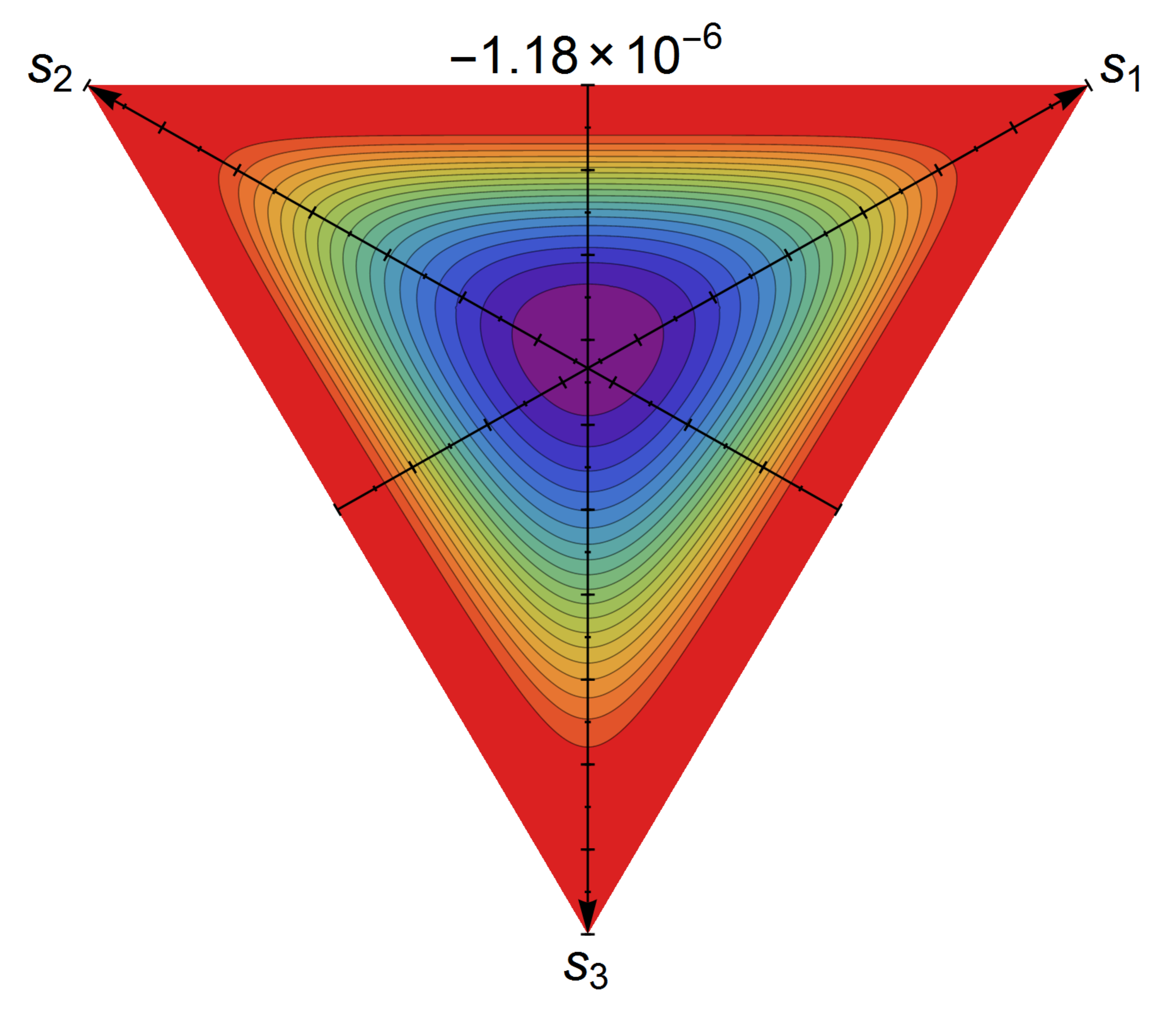}\\
\includegraphics[width=0.32\textwidth]{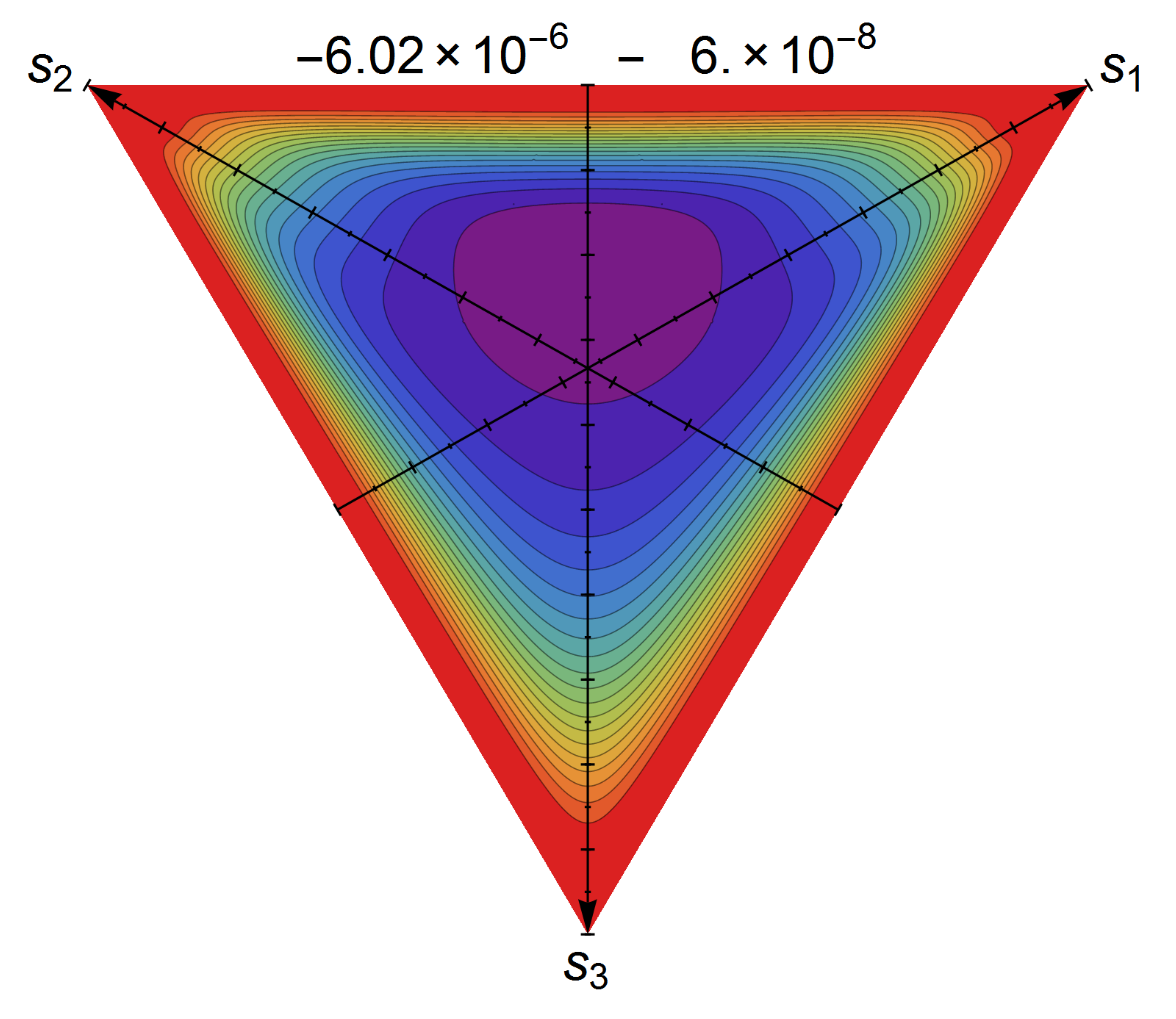}
\includegraphics[width=0.32\textwidth]{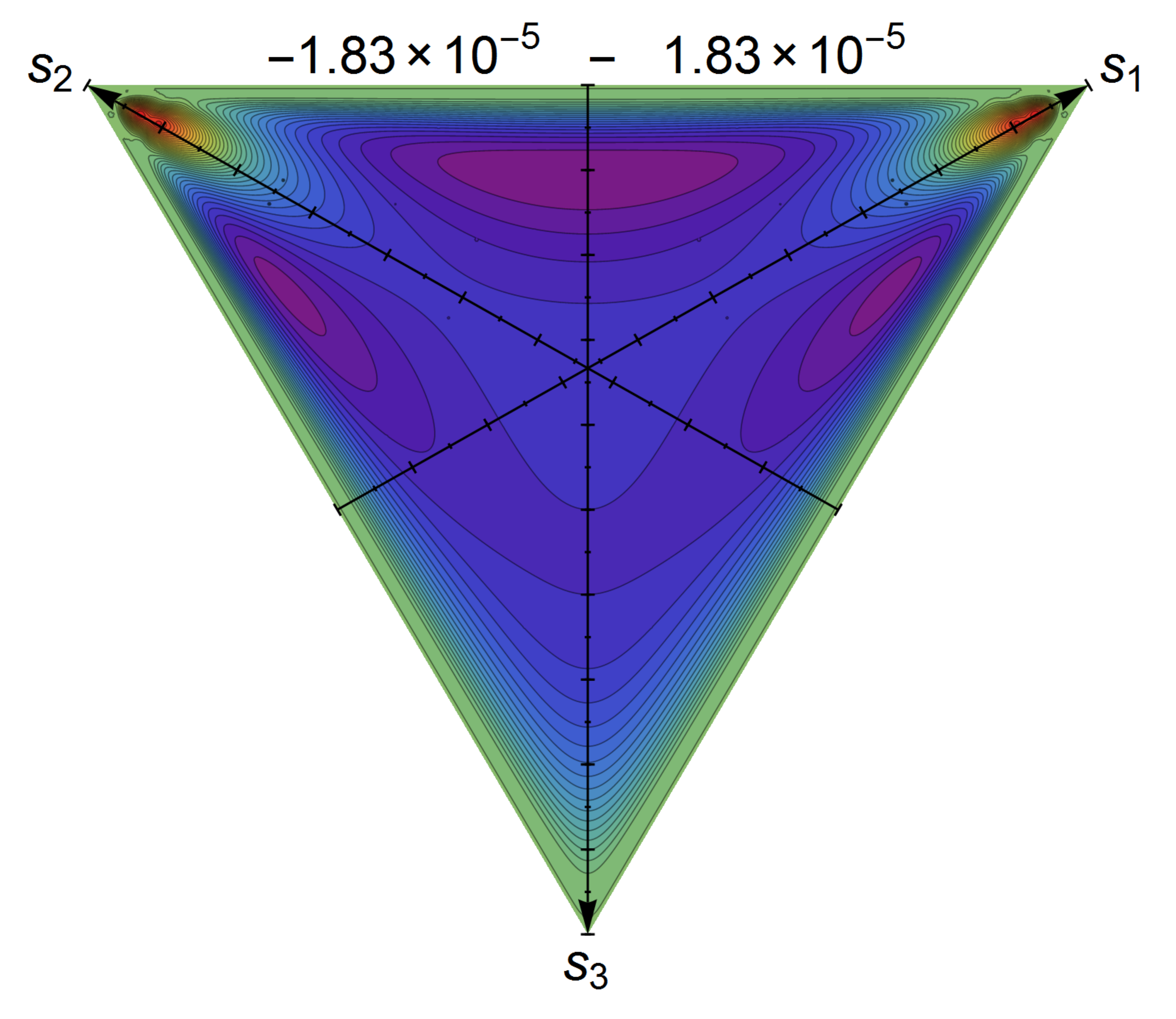}
\includegraphics[width=0.32\textwidth]{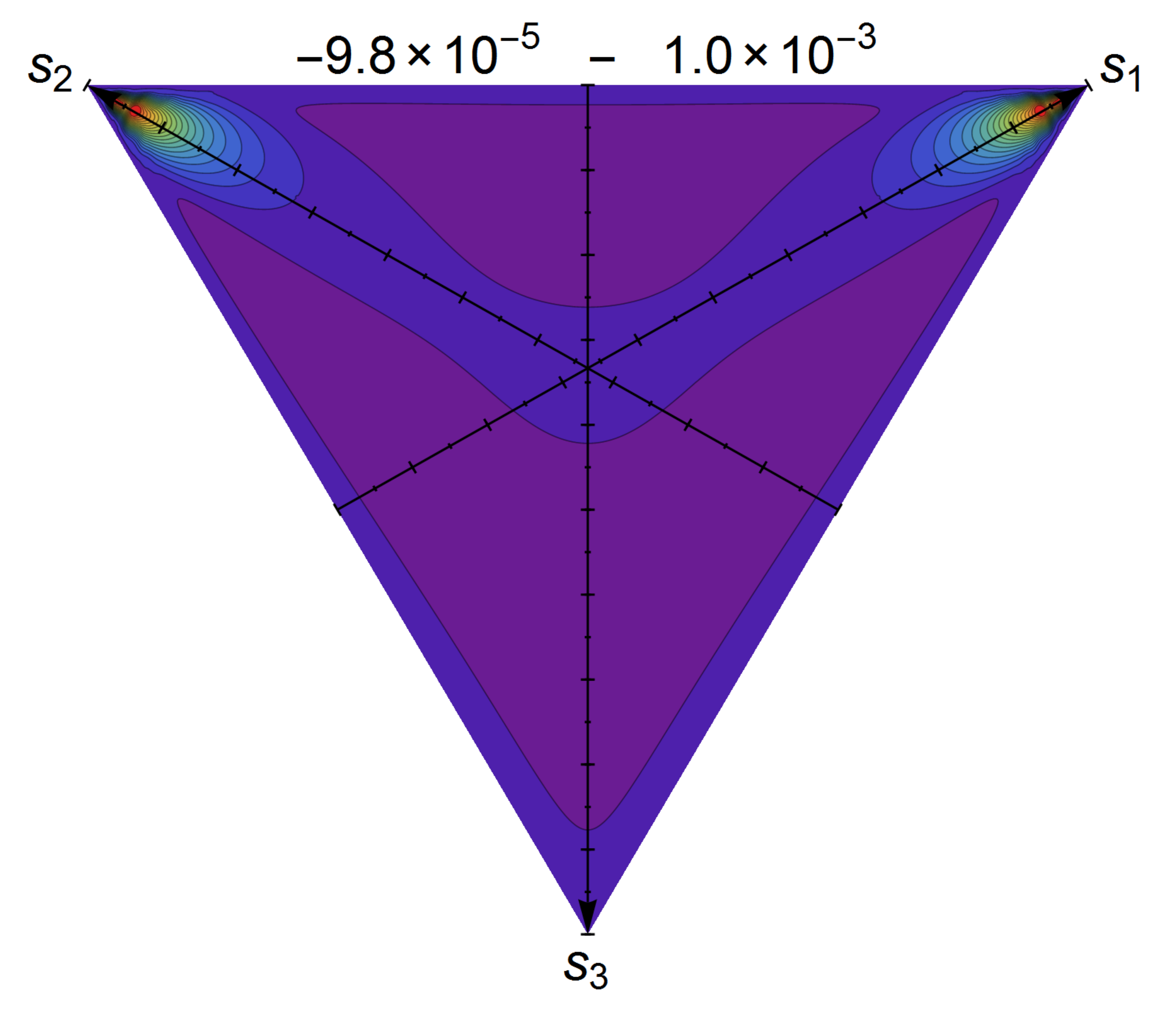}
\caption{These plots show the one-step part of the probability in the LCF approximation, $\mathcal{R}_1^\LCF$ with $\mathcal{R}_1$ defined in~\eqref{PintermsofR} and~\eqref{propertimeRates}, as a function of the longitudinal momenta $s_i$. From top left to bottom right we have $\chi(\phi)=1/2,1,2,4,8,16$. The values on top of each plot give the maximum/minimum value, i.e. $\text{min}<\mathcal{R}_1<\text{max}$, so for $\chi=1/2$ we have $-2.13*10^{-10}<\mathcal{R}_1<0$, and $-9.8*10^{-5}<\mathcal{R}_1<1.0*10^{-3}$ for $\chi=16$. The difference between two neighboring contours is $5\%$ of the larger of $|\text{max}|$ and $|\text{min}|$, and purple and red correspond, respectively, to values close to min and max.}
\label{oneStepTriangles}
\end{figure}
\begin{figure}
\includegraphics[width=0.32\textwidth]{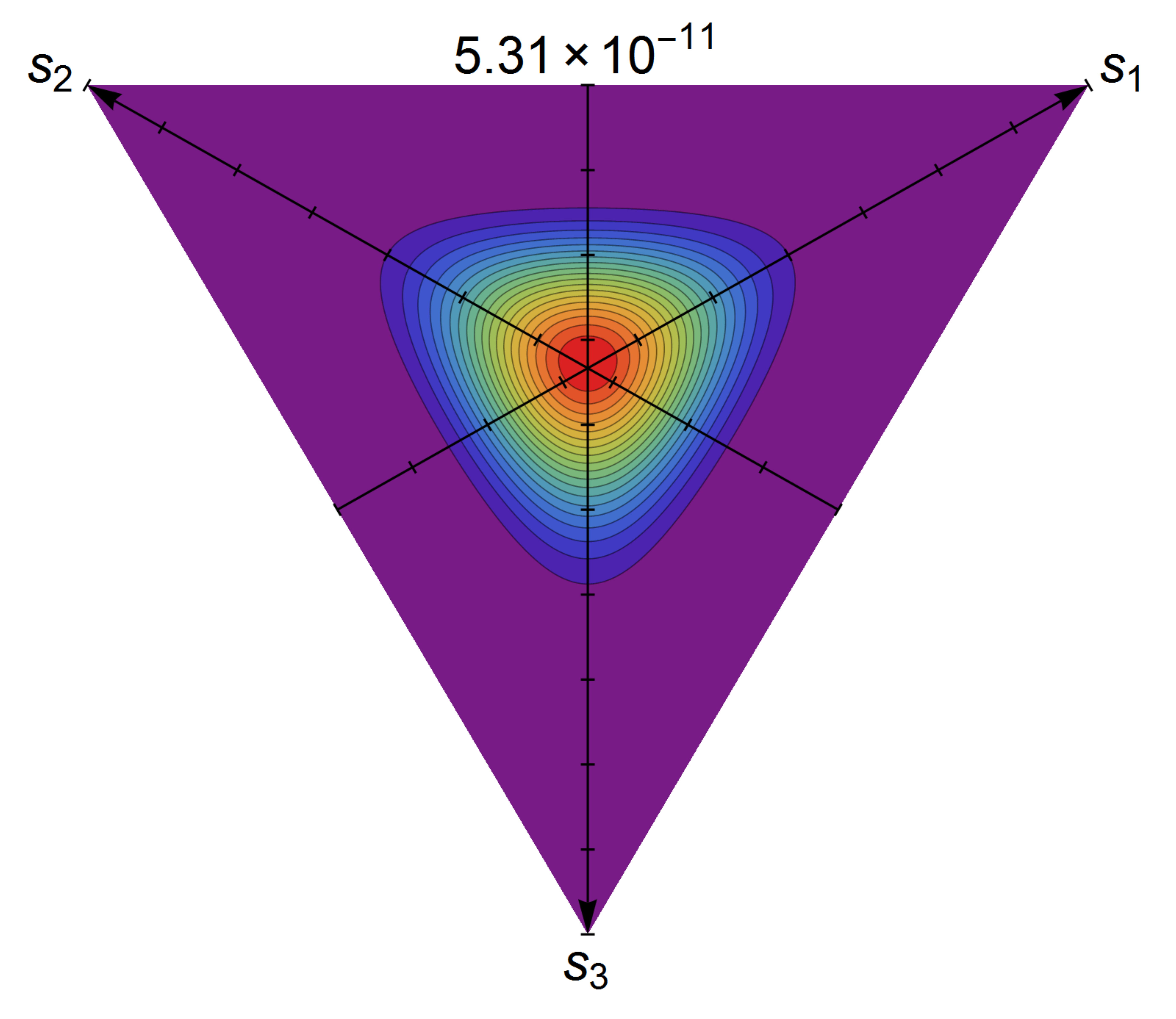}
\includegraphics[width=0.32\textwidth]{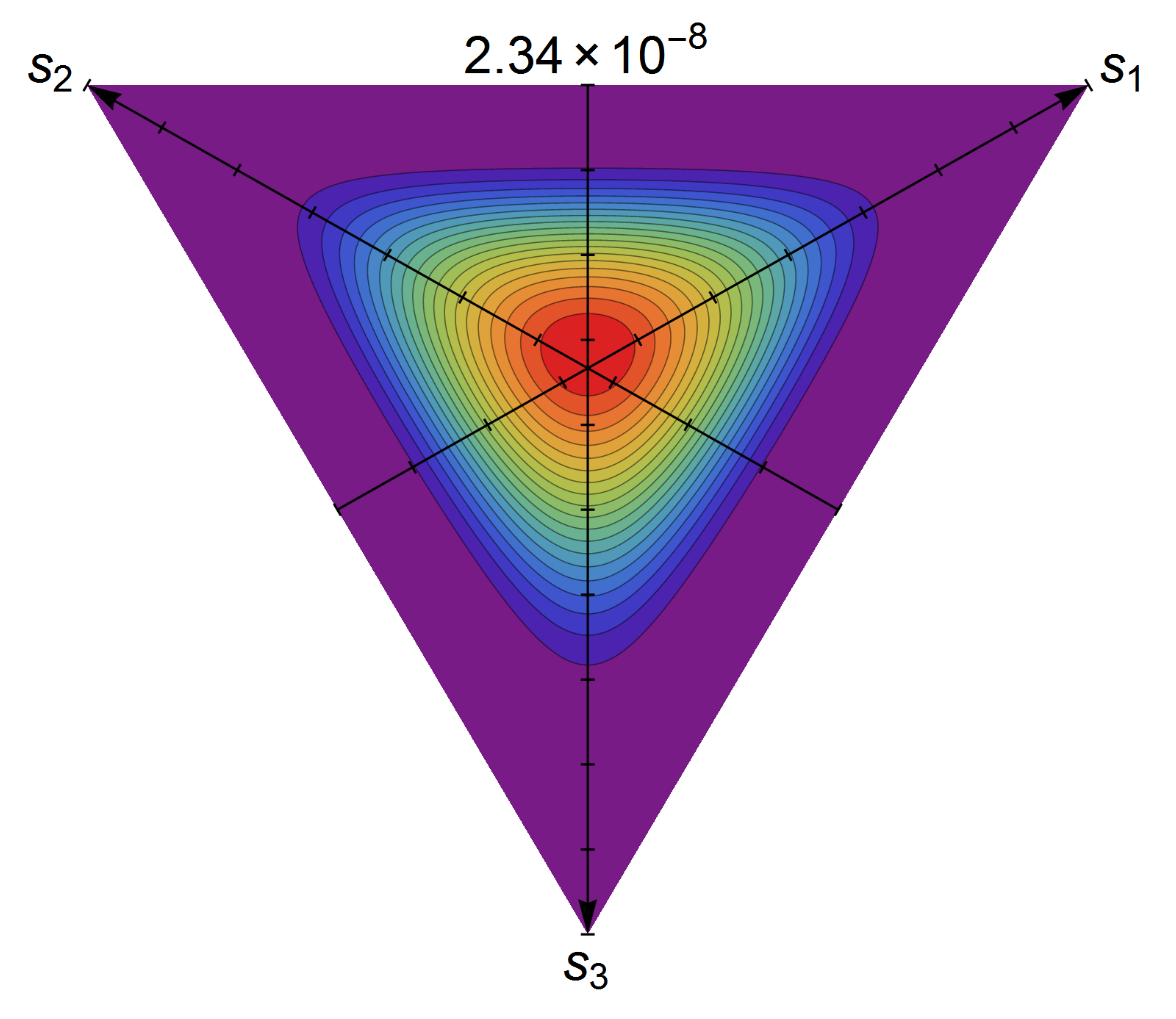}
\includegraphics[width=0.32\textwidth]{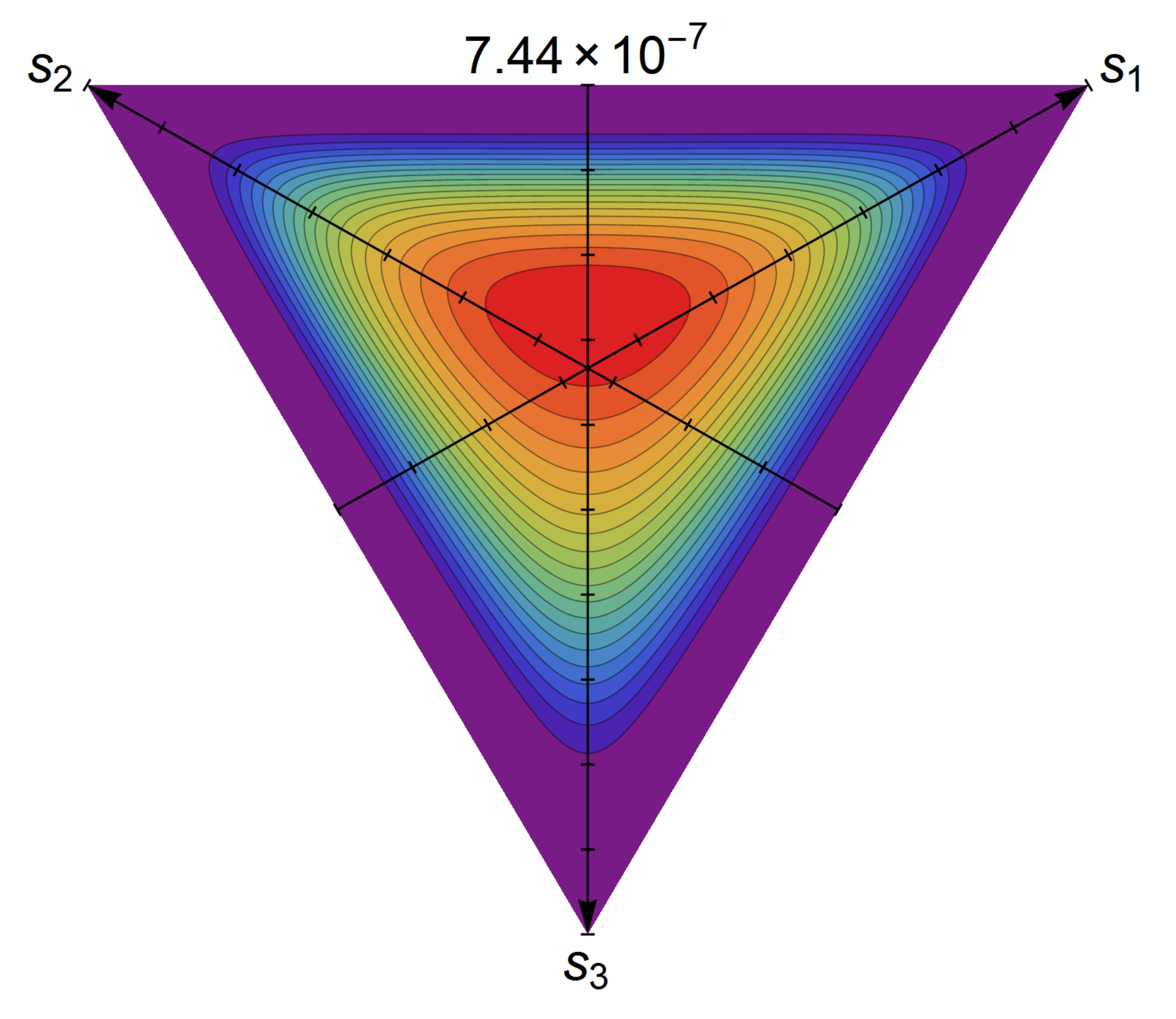}\\
\includegraphics[width=0.32\textwidth]{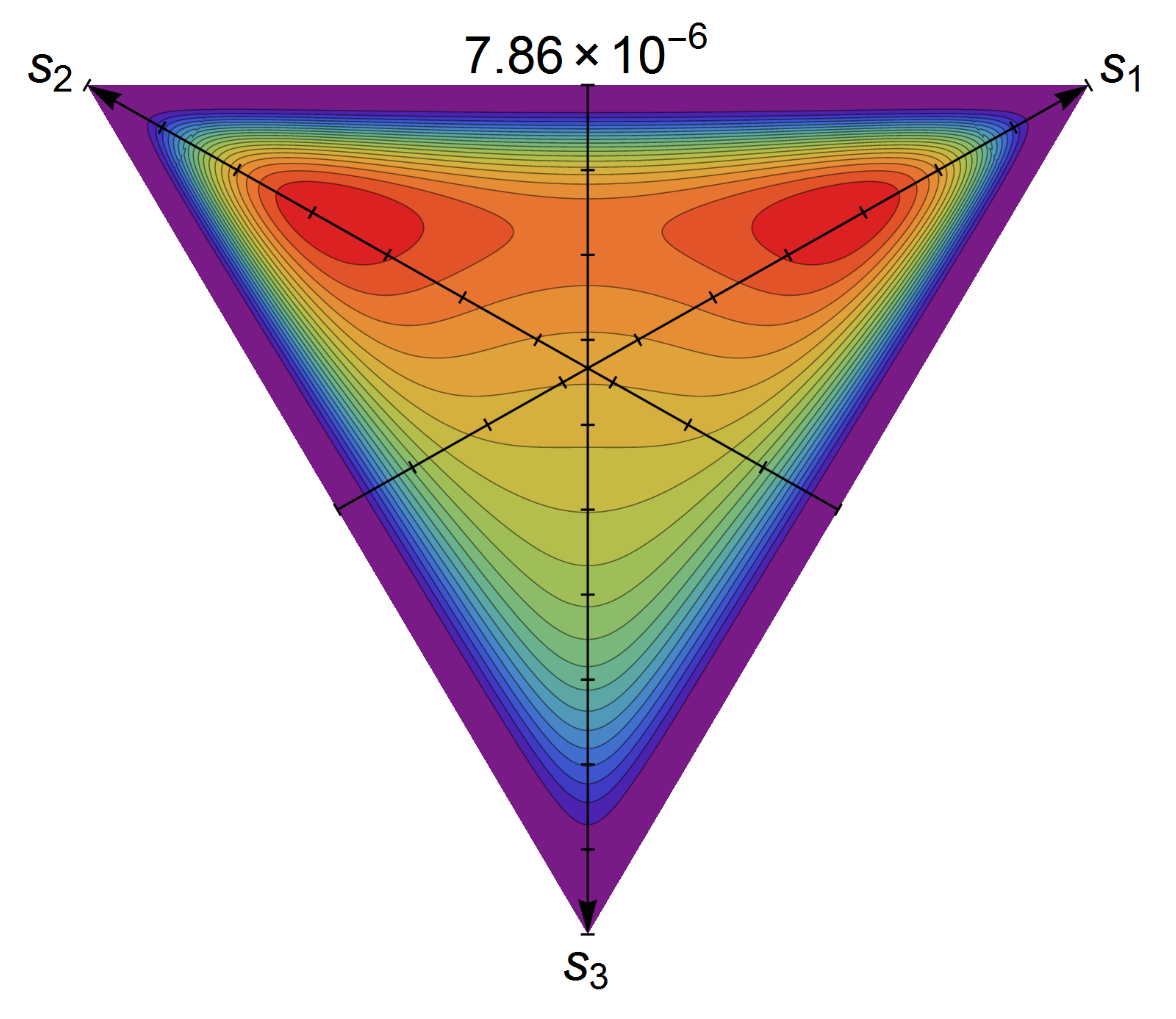}
\includegraphics[width=0.32\textwidth]{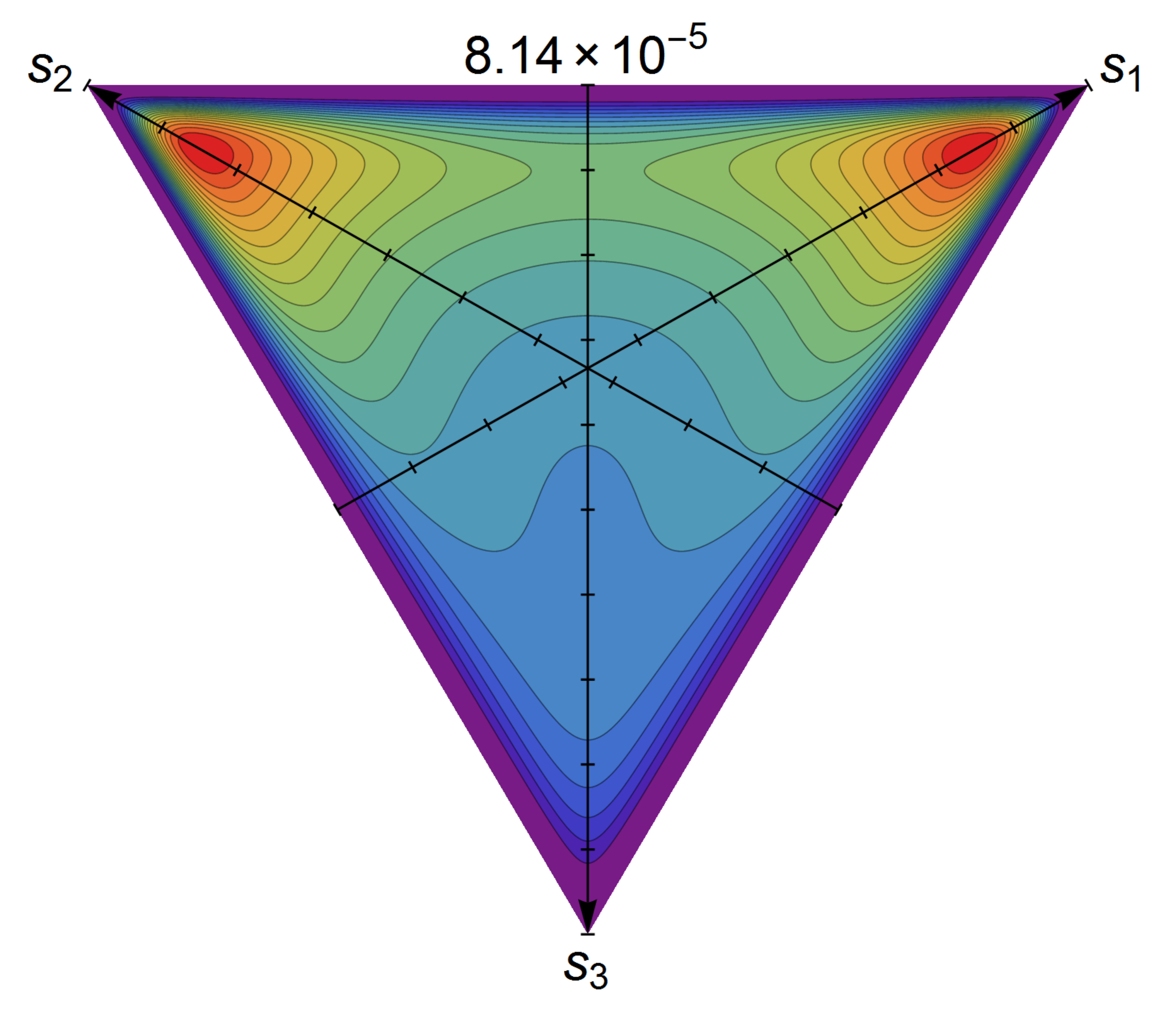}
\includegraphics[width=0.32\textwidth]{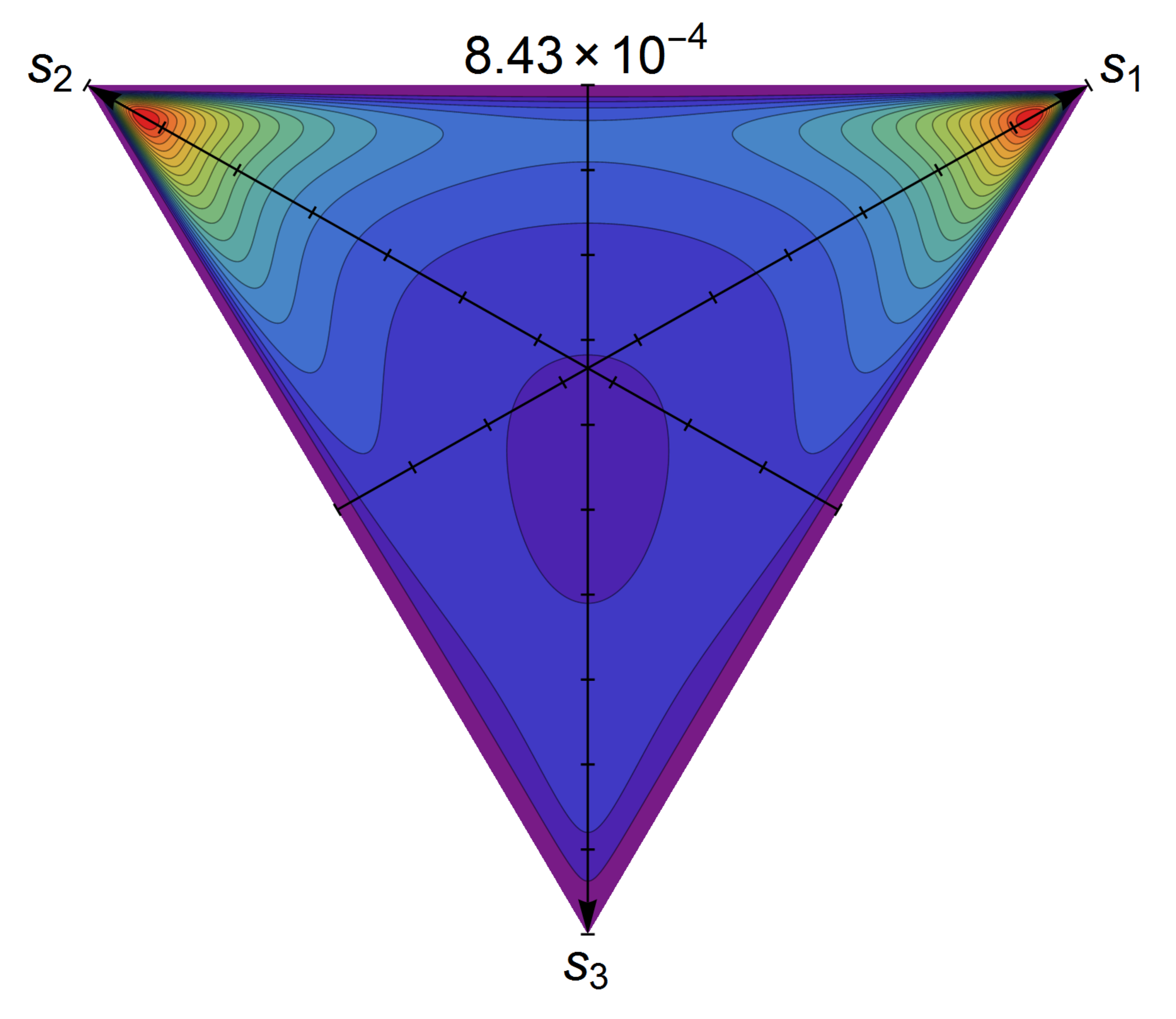}
\caption{These plots show the two-step part of the probability in the LCF approximation, $\mathcal{R}_2^\LCF(s)$ with $\mathcal{R}_2$ defined in~\eqref{R2intermsofRCandRBW} and~\eqref{propertimeRates}, as a function of the longitudinal momenta $s_i$ with $\chi_{\rm C}=\chi_{\rm BW}$. From top left to bottom right we have $\chi_C=1/2,1,2,4,8,16$. The value on top of each plot gives the maximum value. Contours and colors are chosen in the same way as in Fig.~\ref{oneStepTriangles}.}
\label{twoStepTriangles}
\end{figure}
Before turning to analytical approximations, though, let us first plot the one-step and two-step rates, $R_1$ and $R_2$, as defined in~\eqref{PintermsofR}. To leading order in the LCF approximation we have, at constant $\chi$, $R_2^\LCF\propto a_0^2$ and $R_1^\LCF\propto a_0$ (recall that $R_2=R_2^\LCF+\mathcal{O}(a_0^0)$ so the next-to-leading-order correction to $R_2$ does not mix with the leading order in $R_1$). Even in the LCF limit, $R_2^\LCF$ still depends on two different field values, $a'(\sigma_{21})\ne a'(\sigma_{43})$. It is convenient to plot $\mathcal{R}_2^\LCF=b_0^2R_2^\LCF$ since this only depends on $a_0$, $b_0$, $\sigma_{21}$ and $\sigma_{43}$ via $\chi_{\rm C}=b_0|a'(\sigma_{21})|$ and $\chi_{\rm BW}=b_0|a'(\sigma_{43})|$. For this reason we also plot $\mathcal{R}_1^\LCF(\phi)=b_0R_1^\LCF(\phi)$, which only depends on $a_0$, $b_0$, $\phi$ via $\chi(\phi)=b_0|a'(\phi)|$.  
In Fig.~\ref{oneStepTriangles} and~\ref{twoStepTriangles} we have plotted $\mathcal{R}_1^\LCF$ and $\mathcal{R}_2^\LCF$ as triangular contour plots where each of the three $s_i$ values is given by the distance to one of the triangle's sides, for different values of $\chi$ (or rather $\chi(\phi)$ and $\chi(\sigma_{21})=\chi(\sigma_{43})$). As expected e.g. from~\cite{King:2013zw}, the spectrum is peaked at $s_1=s_2=s_3=1/3$ for low $\chi$, and close to the corners, $s_1\approx1$ or $s_2\approx1$, for large $\chi$. Note that these plots contain all the one-step terms, both the direct and the exchange ones.

\subsection{Constant fields and $\chi\ll1$}\label{constant field and low chi}

As Fig.~\ref{oneStepTriangles} and~\ref{twoStepTriangles} show, for $\chi\ll1$ the spectrum is peaked at $s_1=s_2=s_3=1/3$. We can therefore perform these integrals using the saddle-point method. The lightfront time integrals can also be performed by the saddle-point method as described in the next section for non-constant fields. 

For a constant field and $\chi\ll1$ we find
\be\label{P22ConstantSmallChi}
\mathbb{P}_{\rm dir}^{22}=\alpha^2\left\{\frac{(a_0\Delta\phi)^2}{64}-\frac{a_0\Delta\phi\sqrt{\chi}}{16\sqrt{6\pi}}\right\}\exp\left(-\frac{16}{3\chi}\right)
=\mathbb{P}_{\rm dir}^{22\to2}+\mathbb{P}_{\rm dir}^{22\to1} \;.
\ee
We have already shown that the $\mathcal{O}(a_0^2)$-term in $\mathbb{P}_{\rm dir}^{22\to2}$ agrees with the literature for arbitrary $\chi$. By comparing $\mathbb{P}_{\rm dir}^{22\to1}$ in~\eqref{P22ConstantSmallChi} with eq.~(29) in~\cite{Ritus:1972nf} or eq.~(6.57) in~\cite{Baier:1998vh} (see also eq.~(28) in~\cite{King:2013osa} for normalization similar to ours), we see that $\mathbb{P}_{\rm dir}^{22\to1}$ for $\chi\ll1$ also agrees with previous results. So, $\mathbb{P}_{\rm dir}^{22}$ already gives the known $\chi\ll1$ results for both the two-step and the one-step terms, even though $\mathbb{P}_{\rm dir}^{22}$ is only one out of a total of six terms. The reasons for this are:
$\mathbb{P}_{\rm dir}^{12}(s)$ vanishes upon integrating over $s_1$ and $s_2$, $\mathbb{P}_{\rm dir}^{11}$ turns out to be smaller than $\mathbb{P}_{22\to1}$ in~\eqref{P22ConstantSmallChi} (see below), and the remaining terms, $\mathbb{P}_{\rm ex}^{11}$, $\mathbb{P}_{\rm ex}^{12}$ and $\mathbb{P}_{\rm ex}^{22}$,  give the exchange part of the probability, while the previous results just cited only give the direct part. Indeed, to the best of our knowledge, there are no analytical expressions for the exchange terms in this regime to compare with. So, our results for the exchange part below are new.     

From~\eqref{P11GenFin}, \eqref{P12dirGenFin} and~\eqref{P12ExchGenFin} we find
\be\label{P11andP12ConstSmallChi}
\mathbb{P}^{11}=\frac{\alpha^2a_0\Delta\phi\chi^\frac{3}{2}}{384\sqrt{6\pi}}\exp\left(-\frac{16}{3\chi}\right)=\frac{\chi}{24}|\mathbb{P}_{\rm dir}^{22\to1}| 
\qquad
\mathbb{P}^{12}=-\frac{7\alpha^2a_0\Delta\phi\chi^\frac{3}{2}}{1728\sqrt{6\pi}}\exp\left(-\frac{16}{3\chi}\right)=-\frac{7\chi}{108}|\mathbb{P}_{\rm dir}^{22\to1}| \;,
\ee  
where $\mathbb{P}_{\rm dir}^{22\to1}$ is given by the $a_0$ term in~\eqref{P22ConstantSmallChi}.
Eq.~\eqref{P11andP12ConstSmallChi} includes both the direct and the exchange part.
In fact, their contribution is on the same order of magnitude: we have $\mathbb{P}_{\rm ex}^{11}=-(1/2)\mathbb{P}_{\rm dir}^{11}$ (to leading order in $\chi$), and $\mathbb{P}_{\rm dir}^{12}$ is identically zero so $\mathbb{P}^{12}=\mathbb{P}_{\rm ex}^{12}$.
However, both $\mathbb{P}^{11}$ and $\mathbb{P}^{12}$ are smaller than $\mathbb{P}_{22\to1}$ in~\eqref{P22ConstantSmallChi} by a factor of $\chi\ll1$. In deriving~\eqref{P22ConstantSmallChi} we have already thrown away terms on the order of~\eqref{P11andP12ConstSmallChi}, so the terms in~\eqref{P11andP12ConstSmallChi} are only part of the higher-order corrections, see Appendix~\ref{Comparing numerical and analytical}.

For the last term we find
\be\label{P22exchConstSmallChi}
\mathbb{P}_{\rm ex}^{22}=-\frac{13\alpha^2a_0\Delta\phi\sqrt{\chi}}{288\sqrt{6\pi}}\exp\left(-\frac{16}{3\chi}\right)=-\frac{13}{18}|\mathbb{P}_{\rm dir}^{22\to1}| \;.
\ee
Importantly, $\mathbb{P}_{\rm ex}^{22}$ is on the same order of magnitude as $\mathbb{P}_{\rm dir}^{22\to1}$ in~\eqref{P22ConstantSmallChi}, so to leading order in $1/a_0\ll1$ and $\chi\ll1$ the one-step part is given by $\mathbb{P}_1=a_0P_1=\mathbb{P}_{\rm dir}^{22\to1}+\mathbb{P}_{\rm ex}^{22}$, i.e. the exchange and the direct part of $\mathbb{P}_1$ are on the same order of magnitude.
Of course, in the regime we consider here, both these contributions to $\mathbb{P}_1$ are small corrections to the two-step $\mathbb{P}_2$, but at least among the one-step terms, the direct and exchange parts are on the same order of magnitude as long as $\chi$ is not too large. We have checked that the analytical approximations in this section agree with our numerical results for sufficiently small $\chi$, see Appendix~\ref{Comparing numerical and analytical}. In the next section we will consider $a_0\gg1$ and $\chi\ll1$ for non-constant fields.

\section{Pulsed fields with $a_0\gg1$ and $\chi\ll1$}\label{Pulsed fields with large a0 and small chi}

In this section we will generalize the results in the previous section and obtain simple analytical approximations for $a_0\gg1$ and $\chi\ll1$ for non-constant fields. We consider a linearly polarized field $a(\phi)=a_0 f(\phi)$ that has a single maximum at $\phi=0$, i.e. $f''(0)=0$, and expand around this point. All the integrals are finite and there are no volume factors. We fix the field strength and the frequency by normalizing $f'(0)=1$ and $f^{(3)}(0)=-\zeta<0$. We could set $\zeta=1$, but this might not always be convenient. For example, for a Sauter pulse it is natural to choose $f(\phi)=\tanh\phi$ and then $\zeta=2$.

We begin with $\mathbb{P}_{\rm dir}^{22}$ in~\eqref{P22dirGenFin} and change variables as in~\eqref{phiitophivarphithetaeta}.
We rescale $\theta\to\theta/a_0$ and $\eta\to\eta/a_0$, for $\mathbb{P}_{\rm dir}^{22\to 1}$ we also rescale $\varphi\to\varphi/a_0$, and then expand the integrand in powers of $1/a_0$.
For $\chi\ll1$ we can perform all the integrals with the saddle-point method. We have a saddle point at
\be\label{saddlePointLCFzeta}
\phi=\varphi=\eta=0 \qquad \theta=2i \qquad s_1=s_2=\frac{1}{3} \;.
\ee
The second-order variation in the exponent is for the momentum integrals given by
$\exp\left\{-\frac{36}{\chi}(\delta s_1^2+\delta s_1\delta s_2+\delta s_2^2)\right\}$,
where $\delta s_i=s_i-1/3$, so $\delta s_i\sim\sqrt{\chi}\ll1$. The other variations are also $\sim\sqrt{\chi}\ll1$, so we rescale all the integration variables with $\sqrt{\chi}$ and expand the integrand in a series in $\chi$. The resulting integrals are all elementary, and we thus find
\be\label{a0andchiExpansionP22to2}
\mathbb{P}_{\rm dir}^{22\to 2}=\alpha^2\left\{\frac{\pi\sqrt{3}}{128}\frac{a_0^2\chi}{\zeta}\left[1-\frac{\sqrt{\chi}f^{(4)}_0}{3\sqrt{2\pi}\zeta^\frac{3}{2}}+\chi\left[-\frac{25}{27}+\frac{5f^{(4)2}_0}{24\zeta^3}+\frac{f^{(5)}_0}{8\zeta^2}\right]\right]+\frac{\pi}{80\sqrt{3}}+...\right\}\exp\left(-\frac{16}{3\chi}\right) \;,
\ee
\be\label{a0andchiExpansionP22to1}
\mathbb{P}_{\rm dir}^{22\to 1}=\alpha^2\left\{-\frac{a_0\chi}{64\sqrt{\zeta}}\left[1+\chi\left[-\frac{2683}{10368}+\frac{5f_0^{(4)2}}{128\zeta^3}+\frac{3f_0^{(5)}}{128\zeta^2}\right]\right]-\frac{1}{120}\frac{\sqrt{\zeta}}{a_0}+...\right\}\exp\left(-\frac{16}{3\chi}\right) \;,
\ee
where $f^{(n)}_0=\partial_\phi^n f(0)$. We have assumed here that $1/(\chi a_0^2)\ll1$ in order to expand a term in the exponent down to the prefactor. The terms proportional to $a_0^2$ and $a_0$ correspond, respectively, to what are usually referred to as two-step and one-step terms. The $a_0^2$ term can hence be obtained by expanding the Airy functions in~\eqref{TwoStepAiry}, which in turn can be obtained directly from the literature.     
The reason that we have included higher orders in $\chi$ for the $a_0^2$ term is that these can be on the same order as the leading-order-in-$\chi$ contribution to the $a_0^0$ term, and similarly for the first term in~\eqref{a0andchiExpansionP22to1}. How many orders in $\chi$ one should keep at a given order in $a_0$ depends of course on the relative size of $\chi$ and~$a_0$; we have included the higher-order terms in~\eqref{a0andchiExpansionP22to2} and~\eqref{a0andchiExpansionP22to1} mainly as examples of this double expansion. In any case, using Mathematica it is straightforward to calculate higher orders in both $1/a_0$ and $\chi$. Note that there are no volume factors here (cf.~the $\Delta\phi$ terms in the previous section) because of the damping given by $\zeta>0$.

For~\eqref{P11GenFin} and~\eqref{P12ExchGenFin} we find
\be
\mathbb{P}^{11}=\frac{\alpha^2}{1536}\frac{a_0\chi^2}{\sqrt{\zeta}}\exp\left(-\frac{16}{3\chi}\right)+\mathcal{O}(a_0^{-1}) \
\qquad
\mathbb{P}_{\rm ex}^{12}=-\frac{7\alpha^2}{6912}\frac{a_0\chi^2}{\sqrt{\zeta}}\exp\left(-\frac{16}{3\chi}\right)+\mathcal{O}(a_0^{-1}) \;.
\ee
These are both smaller by a factor of $\chi\ll1$ than the leading-order term in $\mathbb{P}_{\rm dir}^{22\to1}$, i.e. the first term in~\eqref{a0andchiExpansionP22to1}. This agrees with what we found above for constant fields, see~\eqref{P11andP12ConstSmallChi}.

Next we consider $\mathbb{P}_{\rm ex}^{22}$ given by~\eqref{P22exchs0}. 
In order to expand in $1/a_0$ we rescale $\theta\to\theta/a_0$, $\eta\to\eta/a_0$ and 
$\varphi\to\varphi/a_0$.
We expand the integrand around the saddle point~\eqref{saddlePointLCFzeta} and perform the resulting integrals.
We find
\be\label{P22exchSmallchizeta}
\mathbb{P}_{\rm ex}^{22}=\frac{13\alpha^2}{18}\left\{-
\frac{a_0\chi}{64\sqrt{\zeta}}\left[1+\chi\left[-\frac{371989}{524160}+\frac{5f_0^{(4)2}}{128\zeta^3}+\frac{3f_0^{(5)}}{128\zeta^2}\right]\right]-\frac{\sqrt{\zeta}}{120a_0}+...\right\}\exp\left(-\frac{16}{3\chi}\right) \;.
\ee
As in the constant-field case in the previous section, $\mathbb{P}_{\rm ex}^{22}$ gives the dominant contribution to the exchange part and it is on the same order of magnitude as the direct one-step term~\eqref{a0andchiExpansionP22to1}. 
Thus, the one-step terms, $a_0P_1=a_0P_1^{\rm dir}+a_0P_1^{\rm ex}$, are to leading order in $\chi\ll1$ given by $a_0P_1^{\rm dir}=\text{lin}_{a_0}\mathbb{P}_{\rm dir}^{22}$ and $a_0P_1^{\rm ex}=\text{lin}_{a_0}\mathbb{P}_{\rm ex}^{22}$, where $\text{lin}_{a_0}$ refers to the terms that are linear in $a_0$.
We find, to leading order in $\chi$, the same relation as in~\eqref{P22exchConstSmallChi} between $P_1^{\rm dir}$ and $P_1^{\rm ex}$, i.e. $P_1^{\rm ex}/P_1^{\rm dir}=13/18$, and both $P_1^{\rm dir}$ and $P_1^{\rm ex}$ are negative. The exchange term is also negative for $a_0\ll1$, which we can confirm by comparing with the literature, see Appendix~\ref{Perturbative limit}.

\section{Sauter pulse with $a_0\sim1$ and $\chi\ll1$}\label{Sauter pulse section}

\begin{figure}
\includegraphics[width=0.4\textwidth]{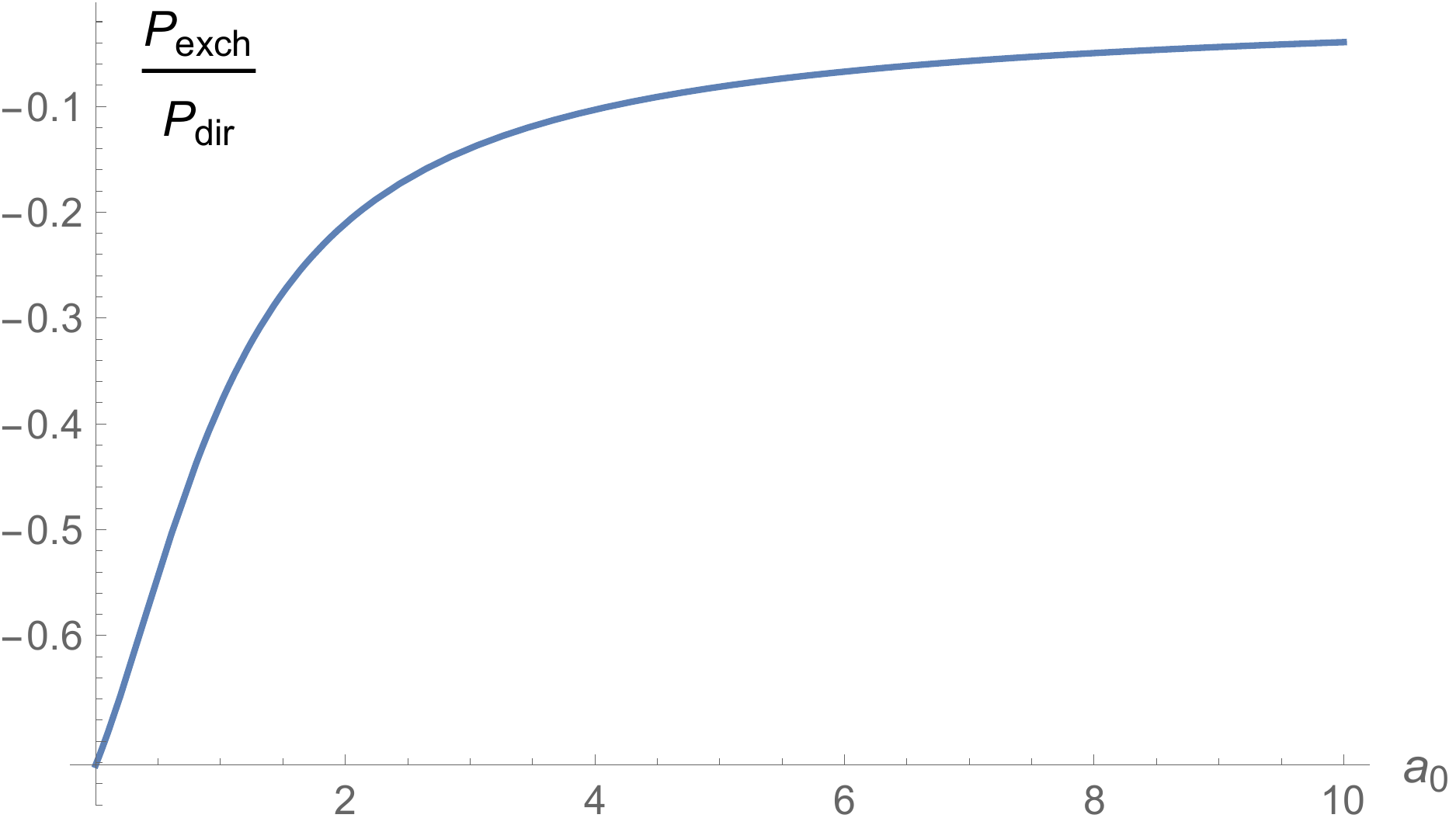}
\caption{This is a plot of $\mathbb{P}_{\rm ex}/\mathbb{P}_{\rm dir}=\eqref{SauterExchSmallChiGena0}/\eqref{SauterDirSmallChiGena0}$ as a function of $a_0$. This ratio is independent on $\chi$ to leading order in $\chi\ll1$.}
\label{SauterExchDirRatioSmallChiFig}
\end{figure}

In this section we will consider a Sauter pulse, $a(\phi)=a_0\tanh\phi$. We again assume $\chi\ll1$, but now we do not assume that $a_0$ is large. The results in this section for $a_0\sim1$ therefore go beyond the LCF approximation. Our starting point is the exact expressions in Sec.~\ref{Exact analytical results}. Recall that the field enters these expressions only via $M^2$ and $\Delta$.    
For the Sauter pulse the integrals in $M^2$ and $\Delta$ can be performed analytically and this gives  
\be
M^2(\phi_2,\phi_1)=1+a_0^2\left\{1-\frac{\tanh\phi_2-\tanh\phi_1}{\phi_2-\phi_1}-\left[\frac{\ln[\cosh\phi_2/\cosh\phi_1]}{\phi_2-\phi_1}\right]^2\right\} 
\ee
\be
\Delta(\phi_2,\phi_1)=a_0\left\{\tanh\phi_2-\frac{\ln[\cosh\phi_2/\cosh\phi_1]}{\phi_2-\phi_1}\right\} \;.
\ee
For $\chi\ll1$ we can perform all the integrals with the saddle-point method. We first change variables as in~\eqref{phiitophivarphithetaeta}. We have a saddle point at
\be
\phi=\varphi=\eta=0 \qquad \theta=2i\text{ arccot }a_0 \qquad s_1=s_2=\frac{1}{3} \;.
\ee
By expanding the integrands around this saddle point we find
\be\label{P2smallchiSauter}
\mathbb{P}_{\rm dir}^{22\to2}=\frac{\pi\alpha^2\chi}{192\sqrt{3}}\frac{a_0\exp\left\{-\frac{8a_0}{\chi}\left[(1+a_0^2)\text{arccot}a_0-a_0\right]\right\}}{(1+a_0^2)\text{arccot}a_0\left[(1+a_0^2)\text{arccot}a_0-a_0\right]^2}
\ee
and
\be\label{P1smallchiSauter}
\mathbb{P}_{\rm dir}^{22\to1}=-\frac{2}{\pi}\arctan\sqrt{1-\frac{a_0}{(1+a_0^2)\text{arccot}a_0}}\mathbb{P}_{22\to2} \qquad
\mathbb{P}_{\rm ex}^{22}=\frac{13}{18}\mathbb{P}_{22\to1} \;.
\ee
As in the previous section, $\mathbb{P}^{11}$ and $\mathbb{P}^{12}$ are smaller by a factor of $\chi\ll1$, so to leading order in $\chi$ we have
(these expressions can also be obtained without much extra work by starting instead with~\eqref{Pdg} and~\eqref{Pexch-g})
\be\label{SauterDirSmallChiGena0}
\mathbb{P}_{\rm dir}=
\mathbb{P}_{\rm dir}^{22\to2}+\mathbb{P}_{\rm dir}^{22\to1}=\frac{\alpha^2\chi a_0}{96\sqrt{3}}\frac{\text{arccot}\sqrt{1-\frac{a_0}{(1+a_0^2)\text{arccot}a_0}}}{(1+a_0^2)\text{arccot}a_0[(1+a_0^2)\text{arccot}a_0-a_0]^2}\exp\left\{-\frac{8a_0}{\chi}\left[(1+a_0^2)\text{arccot}a_0-a_0\right]\right\} 
\ee
and
\be\label{SauterExchSmallChiGena0}
\mathbb{P}_{\rm ex}=\mathbb{P}_{\rm ex}^{22}=-\frac{13\alpha^2\chi a_0}{1728\sqrt{3}}\frac{\text{arctan}\sqrt{1-\frac{a_0}{(1+a_0^2)\text{arccot}a_0}}}{(1+a_0^2)\text{arccot}a_0[(1+a_0^2)\text{arccot}a_0-a_0]^2}\exp\left\{-\frac{8a_0}{\chi}\left[(1+a_0^2)\text{arccot}a_0-a_0\right]\right\} \;.
\ee
In the limit $a_0\gg1$, \eqref{P2smallchiSauter} and~\eqref{P1smallchiSauter} reduce to the expansions in~\eqref{a0andchiExpansionP22to2},~\eqref{a0andchiExpansionP22to1} and~\eqref{P22exchSmallchizeta} (with $\zeta=2$ for this field). Eq.~\eqref{P1smallchiSauter} shows that the exchange term is on the same order of magnitude as $\mathbb{P}_{\rm dir}^{22\to1}$, and their ratio, $\mathbb{P}_{\rm ex}^{22}/\mathbb{P}_{\rm dir}^{22\to1}=13/18$, is the same as the $\mathcal{O}(a_0)$ terms we found in the previous section. In the limit $a_0\gg1$ these two terms gives the one-step term, i.e. $\mathbb{P}_{\rm dir}^{22\to1}+\mathbb{P}_{\rm ex}^{22}\to a_0P_1$, while $\mathbb{P}_{\rm dir}^{22\to2}\to a_0^2P_2$. However, for $a_0\sim1$ the ``one-step'' terms, $\mathbb{P}_{\rm dir}^{22\to1}$ and $\mathbb{P}_{\rm ex}^{22}$, are on the same order of magnitude as the ``two-step'' term, $\mathbb{P}_{\rm dir}^{22\to2}$. For $a_0=1$ we have $\mathbb{P}_{\rm ex}/\mathbb{P}_{\rm dir}\approx-0.4$, while for $a_0\gtrsim4$ the ratio decreases to $-0.1\lesssim\mathbb{P}_{\rm ex}/\mathbb{P}_{\rm dir}<0$, see Fig.~\ref{SauterExchDirRatioSmallChiFig}.

The saddle-point calculation of the pre-exponential factors above breaks down when $a_0$ becomes too small. However, the exponential part of~\eqref{P2smallchiSauter} scales for $a_0\ll1$ as
\be\label{PexpSauterPerturbative}
\mathbb{P}\sim a_0^2\exp\left(-\frac{4\pi}{b_0}\right) \;,
\ee 
which is what one would expect for perturbative trident: In terms of the Fourier frequency, $\hat{\omega}$, of $a(\phi)$, the threshold for pair production is given by $\hat{\omega}_{\rm th}=4/b_0$. For $\hat{\omega}_{\rm th}\gg1$ the Fourier transform of the Sauter pulse is approximately exponential,
\be
\tanh\phi=\int\frac{\ud\hat{\omega}}{2\pi}a(\hat{\omega})e^{-i\hat{\omega}\phi}
\qquad\to\qquad
|a(4/b_0)|^2\sim\exp\left(-\frac{4\pi}{b_0}\right) \;,
\ee
which gives the exponential in~\eqref{PexpSauterPerturbative} (cf. the treatment of Sauter-like pulses in~\cite{Torgrimsson:2017pzs} for Schwinger pair production). 

Interestingly, the exponent in~\eqref{P2smallchiSauter} has the same functional form as the one in photon-stimulated Schwinger pair production in a constant electric field, eq.~(5) in~\cite{Dunne:2009gi}, but with different parameters.

\section{Monochromatic field with $a_0\sim1$ and $\chi\ll1$}\label{Monochromatic field section}

We now consider a monochromatic field, $a(\phi)=a_0\sin\phi$. Consider first $\mathbb{P}_{22\to2}$. We find saddle points at
\be\label{monochromaticSaddles}
\theta_{21}=\theta_{43}=2i\text{arcsinh}\frac{1}{a_0} \qquad \sigma_{21}=n_1\pi \qquad \sigma_{43}=n_2\pi \qquad s_1=s_2=\frac{1}{3} \;,
\ee
where $\sigma_{ij}=(\phi_i+\phi_j)/2$ and $n_1$ and $n_2$ are integers. The saddle points for $\sigma$ correspond to maxima and minima of the field. The step function $\theta(\sigma_{43}-\sigma_{21})$ implies $n_2\geq n_1$, so either photon emission and pair production happen at the same max/min or the photon is emitted at one max/min and travels to a different max/min where it decays into a pair. All $n_1$ and $n_2$ give the same contribution, except the term with $n_1=n_2$ which is, due to $\theta(\sigma_{43}-\sigma_{21})$, a factor of $1/2$ smaller than the other terms. With $2N$ maxima and minima, adding the contribution from all saddle points gives $\mathbb{P}_{\rm dir}^{22\to2}=N^2\mathbb{P}_{22\to2}^{N=1}$, where $\mathbb{P}_{22\to2}^{N=1}$ is the contribution from a single maximum or minimum, which we find to be
\be\label{monochromaticPsaddle}
\mathbb{P}_{22\to2}^{N=1}=\frac{\pi\alpha^2\chi}{96\sqrt{3}a_0}\frac{\exp\left\{-\frac{4a_0}{\chi}\left([2+a_0^2]\Lambda-\sqrt{1+a_0^2}\right)\right\}}{\sqrt{1+a_0^2}\Lambda\left(\sqrt{1+a_0^2}\Lambda-1\right)\left([2+a_0^2]\Lambda-\sqrt{1+a_0^2}\right)} 
\qquad
\Lambda=\text{arcsinh}\frac{1}{a_0}
\;.
\ee
The argument of the exponential in~\eqref{monochromaticPsaddle} has the same functional dependence on $a_0$ as the exponent in~\cite{NikishovRitusVol25Page1135} (see the equation before eq.~15 in~\cite{NikishovRitusVol25Page1135}) for the Breit-Wheeler probability. The only difference is an overall factor of $2kl/kp$, where $l_\mu$ is the photon momentum in the Breit-Wheeler case. This factor is expected in the locally constant field limit, i.e. for $a_0\gg1$, and now we can see the same relation also for general $a_0$.  
For a large number of periods, $N\gg1$, $\mathbb{P}_{\rm dir}^{22\to2}$ gives the dominant contribution, as the other terms only scale as $N$.
For $a_0\gg1$ we recover from~\eqref{monochromaticPsaddle} the $a_0^2$ and $a_0^0$ terms in the general expansion~\eqref{a0andchiExpansionP22to2} (with $\zeta=1$ for this field). For $a_0\ll1$ the exponent in~\eqref{monochromaticPsaddle} scales as
\be\label{multiphotonExp}
\exp\left\{\frac{4}{b_0}\left(\ln\left[\frac{a_0}{2}\right]^2-1\right)\right\}\sim a_0^{8/b_0} \;,
\ee
where $4/b_0$ can be interpreted as the number of photons that need to be absorbed to produce a pair. 

Let us compare with the SLAC experiment~\cite{Bamber:1999zt}. The most straightforward result to compare with is their logarithmic plots of the number of detected positrons as a function of $1/\chi$, which they fit to $\exp(-c/\Upsilon)$\footnote{We thank S.~Meuren for bringing this part of~\cite{Bamber:1999zt} to our attention.}, where $\Upsilon=\chi/\sqrt{2}$. According to eq.~(8.3) in~\cite{KoffasThesis}, for an average of $\langle\eta\rangle=0.2$, where $\eta=a_0/\sqrt{2}$, the SLAC experiment gave
\be\label{SLACcoefficient}
c=2.4\pm0.1(\text{stat.})\genfrac{}{}{0pt}{1}{+0.2}{-0.6}(\text{syst.})
\ee
For this $a_0$ our~\eqref{monochromaticPsaddle} predicts $c\approx2.46$, which is in agreement with~\eqref{SLACcoefficient}. However, the errors in~\eqref{SLACcoefficient} are too large for us to really be able to confirm~\eqref{monochromaticPsaddle}. Indeed, in~\cite{Bamber:1999zt,KoffasThesis} $c$ was also shown to roughly agree with an estimate obtained using a result for Schwinger pair production by a purely time-dependent electric field. Also, this value of $a_0$ is quite small, so the exponent is close to the perturbative one in~\eqref{multiphotonExp}. It would therefore be interesting to compare our~\eqref{monochromaticPsaddle} with future trident experiments with larger $a_0$. 

If the number of oscillations are not large then we should also consider the other terms. We find $\mathbb{P}_{\rm dir}^{22\to1}=N\mathbb{P}_{22\to1}^{N=1}$ and $\mathbb{P}_{\rm ex}^{22}=N\mathbb{P}_{\rm ex}^{22,N=1}$, where
\be
\label{Pmono22to1and22exch}
\mathbb{P}_{22\to1}^{N=1}=-\frac{2}{\pi}\arctan\sqrt{1-\frac{1}{\sqrt{1+a_0^2}\Lambda}}\mathbb{P}_{22\to2}^{N=1}
\qquad
\mathbb{P}_{\rm ex}^{22,N=1}=\frac{13}{18}\mathbb{P}_{22\to1}^{N=1} \;.
\ee   
These relations are similar to the ones we found in the previous section for a Sauter pulse. For $a_0\gg1$ we recover the $a_0$ and $1/a_0$ terms in~\eqref{a0andchiExpansionP22to1} for $\mathbb{P}_{22\to1}^{N=1}$, and~\eqref{P22exchSmallchizeta} for $\mathbb{P}_{\rm ex}^{22,N=1}$ ($\zeta=1$ for this field). Once again $\mathbb{P}^{11}$ and $\mathbb{P}^{12}$ are smaller by a factor of $\chi$ and can therefore be neglected to leading order. 

We can also obtain these expressions by starting instead with~\eqref{Pdg} and~\eqref{Pexch-g}. We have again saddle points given by~\eqref{monochromaticSaddles}: For the exchange term~\eqref{Pexch-g} we have saddle points for $n_1=n_2$, which give $\mathbb{P}_{\rm ex}=\mathbb{P}_{\rm ex}^{22}$ with $\mathbb{P}_{\rm ex}^{22}$ as in \eqref{Pmono22to1and22exch}. For the direct term~\eqref{Pdg} we have saddle points for all $n_1$ and $n_2$, but again because of the step functions only the saddles with $n_1\leq n_2$ contribute. We find
\be
\mathbb{P}_{\rm dir}^{n_1=n_2}=\frac{2}{\pi}\text{arccot}\sqrt{1-\frac{1}{\sqrt{1+a_0^2}\Lambda}}\mathbb{P}_{22\to2}^{N=1} \qquad
\mathbb{P}_{\rm dir}^{n_1<n_2}=2\mathbb{P}_{22\to2}^{N=1} \;,
\ee
where $\mathbb{P}_{22\to2}^{N=1}$ is given by~\eqref{monochromaticPsaddle}.
which agrees with what we found with the first approach.

\section{$a_0\gg1$ and general $\chi$}\label{LCF general chi}

\begin{figure}
\includegraphics[width=0.6\textwidth]{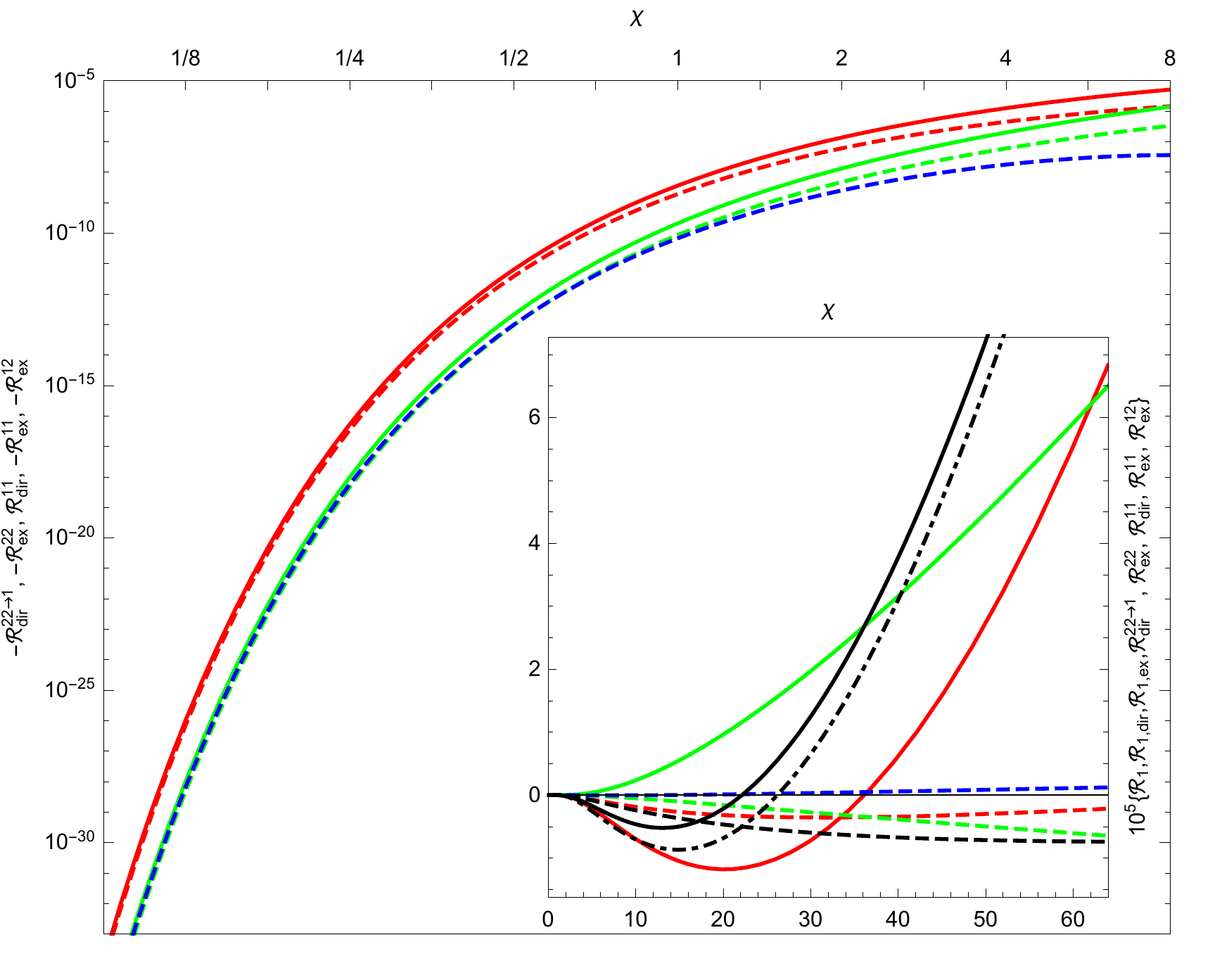}
\caption{This figure shows the five different terms contributing to the total one-step rate $\mathcal{R}_1$. We have solid lines for direct terms and dashed lines for exchange terms, red for $\mathcal{R}_{\rm dir}^{22\to1}$ and $\mathcal{R}_{\rm ex}^{22}$, green for $\mathcal{R}_{\rm dir}^{11}$ and $\mathcal{R}_{\rm ex}^{11}$, blue for $\mathcal{R}_{\rm ex}^{12}$, black for $\mathcal{R}_{1,\rm dir}=\mathcal{R}_{\rm dir}^{11}+\mathcal{R}_{\rm dir}^{22\to1}$ (solid) and $\mathcal{R}_{1,\rm ex}=\mathcal{R}_{\rm ex}^{11}+\mathcal{R}_{\rm ex}^{12}+\mathcal{R}_{\rm ex}^{22}$ (dashed), and black dot-dashed for the total one-step rate $\mathcal{R}_1=\mathcal{R}_{1,\rm dir}+\mathcal{R}_{1,\rm ex}$. All these rates are obtained in the LCF approximation, and $\mathcal{R}_{\rm ex}^{22}$ is the contribution to $\mathcal{R}_1$ in~\eqref{PintermsofR} coming from $\mathbb{P}_{\rm ex}^{22}$ etc.}
\label{R1overChi}
\end{figure}

In the previous couple of sections we have studied analytically the low-$\chi$ regime. In this section we will consider the dependence on $\chi$ up to large values of $\chi$. To do so assume $a_0\gg1$ and integrate the integrals in the LCF approximation using the numerical methods described in Sec.~\ref{Numerical method}. In Fig.~\ref{R1overChi} we have plotted the five terms that contribute to the total one-step rate $\mathcal{R}_1^\LCF$ as a function of $\chi$, where the rate is obtained from the probability as in~\eqref{propertimeRates}. 
\begin{figure}
\includegraphics[width=0.6\textwidth]{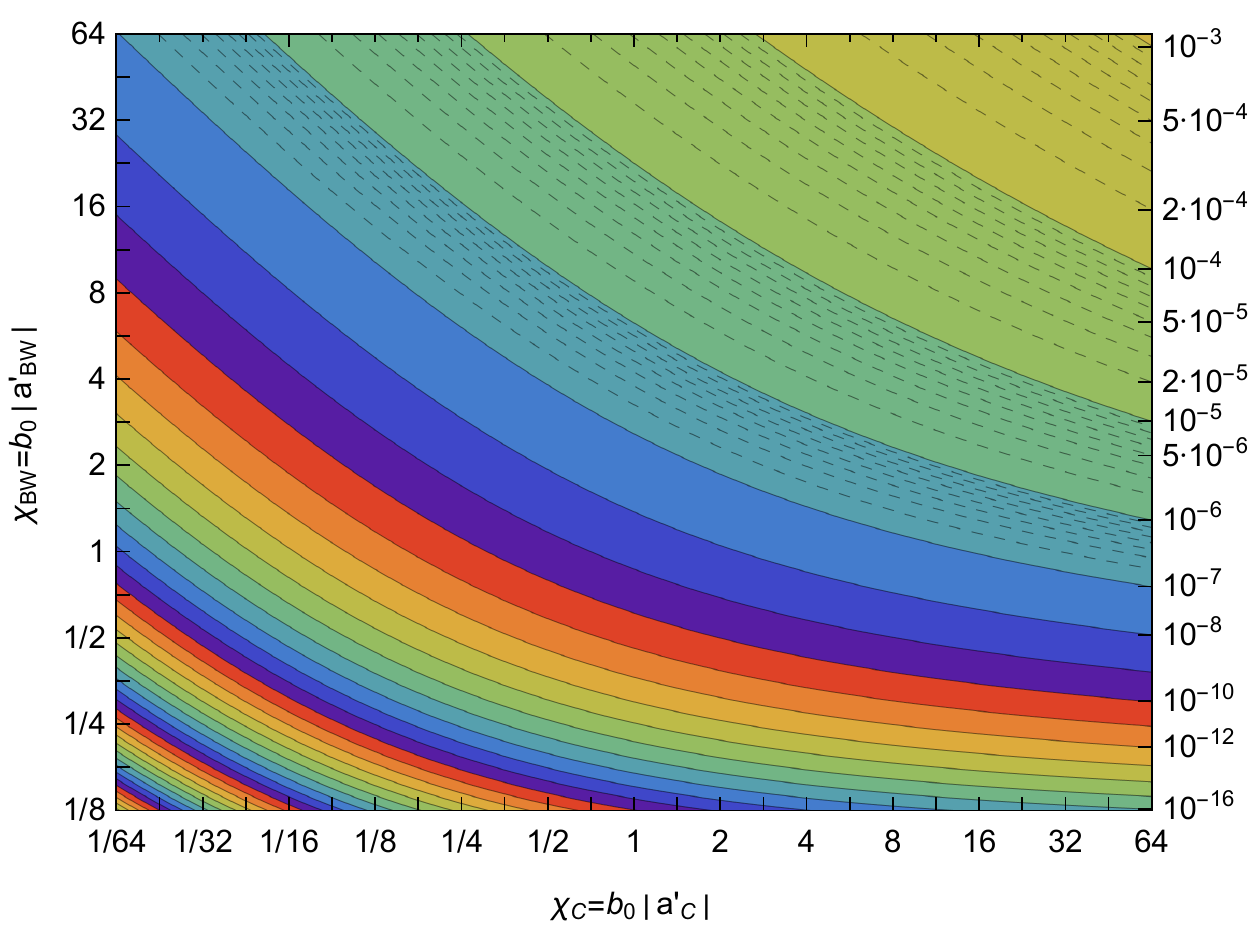}
\caption{This figure shows the two-step rate in the LCF approximation, $\mathcal{R}_2^{\rm LCF}(\chi_{\rm C},\chi_{\rm BW})$, where $\mathcal{R}_2$ is defined in~\eqref{propertimeRates}, as a function of the two different $\chi_{\rm C}\ne\chi_{\rm BW}$.}
\label{R2overChiChi}
\end{figure}
(Recall that there are only five and not six terms because $\mathbb{P}_{\rm dir}^{12}$ vanishes upon integrating over the longitudinal momenta.) 
In Fig.~\ref{R1overChi} we see that the direct part of the one-step, $\mathcal{R}_{1,\rm dir}$, is negative for small $\chi$ (in agreement with the $\chi\ll1$ approximations), grows in magnitude until it reaches a minimum, then starts to increase and eventually becomes positive. This dependence is nothing new, it can already be found in~\cite{King:2013osa}\footnote{When comparing our results with~\cite{King:2013osa}, note that we have plotted $\mathcal{R}_1$ which is related to $R_1/a_0$ by a factor of $\chi$.} and the fact that $\mathcal{R}_{1,\rm dir}$ becomes positive at large $\chi$ can also be seen from the $\chi\gg1$ approximations in e.g.~\cite{Ritus:1972nf}\footnote{We have checked analytically that the $\chi\gg1$ limit of $\mathbb{P}_{\rm dir}^{22\to1}+\mathbb{P}_{\rm dir}^{11}$ agrees with the $\chi\gg1$ approximation in~\cite{Ritus:1972nf}.}. Our results for the exchange part, $\mathcal{R}_{1,\rm ex}$, on the other hand, are new.      
We have shown in the previous sections that the exchange terms can be important at low $\chi$. Fig.~\ref{R1overChi} shows us that the exchange terms continue to be non-negligible even up to quite large $\chi$. In fact, there is a whole range $\{17\lesssim\chi\lesssim26\}$ of moderately large $\chi$, around the region where $\mathcal{R}_{1,\rm dir}$ changes sign, where $|\mathcal{R}_{1,\rm ex}|$ is even larger than $|\mathcal{R}_{1,\rm dir}|$. For $\chi$ above this interval, $\mathcal{R}_{1,\rm dir}$ eventually becomes much larger than $|\mathcal{R}_{1,\rm ex}|$, as expected. If we would increase $\chi$ even further, then at some point the $\alpha$-expansion is expected to break down~\cite{Narozhnyi:1980dc,Fedotov:2016afw}. 

Of course, for $a_0\gg1$ these one-step terms are corrections to the two-step term $\mathcal{R}_2$. In Fig.~\ref{R2overChiChi} we have plotted $\mathcal{R}_2^\LCF$ as a function of the two local values of $\chi$, i.e. $\chi_{\rm C}=b_0|a'(\sigma_{21})|$ and $\chi_{\rm BW}=b_0|a'(\sigma_{43})|$ where $\mathcal{R}_2$ is defined in~\eqref{propertimeRates}.

\section{Numerical method}\label{Numerical method}

In this section we describe the numerical methods we have used to integrate the different terms in $\mathbb{P}(s)$. The terms to be computed involve up to four phase point $\phi_{i}$ integrals whose integrand has the general structure of a relatively slowly varying pre-exponential factor multiplying a fast oscillating phase factor. The phase generally grows with $\phi_2,\phi_4$ and decreases with $\phi_1,\phi_3$. All terms but $\mathbb{P}_{\rm ex}^{22}(s)$ are built up multiplicatively from independent factors describing each individual process. While lightfront-time ordering prevents the full factorization of integrals, factorization of the integrand can still be exploited to greatly reduce the computational effort by the equivalent of up to two quadrature dimensions.
An important issue to tackle is the presence of the $i\epsilon$-prescriptions originating from the infinitesimal damping of the transverse momentum integrals required for their asymptotic convergence. These $i\epsilon$-prescriptions allow for the avoidance of singularities that are present in the remaining integrals. In all terms except $\mathbb{P}_{\rm ex}^{22}$ these singularities are found at the origin of the separations between conjugate phase points, i.e. at $\theta_{21}=0$ and $\theta_{43}=0$. Interestingly, in studying $\mathbb{P}_{\rm ex}^{22}$ one has to deal with a singularity moving with $s$, corresponding to vanishing $d_{0}$, which appears in the pre-exponential factor and also makes the phase diverge. While taking the $\epsilon\searrow0$ limit may be defined in a mathematically sound way, the elimination of the singularity is required for numerical computation. Regularization can be done in several ways. Two classes of methods, both with advantages and disadvantages, that we have
successfully tested are:

A) Subtraction of vanishing or easy-to-compute quantities sharing the same singular part of the Laurent series (or a partial integration to the same purpose), resulting in a pre-exponential factor without any singularity. This requires more analytical effort in order to produce series expansions near the singularity points, where precision loss obviously prevents a direct computation of the difference. For integrals with only one effective mass in the phase, such as $\mathbb{P}^{11}$, $\mathbb{P}_{\rm NLC}$ or $\mathbb{P}_{\rm BW}$, regularization may be achieved by a simple subtraction of the vanishing $a_0=0$ integral.
For factored terms like $\mathbb{P}_{\rm dir}^{22}$ this procedure can also be used, but by now expressing the product of two factors forming the integrand as $AB=(A-A_0)(B-B_0)  +A_0(B-B_0)+(A-A_0)B_0+A_0B_0$, where the subscript $0$ indicates a field-free ($a_0=0$) counterpart. 
Apart from the last, vanishing, fully free term, and the first term, describing interaction with the field at both steps, there are two additional terms which we can think of as ``semi-virtual'', in the sense that one of the processes takes place outside the field with no absorption of photons from the field. 
Note that they only survive theta integration due to the cut-off in the ($\theta_{21}$,  $\theta_{43}$) integrals arising from lightfront time ordering, and hence they only contribute to $\mathbb{P}_1$. 

A similar procedure, which is numerically preferable as it does not introduce different phase factors (corresponding to $a_0\ne0$ and $a_0=0$), is achieved by replacing $\theta_{21}$, $\theta_{43}$ with $\Theta_{21},$ $\Theta_{43}$ as variables, apart from $\phi_{1}$, $\phi_{3}$, and noticing that we can fortunately rewrite the intial $\theta_{ij}$-dependent step functions as a combination of $\Theta_{ij}$-dependent ones, such that the integration domain remains triangular in shape in the new variables. Then the same division of $AB$ is performed where now $A_{0}$ and $B_{0}$ have the same phase factor as $A$ and $B$, and since the $\phi_1$, $\phi_3$ integrals in the first, largest dimension integral do not involve the phase factor, they cost us the resources of only one integral with about as many oscillations as the pulse has. Once these integrals are computed, in a second step one can generate the whole $s$ spectrum (using adequate integration routines that interpolate only the pre-exponential factor in Fourier integrals) and (breaking the integrand into a linear combination of terms factored into products of contributions depending on just one pair of $\phi$ values, with $s$-dependent coefficients) even compute $s$-integrated quantities such as the total probability or energy averages in a very fast way, through mere
two-dimensional quadratures (the $s$ integrals are fast to compute
compute by part analytical, part numerical integration). The additional analytical effort and the numerical effort to change variable to $\Theta_{ij}$ are thus greatly rewarded, by a huge speed-up.

Unfortunately the change of variables works for all terms but $\mathbb{P}_{\rm ex}^{22}(s)$, where the more complex structure of the phase prevents this important additional simplification. Also regularisation by subtraction is very laborious to apply in this case. However, the fact that the dependence of the phase of $\mathbb{P}_{\rm ex}^{22}(s)$ on the field only arises through effective masses and is monotonic can be quite useful, as it may allow here, too, for the fast-oscillating exponential to be separated into only one Fourier integral of a function that involves slowly varying integrals.

B) Complex contour deformation generated by  $\phi_{2,4}\to\phi_{2,4}+i\epsilon/2$ and $\phi_{1,3}\to\phi_{1,3}-i\epsilon/2$ with a finite $\epsilon$. This is essentially the $i\epsilon$-prescription we used to regularize the transverse momentum integrals, except that now we let $\epsilon$ be finite instead of infinitesimal, which means $\epsilon$ potentially appears in all functions of $\phi_i$ and not just in the denominators of pre-exponential factors (like the $1/(\theta_{21}+i\epsilon)^2$ factor in~\eqref{P11GenFin}). Note, though, that this complex deformation does not affect the arguments of the step functions, which hence do not spoil the analyticity.     
If the analytic properties of the field allow for this contour deformation, then this method provides an elegant alternative to A). 

As our formulas are valid not only for pulses and periodic (e.g. monochromatic) fields but also for constant fields, method B) can also be useful in the LCF regime. A suitable choice of $\epsilon$ for the complex deformation makes the convergence of LCF terms fast, as wildly oscillating integrands are converted into fast-converging ones with only a few oscillations, making the drawing of high-resolution plots like Figs.~\ref{oneStepTriangles} and~\ref{twoStepTriangles} a short matter, even on the older-generation personal computer that we have used. When considering pulses, this deformation may not seem like a general enough method as it assumes a certain analyticity of the field, but remember that one can always approximate a given field by a suitable analytic one. For instance we could truncate the pulse's expansion in a basis of Hermite polynomials. 

For a pulse, one finds an important advantage brought by contour deformation,
apart from the great help it brings in offering a comfortable means of
regularization for all terms, $\mathbb{P}_{\rm ex}^{22}(s)$ included. We refer to the fact that it amplifies the integrand in the interaction region and greatly diminishes the ``tails'' seen in the dependence on the separations $\theta_{ij}$ from the asymptotics of the effective mass that follow us even outside the pulse. Still, the low amplitude oscillating tails stay there and their decay is not as fast as for the LCF approximation,
so for a quick and precise integration they are best separated and integrated
as such, by a mix of analytic integration and a 90 degrees complex contour
deformation, turning oscillating exponentials into decaying ones. The
intermingling of all four phase points makes applying this optimization
procedure less straightforward for $\mathbb{P}_{\rm ex}^{22}$.

If one is interested in total quantities, such as the total probability,
average energy of the pair and so on, method A) works efficiently for all terms, but $\mathbb{P}_{\rm ex}^{22}$. For $\mathbb{P}_{\rm ex}^{22}$ one may find it simpler to use a complex contour deformation of the $s$ variables. We found such a deformation, defined piecewise with several cases depending on the relative values of $\phi_{i}$, that makes $s~$integration fast, with few oscillations. One still has then to regularize the remaining singularities appearing when some of the phase points $\phi_{i}$ merge, through either subtraction or complex deformation of $\phi$.

\section{Convergence to the LCF limit}\label{Convergence to the LCF limit}

In this paper we do not present numerical results for total quantities for a pulsed plane wave, for which the methods of type A) and the one just mentioned have great potential; we reserve such investigations for a future, more numerically oriented paper. 
Here we will instead apply these numerical methods to local quantities, namely the rates, and study how the LCF approximation is reached for pulsed plane waves. 

\begin{figure}
\includegraphics[width=0.8\textwidth]{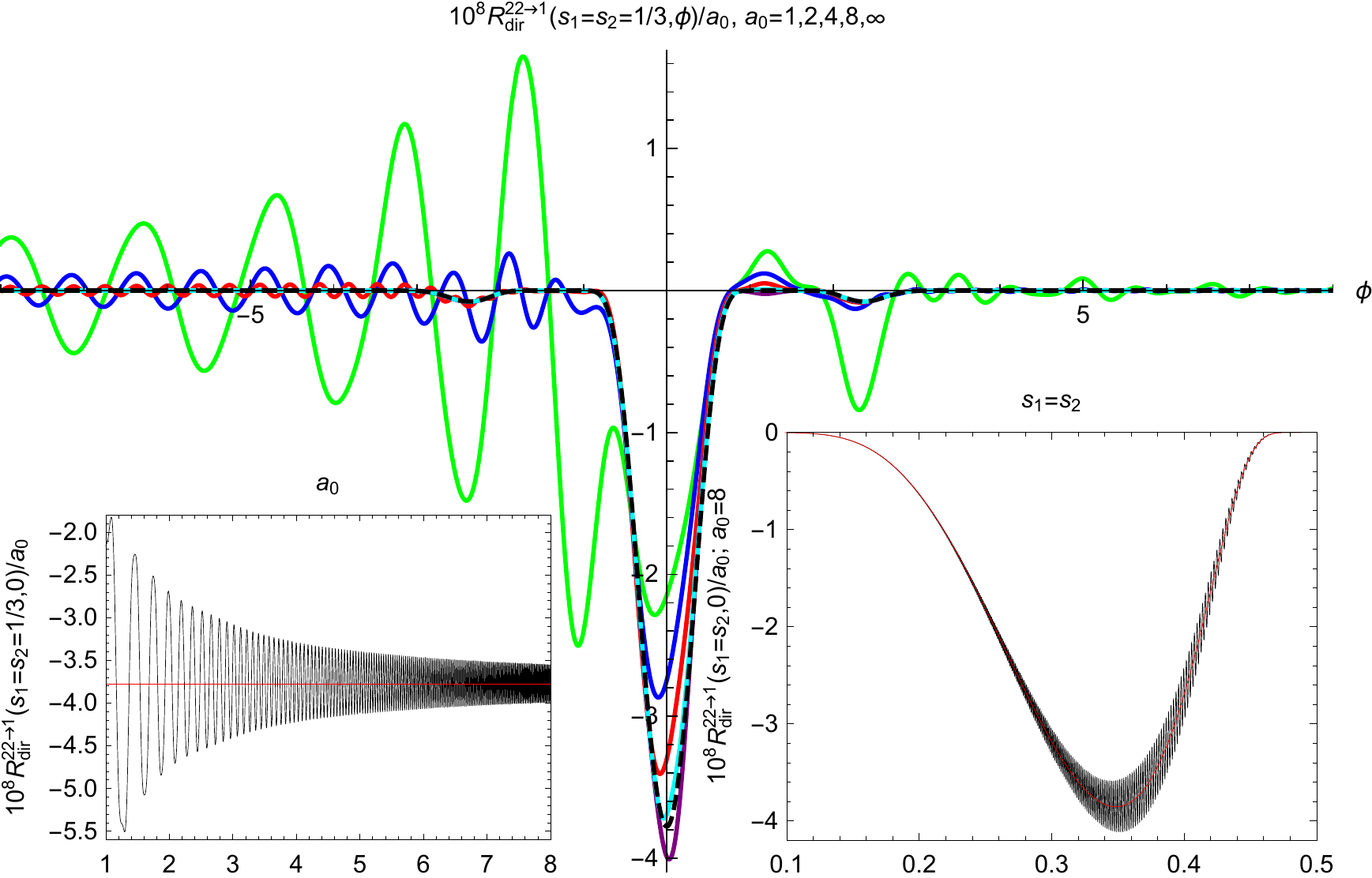}
\caption{This figure illustrates the convergence to the LCF approximation for $\mathbb{P}_{\rm dir}^{22\to1}$ for $\mathcal{T}=\pi$ and $\chi=1$.}
\label{convergencePlot22to1}
\end{figure}
\begin{figure}
\includegraphics[width=0.7\textwidth]{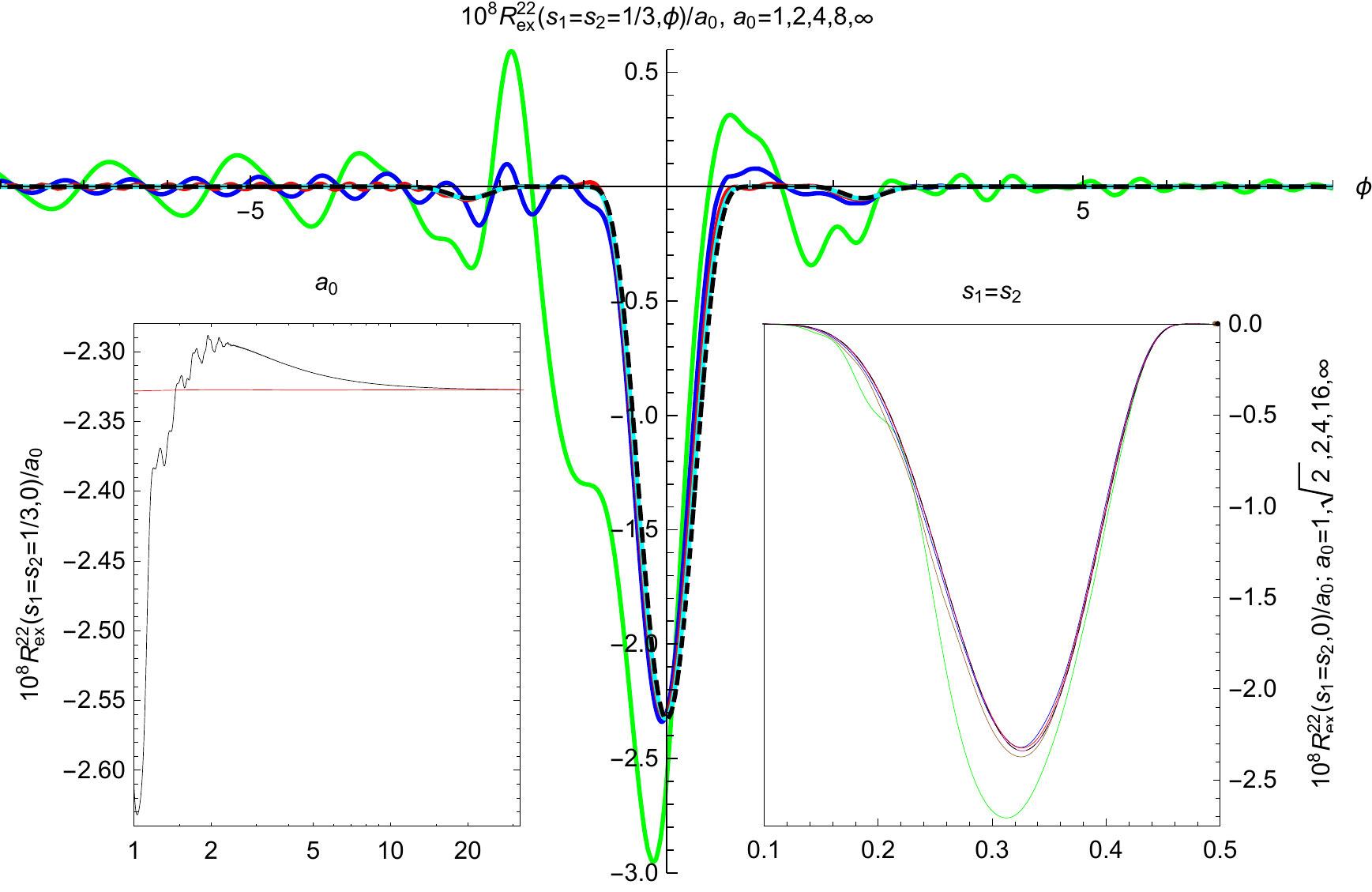}
\caption{This figure illustrates the convergence to the LCF approximation for $\mathbb{P}_{\rm ex}^{22}$ for $\mathcal{T}=\pi$ and $\chi=1$.}
\label{convergencePlot22ex}
\end{figure}
In particular, we consider a short pulse and compare the rates with their LCF
approximations, for a set of increasing $a_0$/decreasing $b_0$ values at constant $a_0b_0$. Both variations contribute to the reduction of what is commonly known as formation length and hence are expected to ensure convergence towards the $a_0=\infty$ limit provided by the LCF approximation.
We choose as pulse model a linearly polarized plane wave with Gaussian envelope:
\be
a(\phi)=a_0e^{-(\phi/\mathcal{T})^2}\sin\phi \;.
\ee
We choose an ultra-short pulse length, given by $\mathcal{T}=\pi$ or $2\pi$. The reason for this is not only that an extremely short pulse maximizes the ratio between the one-step and two-step contributions, but also that it allows us to better see tiny features in the plots detailing the convergence to the LCF limit.
In Fig.~\ref{convergencePlot22to1} and Fig.~\ref{convergencePlot22ex} we plot two contributions to the one-step rate $R_1$, namely $R_{\rm dir}^{22\to1}$ and $R_{\rm ex}^{22}$. 

\begin{figure}
\includegraphics[width=0.7\textwidth]{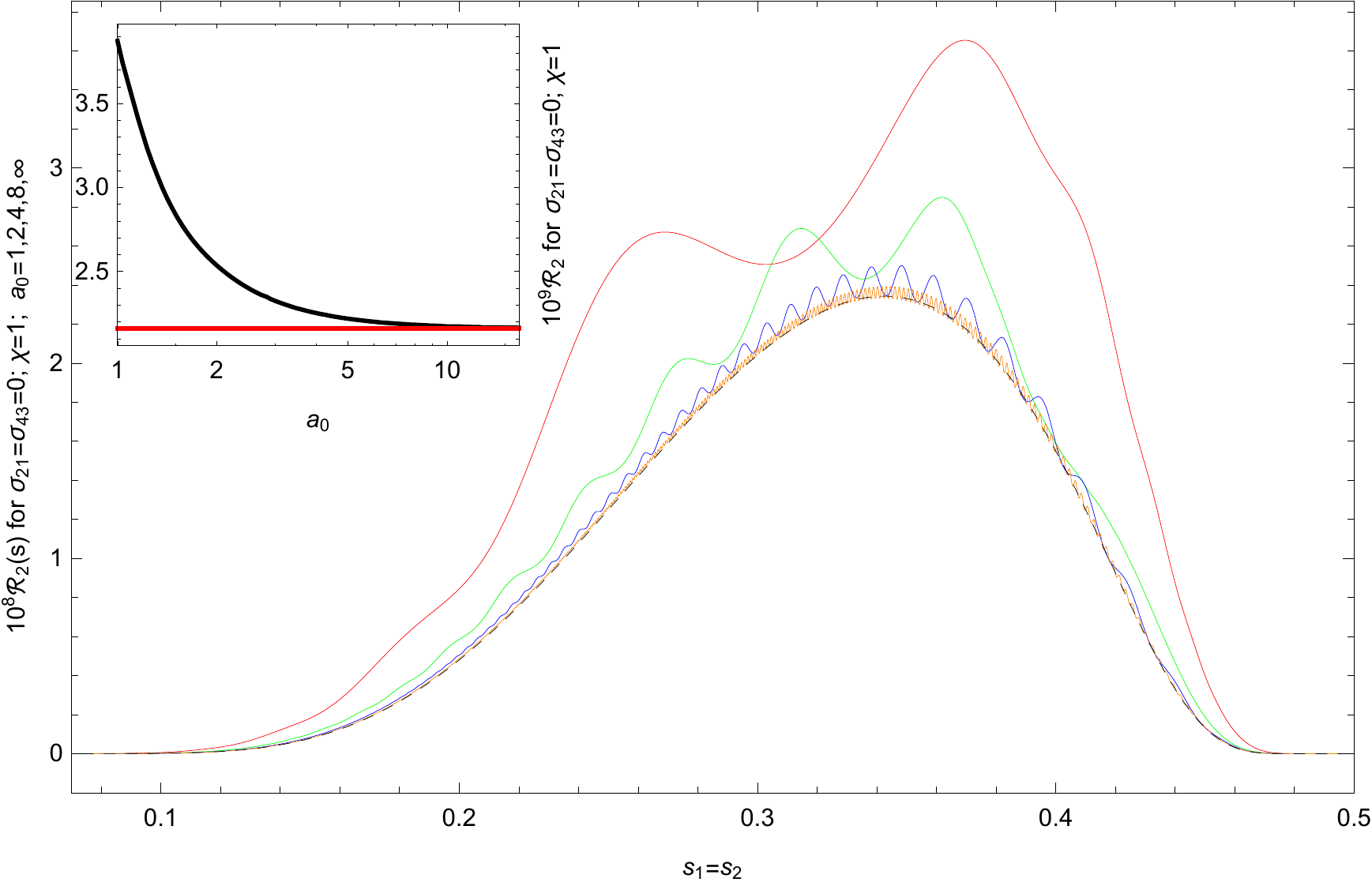}
\caption{This figure describes convergence to LCF for $\mathcal{R}_{2}$ and
$\mathcal{R}_{2}(s_1=s_2)$ at the peak of a Gaussian pulse with $\mathcal{T}=\pi$ as $a_0$ is increased at $\chi=1$. The red line in the inset plot is the LCF limit.}
\label{convergencePlotR2}
\end{figure}
\begin{figure}
\includegraphics[width=0.7\textwidth]{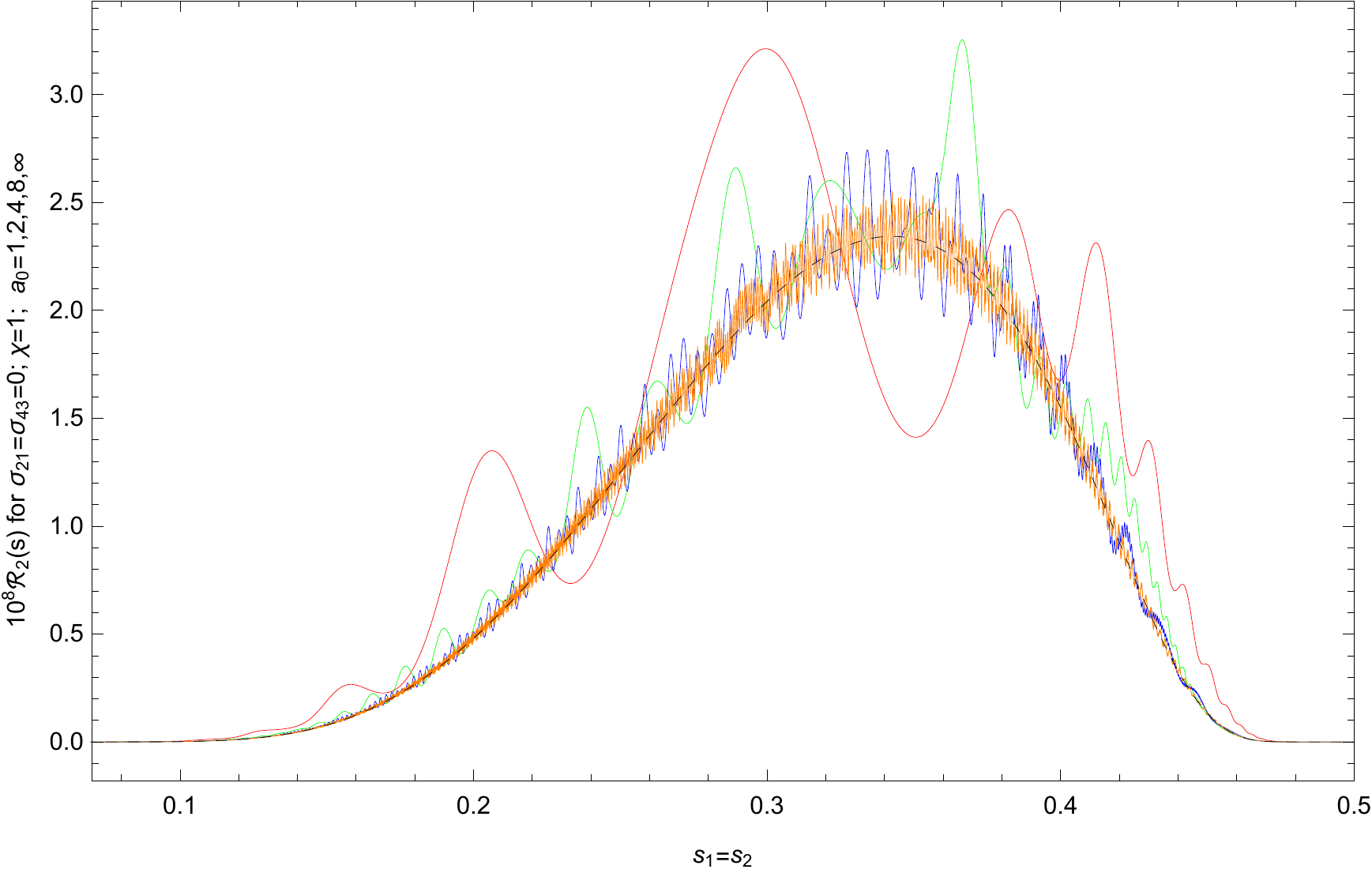}
\caption{This figure describes convergence to LCF for $\mathcal{R}_2(s_1=s_2)$ at the peak of a Gaussian pulse with $\mathcal{T}=2\pi$ as $a_0$
is increased at $\chi=1$.}
\label{beatsFigure}
\end{figure}
\begin{figure}
\includegraphics[width=0.32\textwidth,trim={20cm 7cm 5cm 10cm},clip]{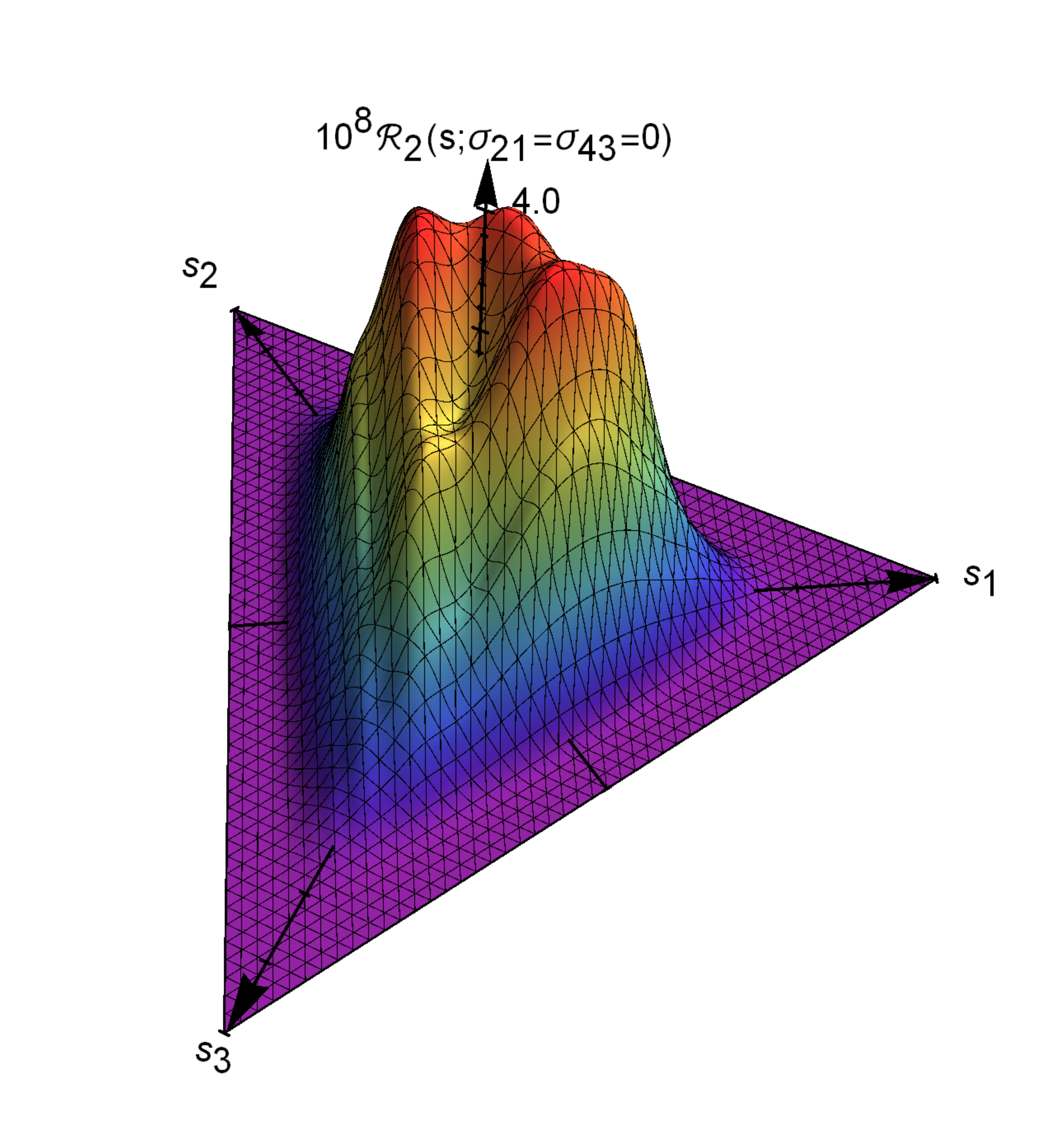}
\includegraphics[width=0.32\textwidth,trim={20cm 7cm 5cm 10cm},clip]{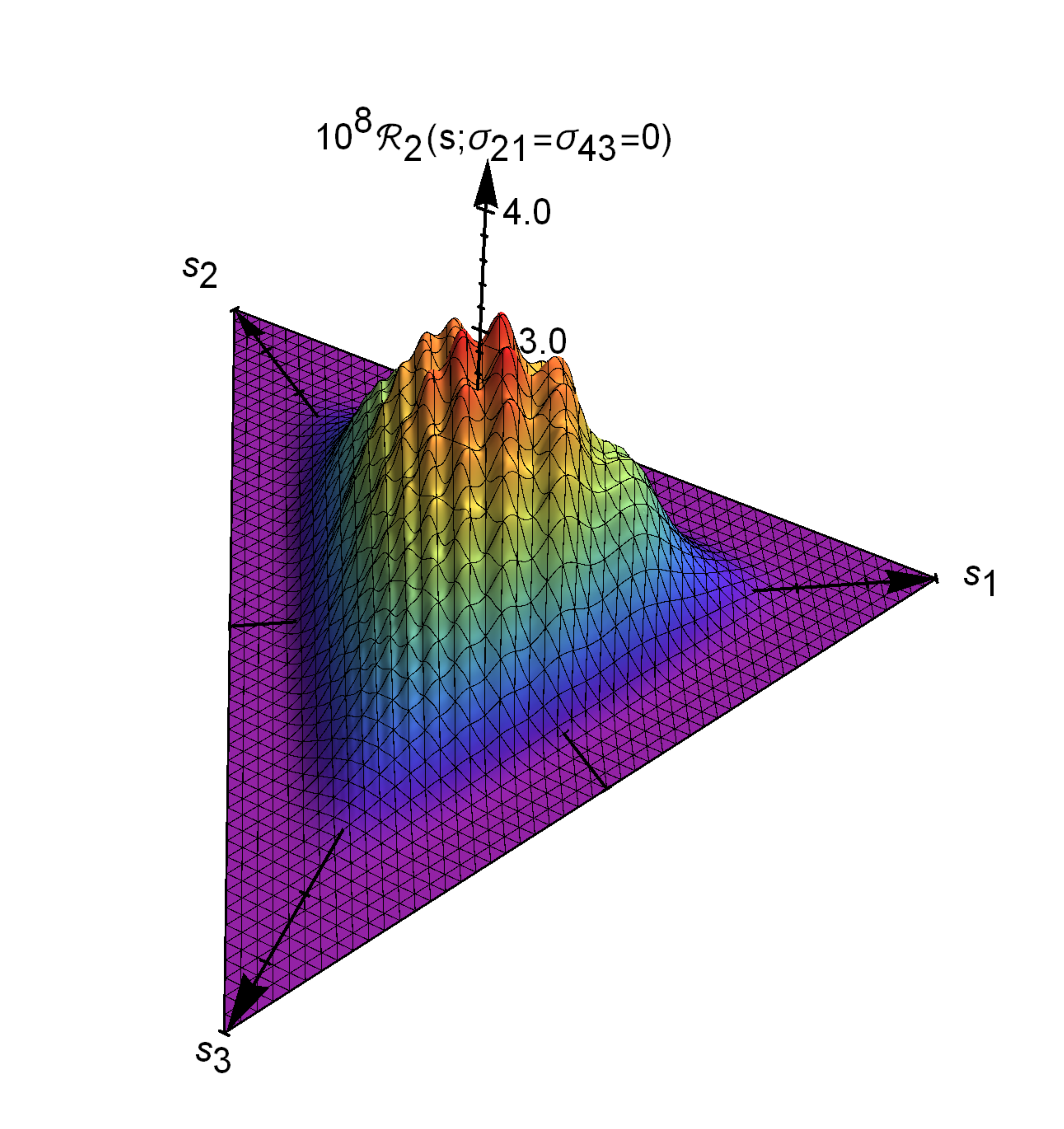}
\includegraphics[width=0.32\textwidth,trim={20cm 7cm 5cm 10cm},clip]{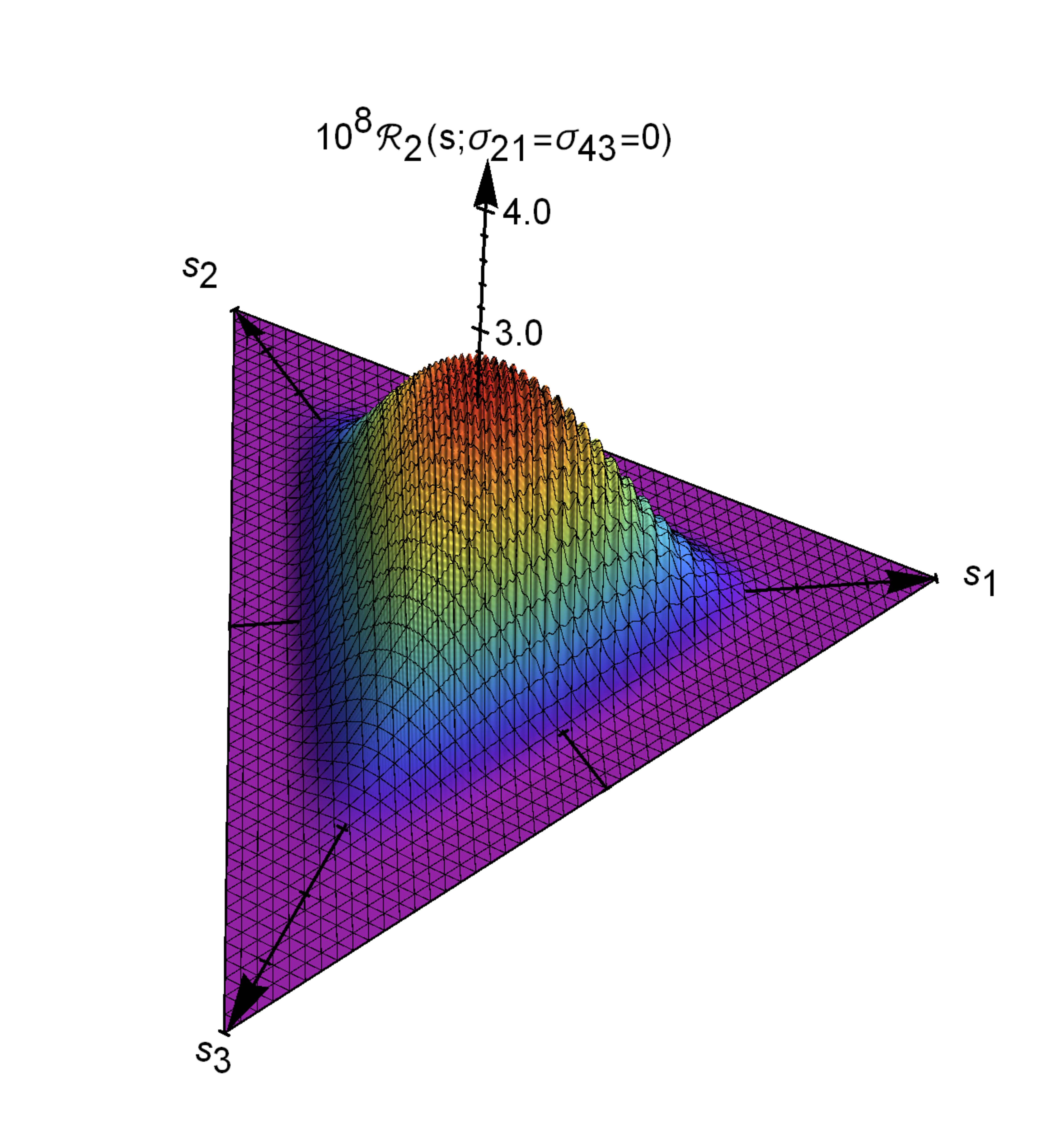}
\caption{This figure describes the dependence on $a_0$ of the $\mathcal{R}_2(s)$ distribution at the peak of a Gaussian pulse with $\mathcal{T}=\pi$ for $\chi=1$ and $a_0=1,2,4$.}
\label{triFigure}
\end{figure}
\begin{figure}
\includegraphics[width=0.3\textwidth]{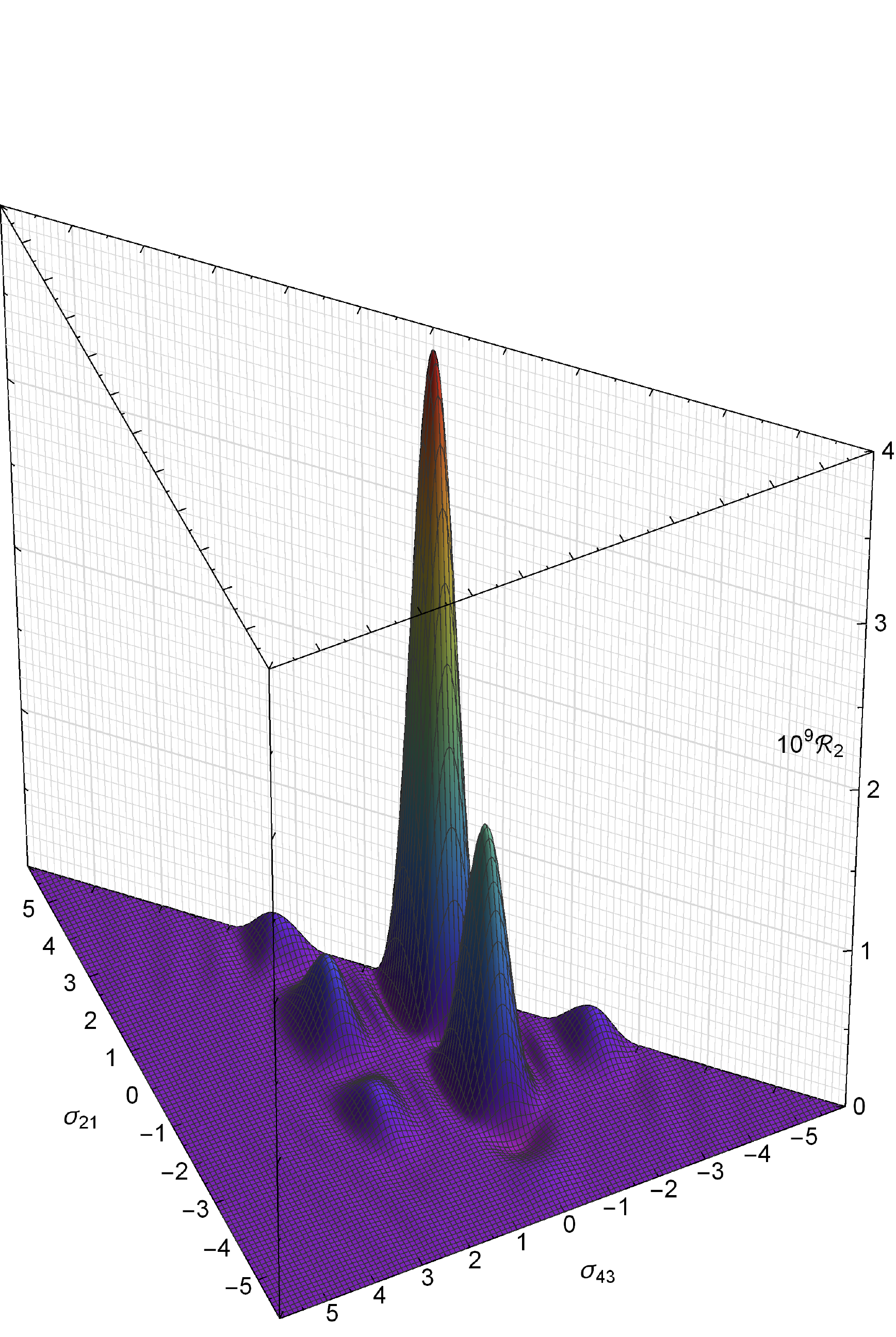}
\qquad
\includegraphics[width=0.3\textwidth]{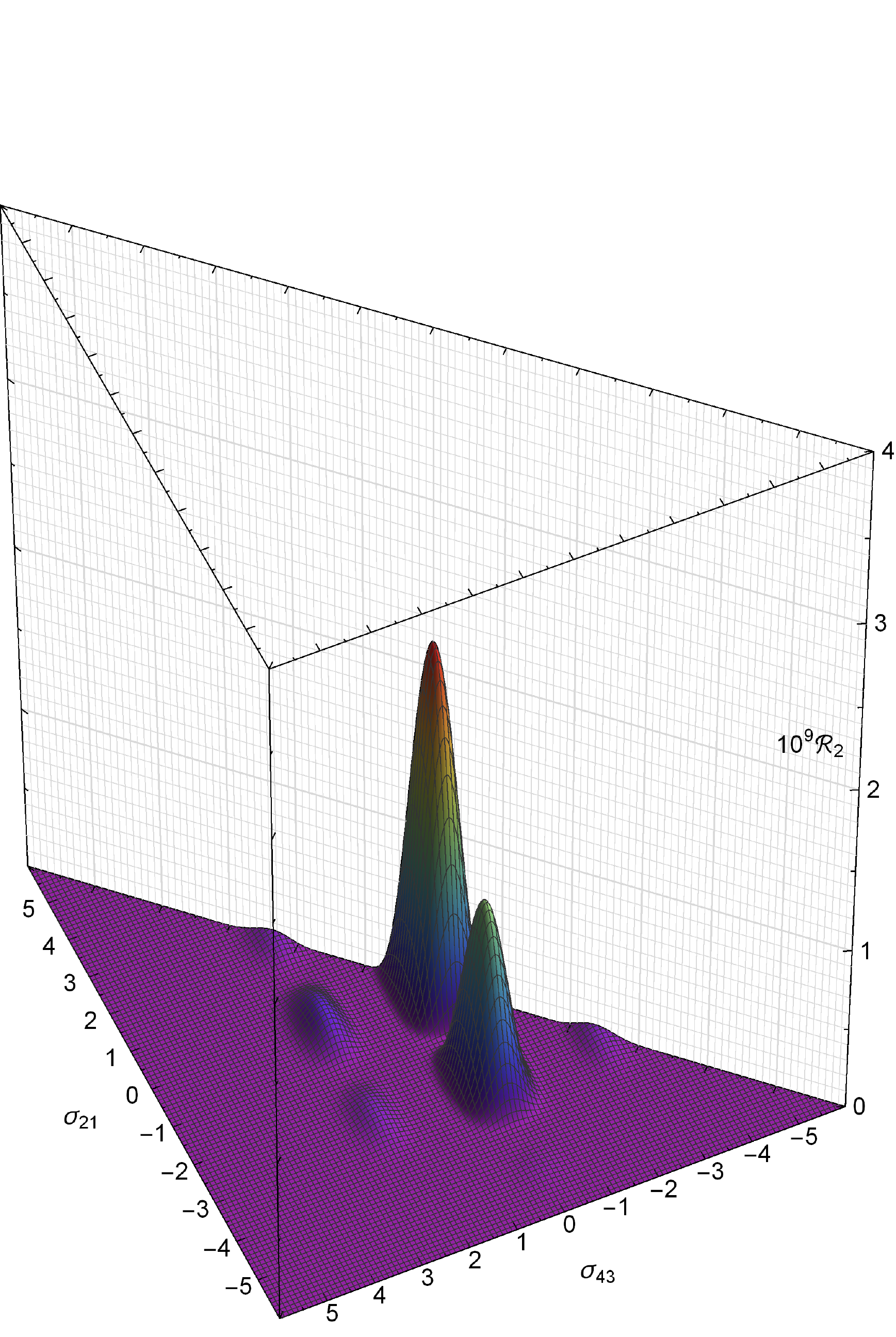}
\caption{This figure describes the dependence of the $s$-integrated $\mathcal{R}_2$ on $(\sigma_{21},\sigma_{43})  $ for a Gaussian pulse with $\mathcal{T}=\pi$,
$\chi=1$ and $a_0=1$ (left) compared to the LCF limit (right).}
\label{stalagmitesFigure}
\end{figure}
In Figs.~\ref{convergencePlotR2}, \ref{beatsFigure}, \ref{triFigure} and~\ref{stalagmitesFigure} we illustrate the effect on the rates $\mathcal{R}_2$ and $\mathcal{R}_2(s)$ produced by raising
$a_{0}$ at constant $\chi=1$, for short Gaussian pulses. We find the remarkable result that coherence may work constructively on this term, increasing the rate on average and therefore the probabilities, unlike what can be
seen in~\cite{Dinu:2013hsd}. Apart from this increase, at spectrum level, we
notice superimposed patterns of oscillations with $s$ and $a_{0}$, getting
more frequent but smaller in amplitude as $a_{0}$ increases. For $\mathcal{T}=2\pi$, the
oscillations show beats, unlike for $\mathcal{T}=\pi$, where they are regular. In the $s$-integrated rate (shown at the peak of the pulse in Fig.~\ref{convergencePlotR2}, as function of $a_{0}$, and all over the pulse in Fig.~\ref{stalagmitesFigure} for $a_{0}=1$) we notice how oscillations have been smoothed out and how convergence to LCF is, therefore,
much faster. Coherence is not constructive everywhere for the rate. In the
ripples seen between the peaks of Fig.~\ref{stalagmitesFigure} (left) there are points where the non-local rate is even slightly negative, unlike the LCF one. However, the global
effect we see is a stark increase of the probability relative to the LCF approximation as $a_{0}$ decreases towards unity and coherence increases. Another thing we notice is the asymmetry between the two processes, already noted in the LCF plot of Fig.~\ref{R2overChiChi}. The rate decreases faster with $\chi_{\rm BW}$ than with $\chi_{\rm C}$.

Conventional wisdom tells us that we should expect the LCF rates to offer good
approximations for the exact ones when the scale at which the field varies
significantly is considerably larger than a so-called coherence length. This
is expected to appear in our integrals as the size of a finite range of values
of the phase differences $\theta_{ij}$ that give a significant contribution.
We mean, for a start, those phase differences that appear in the phase factor
of the integrand and can be seen united by a fermion line in Fig.~\ref{ProbabilityDiagrams}. Of the four diagrams in Fig.~\ref{ProbabilityDiagrams}, $\mathbb{P}_{\rm dir}^{22}$ stands out as the one in which not all points are interconnected by fermion lines, so the contribution of the two pairs $\phi_1,\phi_2$ and $\phi_4,\phi_3$ is not suppressed as they move away from each other, except for the suppression due to the pulse length. However, in breaking the $\mathbb{P}_{\rm dir}^{22\to1}$ term away from $\mathbb{P}_{\rm dir}^{22}$ we have added a step function that imposes an upper limit on the range of separations $|\sigma_{43}-\sigma_{21}|$, adding $\mathbb{P}_{\rm dir}^{22\to1}$ to the list of one-step terms, for which all the phase points $\phi_i$ must be close to one another within the limits of the coherence length. This length depends on $a_0$ and $b_0$ but is unrelated to the pulse length, which constitutes a third scale, larger than the period that gives the scale for the pulse's variation. As this third scale becomes larger, $\mathbb{P}_2$ will increasingly dominate $\mathbb{P}_1$, which is why we find short pulses most interesting.

Before we explain how this suppression outside the coherence length comes into
play, let us mention that previous results tell us not to hold the
blind belief that convergence to the LCF limit must be uniform or even true
for all quantities, see e.g.~\cite{Meuren:2015mra,DiPiazza:2017raw}. For instance, \cite{Dinu:2013hsd} shows that for nonlinear Compton scattering this logic is confirmed for the average energy-momentum (as well as its higher moments) but not for the total
probability. The more important contribution of high-wavelength photons to the latter justifies our approximating the integrand to the leading-order term to its asymptotic, high $\theta_{21}$, expansion (see equation~(23) in~\cite{Dinu:2013hsd}), rather than the
low $\theta_{21}$ one that emphasizes the dominant role of small wavelengths.
For trident, however, the existence of a low energy threshold makes soft photons irrelevant, allowing for the LCF approximation to apply also at probability level.

Another issue is that, in general, when looking at totally integrated
quantities we turn the oscillating phase factor present in e.g.~\eqref{PNLC} and~\eqref{PBW} into a decaying function of $\theta_{ij}$ (not oscillating
for Compton/oscillating for BW). This is likely to make convergence to LCF much
faster than before $s$ integration.

As explained before, after proper regularizations of prefactors and
manipulations of step functions, all terms but $R_{\rm ex}^{22}(s)$ can be written as Fourier transforms of some function of one or two $\Theta_{ij}$ values. This
is achieved by a change of variable, $\theta_{ij}\to\Theta_{ij}$,
which for large $a_0$ is highly non-uniform, with sudden leaps at particular
points. At these points the function to be Fourier transformed has sharp
variation with $\Theta_{ij}$. When the frequencies $\frac{r_k}{2b_0}$ are high, which happens
at constant $\chi$ as we increase $a_0$, only these points give a
significant contribution. Of the aforementioned sharp variations the most
significant is the peak one at the origin of $\Theta_{ij}$, but there are other
such points for linear polarization, see the appendix of~\cite{Dinu:2013gaa}. In
Fig.~\ref{thetaThetaFigure} we see the derivative associated with this change of variable as
function of the new variables for two very short Gaussian pulses, with $\mathcal{T}
=\pi,2\pi$ and $a_0=4$. 
\begin{figure}
\includegraphics[width=0.35\textwidth]{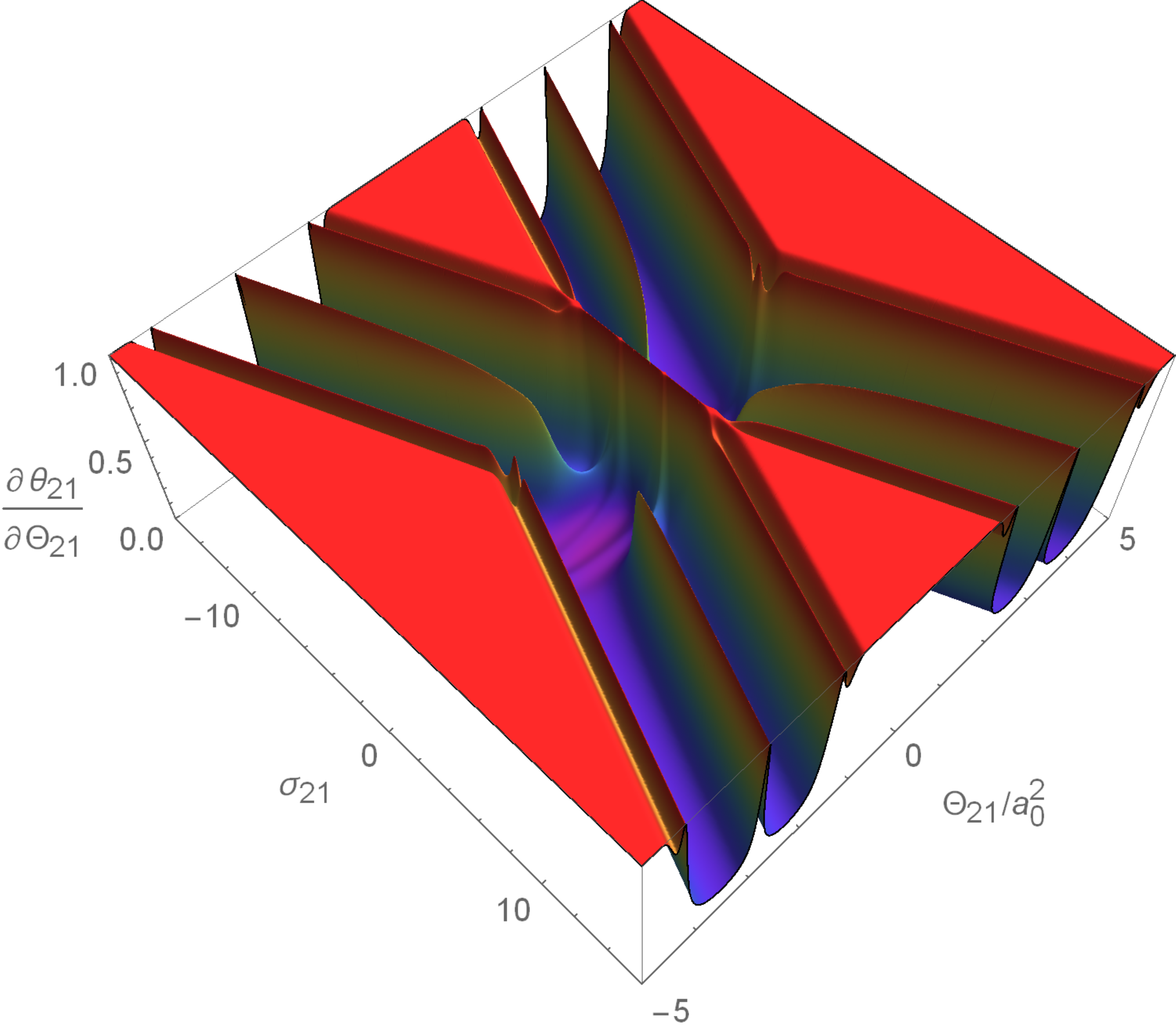}
\qquad
\includegraphics[width=0.35\textwidth]{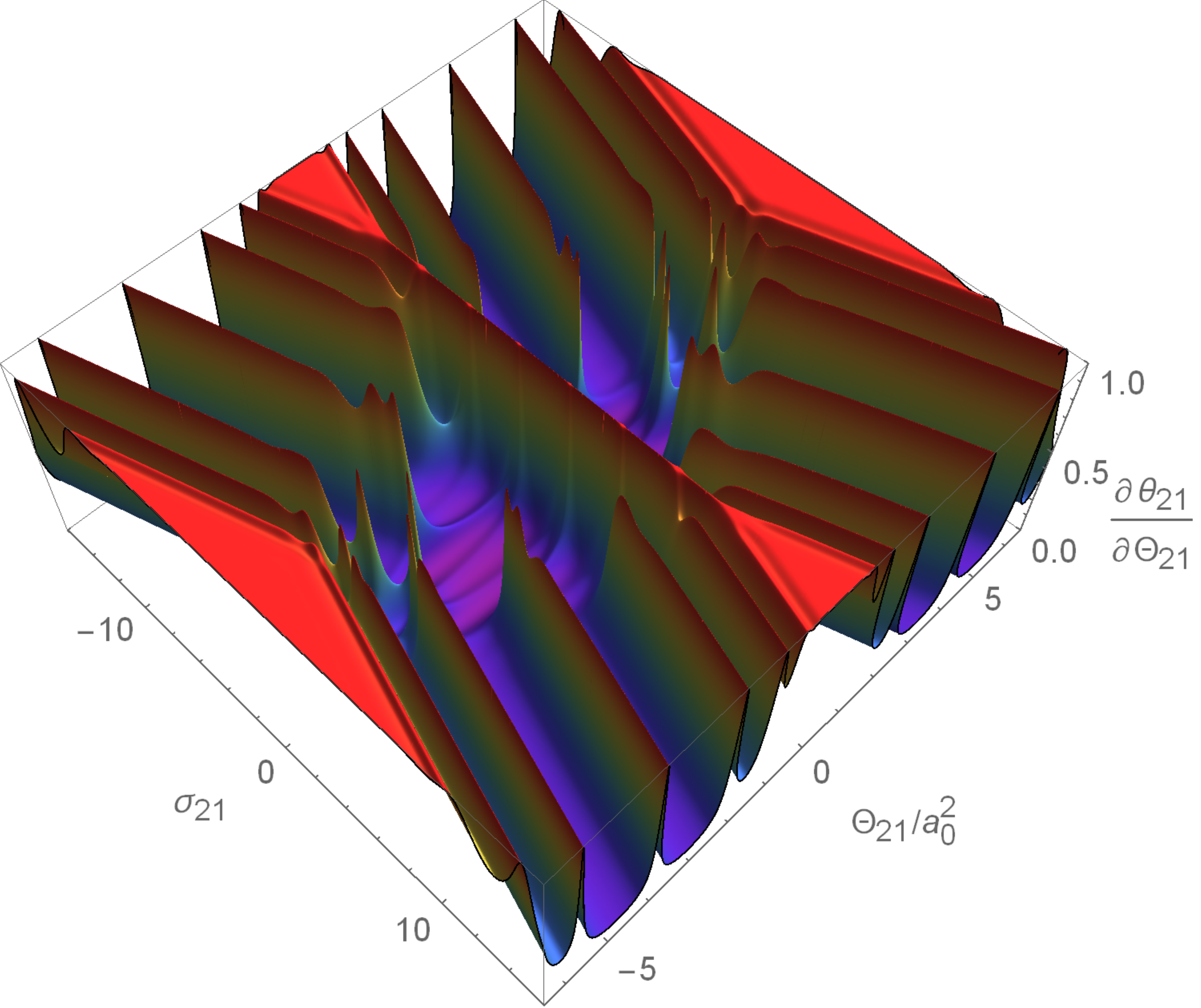}
\caption{This figure shows the derivative associated with the change of variable $\theta_{ij}\to\Theta_{ij}$ for two very short Gaussian pulses with $\mathcal{T}=\pi$ (left) and $\mathcal{T}
=2\pi$ (right), and $a_0=4$.}
\label{thetaThetaFigure}
\end{figure}
We notice a central ridge, located at at the origin of $\Theta_{ij}$ and some sharp
knife-blade-like peaks, located around the points where $a_i=a_j=\langle a\rangle _{ij}$. 
Had we studied a monochromatic field, these peaks would have formed an infinite set, extending indefinitely in both directions~\cite{Dinu:2013gaa}. For a pulse only a finite number of peaks are visible, lining up into a finite number of ridges, which increases with the pulse length. The height of the ridges between the peaks increases and eventually converges to unity as $\sigma_{ij}$ goes outside the pulse; hence for the short pulses in Fig.~\ref{thetaThetaFigure}, only a few peaks are distinguishable from the corresponding ridges.
As the pulse length increases, more and more peaks appear in both $\Theta_{ij}$ and
$\sigma_{ij}$ directions. The non-central ridges line up  into two diverging bundles, oblique to the $\sigma_{ij}$ axis, forming the X-shaped image seen in Fig.~\ref{thetaThetaFigure}. They correspond to the case where one of the pair $\phi_i$, $\phi_j$ stays outside the pulse and the other, say $\phi_j$, is inside the pulse but the condition $a_{j}=\langle a\rangle_{ij}=0$ is approximately fulfilled. Thus, the number of ridges equals the number of zeros of $a_j$ in the pulse and, therefore, increases linearly with pulse length.

When both phase points fall outside the pulse, $\partial\Theta
_{ij}/\partial\theta_{ij}$ forms a plateau at unity, into which the central ridge merges as $\sigma_{ij}$ leaves the pulse. The other ridges
extend to infinity ($\Theta_{ij}\rightarrow\pm\infty$ when $\left\vert
\sigma_{ij}\right\vert \rightarrow\infty$). 
Of course, we must not
forget that we still have to multiply with the initial integrand, which will
bring asymptotic decay in both directions, through its pre-exponential factor.
If we increase $a_{0}$ we notice that, for a given
$\sigma_{ij}$, the comb-like dependence on $\Theta_{ij}$ spreads away (it has
to be scaled by $1/a_{0}^{2}$ to keep the plot about the same size) and becomes extremely sharp at the tip in the center of the bundle, even
for moderately large $a_{0}$. Taking into account just the central ridge gives
us the semi-classical LCF approximation. Including the effect of the other
features will add oscillatory terms, as seen in e.g. Fig.~\ref{convergencePlot22to1} and Fig.~\ref{convergencePlot22ex}. One
notices two types of oscillations in the rates: a) those at fixed $\phi$ or $(\sigma_{21},\sigma_{43})$ when $s$ or $a_0$ is varied, b) those at fixed $s$ and $a_0$ as $\phi$ or $(\sigma_{21},\sigma_{43})$ is varied.

Say the pulse has finite length, lasting for $\phi\in(-L/2,$ $L/2)$
and let $c_{2}^{\infty}=\int_{-\infty}^{\infty}a^{2}$. The bundles are
located between the intervals $\pm\Theta_{ij}\in$ $\left(  2\left\vert
\sigma_{ij}\right\vert -L,2\left\vert \sigma_{ij}\right\vert +L+c_{2}^{\infty
}\right)  $, as $\sigma_{ij}$ moves away from the pulse. They are at the origin of the regularly oscillating ``tail'' seen in the rates as their
arguments, $\phi$ or $\left(  \sigma_{21},\sigma_{43}\right)  $, distance
themselves from the pulse center. This tail's amplitude exhibits polynomial
decay, introduced by the pre-exponential factor combined with the ridges'
obliqueness. In addition, we notice localized type a) oscillations, coming
from the peaks, such as the one at the origin of $\phi$, seen in Fig.~\ref{convergencePlot22to1} or~\ref{convergencePlotR2}.

All oscillations described decrease in amplitude as their frequency increases
with $a_{0}$, since in unscaled coordinates their $\Theta_{ij}$ position moves
away from the central ridge. This is mathematically interesting, as the LCF
limit is reached in a non-uniform way, and thus convergence may not translate
to the rate's derivatives.

For a pulse, superimposed (same $\sigma_{ij}$) peaks and ridges  will
contribute to the rate with terms of different frequencies, adding up to
complicated oscillations in the spectrum after $\sigma_{ij}$ integration, cf.~\cite{Meuren:2015mra}. Compare the regular
oscillations in Fig.~\ref{convergencePlotR2}, ($T=\pi$, one peak at $\sigma_{ij}=0$)
with the beats exhibited in Fig.~\ref{beatsFigure} ($T=2\pi$, two peaks). When considering a
totally integrated quantity or less sharply defined momenta, the non-local oscillations of the rates wash out. In Fig.~\ref{stalagmitesFigure} we have after $s$-integration only smooth ripples left to remind us of the non-local, not always positive, character of the rate $R_{2}$.

In conclusion, as soon as we increase $a_0$ above
unity to even a moderately high value, we expect LCF to work better on
average, but not at spectrum level. The transition from the fully coherent
regime of $a_0=b_0=1$ to LCF passes by an intermediate regime, where
quantum correlations generate important oscillatory terms in the rates and
spectra, stemming from resonant structures originating in the effective mass.
This is true for all terms but $\mathbb{P}_{\rm ex}^{22}(s)$, where the oscillations with $s$ or $a_0$ are much reduced even at spectrum level, due to the way the
exponent mixes up $s$-dependent quantities and effective masses, so the
integrand cannot be just written as a linear combination of Fourier transforms
of $s$-independent terms with $s$-dependent frequencies.

If instead of keeping $\chi$ constant, we increase the frequencies
$\frac{r_k}{2b_0}$ at constant $a_0$, the result decays exponentially as
the corresponding period decreases below the minimum ``coarseness'' scale at
which the prefactor expressed in the variables $\Theta_{ij}$ varies
significantly. This makes the distribution concentrate near the central point
$s_1=s_2=s_3$ and decay exponentially as a whole with the decrease of
$b_0$, as seen in the LCF plots and the low $\chi$ approximations.

A useful step for the $\sigma_{ij}$/$\phi$ integration of terms in the
coherent regime would be to separate oscillatory parts that extend outside a
finite pulse. This can be done noticing that they arise from the integration
regions where at least one of the points is outside the pulse. As the simplest
example, for a 2D term like $\mathbb{P}_{\rm C}(s)$, it is natural to perform a piecewise
change of variable from $\theta_{ij}$ to the one of $\phi_i$ or $\phi_j$ that is inside the pulse. Say $\phi_i$ is inside the pulse. Then the phase factor looks like
\be
\exp\left\{i\frac{r_{k}}{2b_{0}}\left[  2\left(  \phi_{i}-\sigma_{ij}\right)
+\int_{-\infty}^{\phi_{i}}a^{2}\left(\phi\right)\ud\phi-\frac{1}{2\left(
\phi_{i}-\sigma_{ij}\right)}\left(
\int_{-\infty}^{\phi_{i}}a\left(  \phi\right)\ud\phi\right)^2\right]\right\}
\ee
A similar procedure can be applied to the regularized pre-exponential factor
and, at large $\sigma$ distances, the condition $|\sigma|\gg\mathcal{T}>|\phi_i|$ allows us to write the integral as an expansion in powers of $1/\sigma_{ij}$ times oscillations of frequency $\frac{r_k}{b_0}$.

\section{Conclusions}\label{Conclusions section}

We have studied the trident process in pulsed, constant and oscillating plane wave backgrounds. We have derived compact expressions for the exact probability for general pulse shapes. We have used these expressions to obtain various analytical approximations that go beyond the locally constant field (LCF) approximation. The formulas presented in this paper also offer a great numerical advantage due to the reduction of the number of successive quadratures needed, achieved through partial analytical integration. Their simple analytic structure thus not only allows for a more insightful view of the process and for an easy comparison between its components and with their various approximations, but also provides a good starting point for applying efficient methods to reduce the numerical complexity, as we have explained.

The trident probability in a plane wave is separated into direct and exchange terms as well as into terms characterized by the number of lightfront time ($x^\LCp$) integrals. For a constant field, such $x^\LCp$-integrals give volume factors, $\Delta x^\LCp$, and then the terms proportional to $(\Delta x^\LCp)^2$ are referred to as two-step terms, while the ones proportional to $\Delta x^\LCp$ are referred to as one-step terms~\cite{Ritus:1972nf,Baier,King:2013osa}. For general field shapes, our lightfront separation of the probability leads to three direct and three exchange terms, which have either two, three or four $x^\LCp$-integrals. These terms come from squaring an amplitude with two $x^\LCp$-integrals and with a photon propagator in lightfront gauge that leads to one term with $\theta(x_2^\LCp-x_1^\LCp)$ and another one with $\delta(x_2^\LCp-x_1^\LCp)$. In the lightfront quantization formalism, the term with $\delta(x_2^\LCp-x_1^\LCp)$ comes from an ``instantaneous'' term in the lightfront Hamiltonian. The word ``instantaneous'' might suggest that the corresponding terms in the probability should be related to the standard one-step terms, at least for constant fields. However, this is not the case; we found instead that all six lightfront terms contribute to the standard one-step term. We therefore group together the six lightfront terms into $\mathbb{P}_2$ and $\mathbb{P}_1$, where $\mathbb{P}_2$ is simply related to the product of the probabilities of nonlinear Compton scattering and Breit-Wheeler pair production. In the limit of a constant field, the direct parts of $\mathbb{P}_2$ and $\mathbb{P}_1$ agree with the literature results for the two-step and one-step terms, respectively, while our results for the exchange part are new as it has previously been neglected (see~\cite{Hu:2010ye} though).

In addition to checking our results in various limits by comparing with the literature, we have derived our results using both the Hamiltonian-based lightfront formalism as well as the standard covariant formalism. While these two formulations are expected to give the same results, this equivalence might not always be trivial or obvious, see e.g.~\cite{Mantovani:2016uxq}. So, our results provide one more explicit example of this equivalence.  

In addition to recovering previous analytical results for the direct terms for constant fields, our approach has also allowed us to go beyond these known results: We have obtained various simple analytical approximations for both the direct and the exchange terms for non-constant fields. By considering non-constant fields all the integrals are finite and well behaved, and so we avoid large volume factors $\Delta x^\LCp$. The terms that for constant fields would be proportional to $(\Delta x^\LCp)^2$ and $\Delta x^\LCp$, are instead distinguished as the terms that scale as $a_0^2$ and $a_0$, respectively, for $a_0\gg1$ and constant $\chi$. These can be seen as the first two terms in a derivative expansion, and our approach allows us to calculate higher-order corrections.     

For large $a_0$, the dominant contribution comes from the two-step term, which is simply obtained by gluing together the individual probabilities of nonlinear Compton scattering and Breit-Wheeler pair production, and then the exchange and one-step parts of the probability only give small corrections. This is of course what makes PIC codes based on three-level processes useful in describing e.g. cascades in high-intensity lasers. However, the trident process itself might be most interesting in regimes where one has to take into account corrections to the two-step part, e.g. from the exchange terms. For sufficiently large $\chi$ one might expect the exchange terms to be small. However, for small $\chi$ and large $a_0$ we have shown analytically that the exchange part is in general on the same order as the direct part of the one-step $\mathbb{P}_1$. By considering some simple field shapes we have also shown that for $\chi\ll1$ and $a_0\sim1$ the exchange part can even be on the same order of magnitude as the total probability. Further, by numerical integration using complex deformation of $x^\LCp$-integrals we have also shown that the exchange terms continue to be important for $\mathbb{P}_1$ even for quite large~$\chi$.      

We have also studied how the exact probability converges to the LCF approximation in the limit of large $a_0$. In the rate we found oscillations around the LCF approximation with decreasing amplitude but increasing frequency. 

In a follow-up paper we plan on exploring trident for more general fields and parameter regimes, in particular by applying the numerical methods described here to the total probability.     
We believe that the methods that we have used here could also be useful for other higher-order processes in strong laser fields, such as nonlinear double Compton scattering~\cite{Seipt:2012tn,Mackenroth:2012rb}.

\section{acknowledgement}

We thank Anton Ilderton for many valuable and extensive discussions. We also thank Ben King for useful discussions.
For this work V. Dinu has been supported by the 111 project 29 ELI RO financed by the Institute of Atomic Physics~A. G.~Torgrimsson is supported by the Alexander von Humboldt foundation.

\appendix
\section{Perturbative limit}\label{Perturbative limit}

While we are mostly interested in strong fields, we consider here the perturbative limit in order to check e.g. our exchange terms against the literature. We first change variables according to~\eqref{phiitophivarphithetaeta} and expand to second order in the field strength. We express the field in terms of its Fourier transform
\be
a(x^\LCp)=\int\frac{\ud\omega}{2\pi}a(\omega)e^{-i\omega x^\LCp} \;.
\ee 
At second order this gives us integrals over two Fourier frequencies, $\omega_1$ and $\omega_2$, but the $\phi$-integral gives a delta function, $\delta(\omega_1+\omega_2)$, which we use to perform one of the Fourier integrals. We then perform the $\theta$-integral with Cauchy's residue theorem (if we instead start with~\eqref{Pdg} and~\eqref{Pexch-g}, then the $\theta$-integral gives a delta function). We keep either $\omega_1$ or $\omega_2$ such that the step function that arises from the $\theta$-integral can be expressed as $\theta(2pk_\omega-r_1-r_2)$, where $k_\omega^\mu=\omega k^\mu/k_\LCp$. We now go to the rest frame of the initial electron, since the literature results we are about to compare with are written in that frame. We also assume frequencies close to the threshold, i.e. $0<\omega-4\ll1$, since this leads to simple analytical results. We expand the integrand to leading order and perform the $\varphi$ and $\eta$ integrals. The step function $\theta(2\omega-r_1-r_2)$ restricts the momentum variables to be close to $s_1=s_2=1/3$.
We thus find
\be
\{\mathbb{P}_{\rm dir},\mathbb{P}_{\rm ex},\mathbb{P}_{\rm dir}+\mathbb{P}_{\rm ex}\}=\alpha^2\left\{\frac{7}{2\!\cdot\!3^4\sqrt{3}},-\frac{1}{27\sqrt{3}},\frac{1}{2\!\cdot\!3^4\sqrt{3}}\right\}\int_4\frac{\ud\omega}{2\pi}|a(\omega)|^2(\omega-4)^2 \;,
\ee
where the Fourier transform is assumed to restrict the integral to $\omega-4\ll1$. Note that the exchange term again gives a negative contribution. The direct part is a factor of $7$ times larger than the sum, i.e. $\mathbb{P}_{\rm dir}=7(\mathbb{P}_{\rm dir}+\mathbb{P}_{\rm ex})$, which agrees exactly with what is stated in~\cite{JosephRohrlich} for the cross section for single-photon trident. To recover also the overall coefficient for the cross section, $\sigma$, we replace the field according to (recall that a factor of $e$ has been absorbed in our definition of the field) $a(\omega)\to e\epsilon_\mu2\pi\delta(\omega-\omega')/\sqrt{2\omega V_3}$ and divide the probability by the temporal volume and the initial flux density, which is given by $1/V_3$ in this case. We obtain
\be
\sigma=\frac{\pi\alpha^3}{4\!\cdot\!3^4\sqrt{3}}(\omega-4)^2 \;,
\ee  
which is exactly the result in~\cite{Votruba,Mork,JauchRohrlichBook}.

\section{Feynman gauge}\label{Feynman gauge section}

In the Feynman gauge we have $D_{\nu\mu}=g_{\nu\mu}$ and from~\eqref{S12} we find
\be\label{FeynmanM12}
M^{12}=-i\pi\alpha\int\!\ud\phi_x\ud\phi_y\frac{1}{kl}g_{\mu\nu}\theta(\phi_y-\phi_x) e^{-\frac{il_\LCp}{k_\LCp}(\phi_y-\phi_x)}\overbrace{\underset{p_2}{\bar{\psi}}\gamma^\mu\underset{p_3}{\psi_\LCm}}^{\phi_y}\overbrace{\underset{p_1}{\bar{\psi}}\gamma^\nu\underset{p}{\psi}}^{\phi_x} \;,
\ee
where now $l_\LCp=l_\LCperp^2/4l_\LCm$. 
From~\eqref{FeynmanM12} we find the direct contribution as
\be\label{Pdg}
\begin{split}
\mathbb{P}_{\rm dir}=\frac{\alpha^2\pi^2}{4kp}&\int\ud\tilde{p}_1\ud\tilde{p}_2\frac{\theta(kp_3)}{kp_3}\frac{1}{kl^2}\int\ud\phi_{1234}\theta(\theta_{42})\theta(\theta_{31})\exp\frac{i}{k_\LCp}\Big[\int_{\phi_3}^{\phi_4}(\underset{p_2}{\pi_\LCp}-\underset{-p_3}{\pi_\LCp}-l_\LCp)+\int_{\phi_1}^{\phi_2}(\underset{p_1}{\pi_\LCp}-\underset{p}{\pi_\LCp}+l_\LCp)\Big]\\
&\text{Tr }(\slashed{p}_3-1)\overbrace{\underset{-p_3}{\bar{K}}\gamma^\nu\underset{p_2}{K}}^{\phi_3}(\slashed{p}_2+1)\overbrace{\underset{p_2}{\bar{K}}\gamma^\mu\underset{-p_3}{K}}^{\phi_4}\text{Tr }(\slashed{p}+1)\overbrace{\underset{p}{\bar{K}}\gamma_\nu\underset{p_1}{K}}^{\phi_1}(\slashed{p}_1+1)\overbrace{\underset{p_1}{\bar{K}}\gamma_\mu\underset{p}{K}}^{\phi_2}+(1\leftrightarrow2) \;,
\end{split}
\ee
and the exchange term as
\be\label{Pexch-g}
\begin{split}
\mathbb{P}_{\rm ex}=&-\frac{\alpha^2\pi^2}{2kp}\text{Re}\int\!\ud\tilde{p}_1\ud\tilde{p}_2\frac{\theta(kp_3)}{kp_3}\frac{1}{klkl'}\int\ud\phi_{1234}\theta(\theta_{42})\theta(\theta_{31}) \\
&\exp\frac{i}{k_\LCp}\left\{l_\LCp\theta_{31}-l'_\LCp\theta_{42}+\int_{\phi_2}^{\phi_1}\!\underset{p}{\pi_\LCp}+\int_{\phi_1}^{\phi_4}\underset{p_1}{\pi_\LCp}+\int_{\phi_3}^{\phi_2}\underset{p_2}{\pi_\LCp}+\int_{\phi_4}^{\phi_3}\underset{-p_3}{\pi_\LCp}\right\} \\
&\text{Tr }(\slashed{p}_3-1)\overbrace{\underset{-p_3}{\bar{K}}\gamma^\mu\underset{p_2}{K}}^{\phi_3}(\slashed{p}_2+1)\overbrace{\underset{p_2}{\bar{K}}\gamma^\nu\underset{p}{K}}^{\phi_2}(\slashed{p}+1)\overbrace{\underset{p}{\bar{K}}\gamma_\mu\underset{p_1}{K}}^{\phi_1}(\slashed{p}_1+1)\overbrace{\underset{p_1}{\bar{K}}\gamma_\nu\underset{-p_3}{K}}^{\phi_4} \;,
\end{split}
\ee
where $l_\mu=(p-p_1+c_1k)_\mu$ and $l'_\mu=(p-p_2+c_2k)_\mu$, with $c_{1,2}$ such that $l^2=l'^2=0$.
Note that~\eqref{Pdg} and~\eqref{Pexch-g} give the total direct and exchange parts of the probability, i.e. they have not been separated into one-step and two-step parts, and compared with the lightfront terms we have $\eqref{Pdg}=\mathbb{P}_{\rm dir}^{11}+\mathbb{P}_{\rm dir}^{12}+\mathbb{P}_{\rm dir}^{22}$ and $\eqref{Pexch-g}=\mathbb{P}_{\rm ex}^{11}+\mathbb{P}_{\rm ex}^{12}+\mathbb{P}_{\rm ex}^{22}$. Although our focus is on expressions derived using the lightfront separation, we have found \eqref{Pdg} and~\eqref{Pexch-g} useful for checking various analytical approximations, as explained in the main text.

\section{Comparing numerics with analytical expansion}\label{Comparing numerical and analytical}

In this section we compare our numerical and analytical results in LCF for $\chi\ll1$. To illustrate the accuracy of our numerical method as well as the convergence of our analytical approximations, we include here higher-order corrections to the leading orders in Sec.~\ref{constant field and low chi},
\be
\begin{split}
\mathbb{P}_{\rm dir}^{22\to1}&=-\frac{\alpha^2 a_0\Delta\phi\sqrt{\chi}}{16\sqrt{6\pi}}\left(1+\frac{233\chi}{10368}-\frac{7838317\chi^2}{71663616}+\frac{8759558921\chi^3}{82556485632}-\frac{42089593753511\chi^4}{380420285792256}\right)\exp\left(-\frac{16}{3\chi}\right) \\
&\approx-\frac{\alpha^2 a_0\Delta\phi\sqrt{\chi}}{16\sqrt{6\pi}}\left(1+0.02\chi-0.1\chi^2+0.1\chi^3-0.1\chi^4\right)\exp\left(-\frac{16}{3\chi}\right)
\end{split}
\ee
\be
\begin{split}
\mathbb{P}_{\rm dir}^{11}&=2\frac{\alpha^2 a_0\Delta\phi\chi^\frac{3}{2}}{384\sqrt{6\pi}}\left(1-\frac{217\chi}{384}+\frac{15473\chi^2}{32768}-\frac{177928745\chi^3}{339738624}\right)\exp\left(-\frac{16}{3\chi}\right) \\
&\approx 2\frac{\alpha^2 a_0\Delta\phi\chi^\frac{3}{2}}{384\sqrt{6\pi}}\left(1-0.6\chi+0.5\chi^2-0.5\chi^3\right)\exp\left(-\frac{16}{3\chi}\right)
\end{split}
\ee
\be
\begin{split}
\mathbb{P}_{\rm ex}^{22}=&-\frac{13\alpha^2 a_0\Delta\phi\sqrt{\chi}}{288\sqrt{6\pi}}\exp\left(-\frac{16}{3\chi}\right) \\
&\left(1-\frac{224569\chi}{524160}+\frac{697830139\chi^2}{3065610240}-\frac{32602208003792677\chi^3}{192778521028853760}+\frac{3793200723359191955933\chi^4}{20431438772722036899840}\right) \\
\approx&-\frac{13\alpha^2 a_0\Delta\phi\sqrt{\chi}}{288\sqrt{6\pi}}(1-0.4\chi+0.2\chi^2-0.2\chi^3+0.2\chi^4)\exp\left(-\frac{16}{3\chi}\right)
\end{split}
\ee
\be\label{Pex12higherOrders}
\begin{split}
\mathbb{P}_{\rm ex}^{12}=&-\frac{7\alpha^2 a_0\Delta\phi\chi^\frac{3}{2}}{1728\sqrt{6\pi}}\exp\left(-\frac{16}{3\chi}\right) \\
&\left(1-\frac{46471\chi}{24192}+\frac{177397141\chi^2}{55738368}-
\frac{351490158181\chi^3}{64210599936}+\frac{27085721944641151\chi^4}{2662942000545792}-\frac{63206521085663491937\chi^5}{3067709184628752384}\right) \\
\approx&-\frac{7\alpha^2 a_0\Delta\phi\chi^\frac{3}{2}}{1728\sqrt{6\pi}}(1-1.9\chi+3.2\chi^2-5.5\chi^3+10.2\chi^4-20.6\chi^5)\exp\left(-\frac{16}{3\chi}\right)
\end{split}
\ee
\be
\begin{split}
\mathbb{P}_{\rm ex}^{11}&=-\frac{\alpha^2 a_0\Delta\phi\chi^\frac{3}{2}}{384\sqrt{6\pi}}\left(1-\frac{313\chi}{384}+\frac{70547\chi^2}{98304}-\frac{249364553\chi^3}{339738624}\right)\exp\left(-\frac{16}{3\chi}\right) \\
&\approx-\frac{\alpha^2 a_0\Delta\phi\chi^\frac{3}{2}}{384\sqrt{6\pi}}(1-0.8\chi+0.7\chi^2-0.7\chi^3)\exp\left(-\frac{16}{3\chi}\right) \;.
\end{split}
\ee
In~\eqref{Pex12higherOrders} we have included two orders more than for the other terms; these extra terms have not been used when comparing with the numerics, but have been included in order to illustrate the growth of the series coefficients for this term.
\begin{figure}
\includegraphics[width=0.45\textwidth]{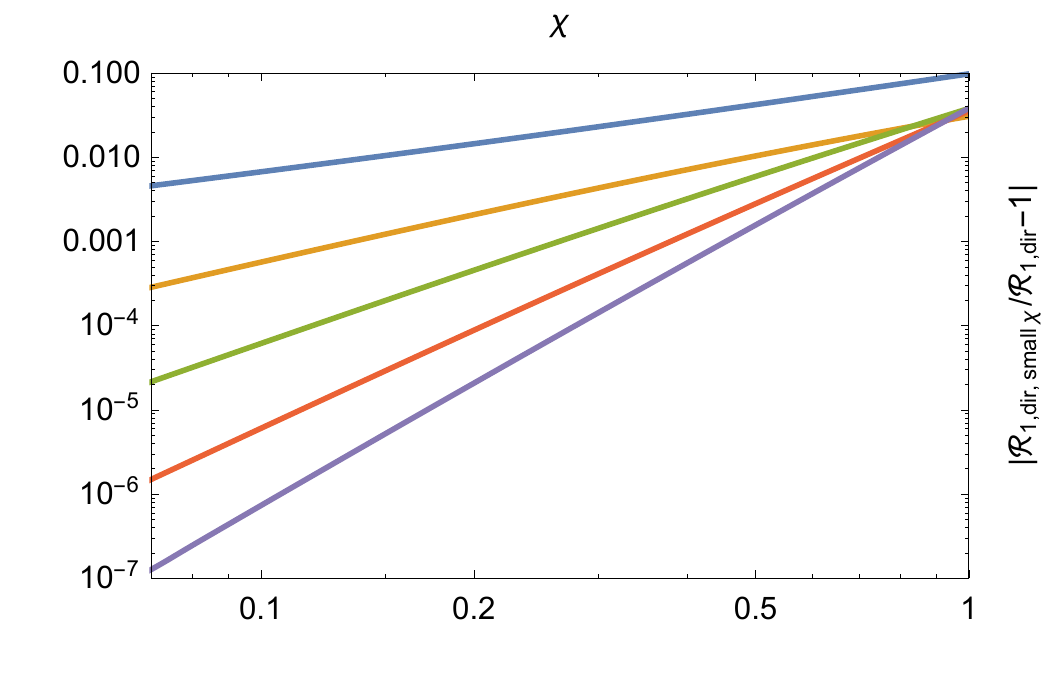}
\includegraphics[width=0.45\textwidth]{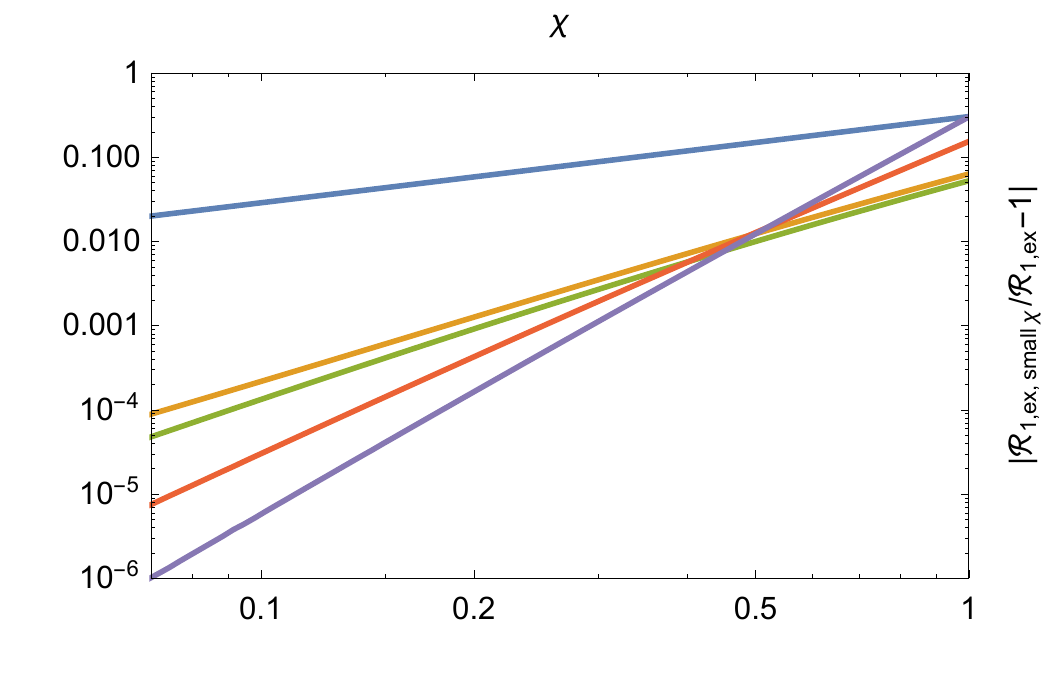}
\caption{These figures show the relative difference between our numerical results and analytical approximations with one (blue), two (orange), three (green), four (red) and five (purple) terms in the $\chi\ll1$ expansion.}
\label{comparingNumAndAna}
\end{figure}
In Fig.~\ref{comparingNumAndAna} we have compared these approximations with our numerical results. These plots show that adding higher orders improves the approximation, and also demonstrate the accuracy of our numerical method. For the (total) direct term, the approximations are good all the way up to $\chi\sim1$ and even by only including the first two orders: At $\chi=1$ the relative error is still only $|\mathcal{R}_{1,\text{dir,small }\chi}/\mathcal{R}_{1,\text{dir}}-1|=0.1,0.03$ for the leading order and the leading order plus the next-to-leading-order correction, respectively. The corresponding values for the exchange term are $|\mathcal{R}_{1,\text{ex,small }\chi}/\mathcal{R}_{1,\text{ex}}-1|=0.3,0.06$. The higher-order corrections for the exchange terms intersect at $\chi\sim0.5$ where the relative error is $|\mathcal{R}_{1,\text{ex,small }\chi}/\mathcal{R}_{1,\text{ex}}-1|\approx0.01$. We conclude that even the leading-order approximations give good order-of-magnitude estimates even for $\chi\sim1$.

\end{document}